%% file: LSST_cadence.tex
\documentclass[traditabstract]{aa} 
\usepackage{graphicx}
\usepackage{siunitx}
\usepackage{txfonts}
\usepackage{graphicx}
\usepackage{caption}
\usepackage{natbib}
\usepackage{subfigure} 
\usepackage{hyperref}
\usepackage[utf8]{inputenc}
\usepackage{placeins}
\usepackage{scrextend}

\newcommand{\baselinelike}{``\mathrm{baseline \, like}"}
\newcommand{\rolling}{``\mathrm{higher \, cadence \, \& \, fewer \, seasons}"}
\newcommand{\better}{``\mathrm{higher \, cadence}"}

\input{LSST_cadence_macro}

\begin{document}
   \title{Strongly lensed SNe Ia in the era of LSST:  observing cadence for lens discoveries and time-delay measurements}

  \titlerunning{Strongly lensed SNe Ia in the era of LSST}

   \author{S. Huber\inst{1,2}
   		 \and    
          S. H. Suyu\inst{1,2,3}
                    \and
          U. M. Noebauer \inst{1,4}  
                  \and 
          V. Bonvin \inst{5}         
               \and
                D. Rothchild \inst{6}
             \and 
             J. H. H. Chan\inst{5}
             \and  
             H. Awan \inst{7} 
             \and          
           F. Courbin\inst{5}
             \and 
             M. Kromer\inst{8,9}
             \and
             P. Marshall\inst{10}
             \and
             M. Oguri\inst{11,12,13}
             \and
             T. Ribeiro\inst{14}, 
             The
LSST Dark Energy Science Collaboration
          }

   \institute{Max-Planck-Institut f\"ur Astrophysik, Karl-Schwarzschild Str. 1, 85741 Garching, Germany\\
              \email{shuber@MPA-Garching.MPG.DE}
         \and 
           Physik-Department, Technische Universit\"at M\"unchen, James-Franck-Stra\ss{}e~1, 85748 Garching, Germany
           \and
           Institute of Astronomy and Astrophysics, Academia Sinica, 11F of ASMAB, No.1, Section 4, Roosevelt Road, Taipei 10617, Taiwan
           \and
           Munich Re, IT1.6.1.1, Königinstraße 107, 80802 München
			\and
           Institute of Physics, Laboratory of Astrophysics, Ecole Polytechnique F\'ed\'erale de Lausanne (EPFL), Observatoire de Sauverny, 1290 Versoix, Switzerland
           \and
Department of Electrical Engineering and Computer Science, University of California, 253 Cory Hall, Berkeley, CA 94720-1770, USA
            \and
Department of Physics and Astronomy, Rutgers University, Piscataway, NJ 08854, USA 
\and
           Zentrum f\"ur Astronomie der Universit\"at Heidelberg, Institut f\"ur
Theoretische Astrophysik, Philosophenweg 12, 69120, Heidelberg, Germany
			\and
			Heidelberger Institut f\"ur Theoretische Studien,
Schloss-Wolfsbrunnenweg 35, 69118 Heidelberg, Germany
\and
Kavli Institute for Particle Astrophysics and Cosmology, P.O. Box 20450, MS29, Stanford, CA 94309, USA
			\and
           Research Center for the Early Universe, University of Tokyo, 7-3-1 Hongo, Bunkyo-ku, Tokyo 113-0033, Japan
           \and
           Department of Physics, University of Tokyo, Tokyo 113-0033, Japan   
           \and
           Kavli Institute for the Physics and Mathematics of the Universe (Kavli IPMU, WPI), The University of Tokyo, Chiba 277-8582, Japan
           \and
           LSST, 933 N. Cherry Ave., Tucson, AZ 85721, USA
             }

   \date{Received --; accepted --}

 
  \abstract
  {The upcoming Large Synoptic Survey Telescope (LSST) will detect
    many strongly lensed Type Ia supernovae (LSNe Ia) for time-delay
    cosmography. This will provide an independent and direct way for
    measuring the Hubble constant $H_0$, which is necessary to address
    the current $4.4 \sigma$ tension in $H_0$ between the local
    distance ladder and the early Universe measurements. We present a detailed
    analysis of different observing strategies (also referred to as cadence strategy) for the LSST, and
    quantify their impact on time-delay 
    measurement between multiple images of LSNe Ia. For this, we simulated observations by using mock LSNe Ia for which we
    produced mock-LSST light curves that account for
    microlensing. Furthermore, we used the free-knot splines estimator from the software {\tt PyCS} to measure the time delay from the simulated
    observations. We find
    that using only LSST data for time-delay cosmography is not
    ideal. Instead, we advocate using LSST as a discovery machine for LSNe Ia, enabling time delay measurements from follow-up observations from other instruments in order to increase the number of systems by a factor of 2 to 16 depending on the observing strategy. Furthermore, we find that LSST observing strategies,
    which provide a good sampling frequency (the mean inter-night gap is around two days) and high cumulative season
    length (ten seasons with a season length of around 170 days per season), are favored. Rolling cadences subdivide the survey and focus on different parts in different years; these observing strategies trade the number of seasons for better sampling frequency. In our investigation, this leads to half the number of systems in comparison to the best observing strategy. Therefore rolling cadences are disfavored because the gain from the increased sampling frequency cannot compensate for the shortened cumulative season length.
We anticipate that the sample of lensed SNe Ia from our preferred LSST cadence strategies with rapid follow-up observations would yield an independent percent-level constraint on $H_0$.
}
   {}
   {}
   {}
   {}


   \maketitle
%

\section{Introduction}
\label{sec:intro}

The Hubble constant ($H_0$) is one of the key parameters to describe the
Universe. Current observations of the cosmic microwave background (CMB) imply $H_0 =
\SI[separate-uncertainty = true]{67.36 \pm
  0.54}{\kilo\meter\per\s\per\mega\parsec}$, assuming a flat $\Lambda$CDM cosmology and the
standard model of particle physics \citep{Planck:2018vks}. This is in tension to $H_0 = \SI[separate-uncertainty = true]{74.03
  \pm 1.42}{\kilo\meter\per\s\per\mega\parsec}$, which is measured from the local
distance ladder \citep{Riess:2016jrr,Riess:2018byc,Riess:2019cxk}. In order to
verify or refute this $4.4 \sigma$ tension, independent methods are
needed.

One such method is lensing time-delay cosmography, which can determine 
$H_0$ in a single step. The basic idea is to measure the time delays
between multiple images of a strongly lensed variable source
\citep{Refsdal:1964}. This time delay, in combination with
reconstructions of the lens mass distributions and line-of-sight mass structure, directly yields a ``time-delay distance"  which is inversely
proportional to $H_0$ (i.e., $t \propto D_{\Delta t} \propto
H_0^{-1}$). 
While the time-delay distance primarily constrains $H_0$, it also provides information about other cosmological parameters \citep[e.g.,][]{Linder:2011, JeeEtal16, ShajibEtal18, Grillo:2018ume}. 
 Applying this method to four lensed quasar systems, the
H0LiCOW collaboration\footnote{\url{http://h0licow.org}} \citep{Suyu:2016qxx} together with the
COSMOGRAIL collaboration\footnote{\url{http://cosmograil.org}}
\citep{Eigenbrod:2005ie,2017Courbin,Bonvin:2018dcc} measured $H_0 =
72.5^{+2.1}_{-2.3} \, \si{\kilo\meter\per\s\per\mega\parsec}$ in flat
$\Lambda$CDM \citep{Birrer:2018vtm}, which is in agreement with the measurements using,
a local distance ladder, but larger than CMB measurements.

Another promising approach goes back to the initial idea in
\cite{Refsdal:1964} that uses lensed supernovae (LSNe) instead of quasars
for time-delay cosmography. So far only two LSNe systems with resolved
multiple images have been observed. The first one, called SN ``Refsdal"
discovered by \cite{Kelly:2015xvu,Kelly:2015vjq}, was a 1987A-like Type
II SN, which was strongly lensed by the galaxy cluster MACS
J1149.5+222.3. As shown in \cite{Grillo:2018ume}, with SN Refsdal
one can measure $H_0$ with a $1 \sigma$ statistical
error of $7\%$. The second LSNe with resolved images is iPTF16geu
reported by \cite{Goobar:2016uuf} from the intermediate Palomar
Transient Factory (iPTF). The system is a SNe Ia at redshift
$0.409$ and strongly lensed by an intervening galaxy at a redshift of
$0.216$.  Strong lens mass models of the system from
\cite{More:2016sys} yield SN image fluxes that are discrepant with
the observations, which might be partly an effect of microlensing
\citep{Yahalomi:2017ihe,Foxley-Marrable:2018dzu,Dhawan:2019vof}. Additionally, \cite{Mortsell:2019auy} show that the flux anomalies are within stellar microlensing predictions for certain values of the slope of the projected surface density of the lens galaxy. The models in \cite{More:2016sys} and \cite{Goobar:2016uuf} also predict very short time delays ($\approx
\SI{0.5}{\day}$) that can thus be significantly biased by a microlensing time delay \citep{Bonvin:2018b}. Therefore it is important to include microlensing in LSNe studies.

Even though the number of LSNe is a factor of approximately $60$ \citep{Oguri:2010} lower than the number
of lensed quasars, there are important advantages in using LSNe when measuring time delays. First, if they are observed before the peak,
the characteristic SN light curves make time-delay measurements easier and possible
on shorter time scales in comparison to stochastically 
varying quasars. Second, supernova images fade away with time, which facilitates
measurements of lens stellar kinematics and therefore enables the
combination of dynamics 
\citep{Barnabe2011,2017:Yildirim,Shajib:2018} and lens mass modeling. This helps to overcome
degeneracies like the mass-sheet degeneracy
\citep{Falco:1985,Schneider:2013wga}. The intrinsic luminosity of the source can also be another way in avoiding mass-sheet degeneracy. Since SNe Ia are standardizable candles, LSNe Ia are very
promising in breaking the model degeneracies in two independent ways.

Even though only two LSNe with resolved images are currently known, the Large Synoptic Survey Telescope (LSST) will
play a key role in detecting many more LSNe. 
From investigations done by \cite{Oguri:2010} assuming detections based on image multiplicity, we expect to find $45$ LSNe Ia over the ten year survey. A different approach, using strong lensing magnification for detection \citep{GoldsteinNugent:2017,Goldstein:2017bny}, leads to $500-900$ LSNe Ia in ten years (see also \citealp{Quimby:2014}). The differences in the expected number of LSNe Ia arise from different assumptions about the limiting magnitude and cumulative season length, as pointed out by \cite{Wojtak:2019hsc}.
A remaining question, however, 
is how many of the detected systems are valuable for measuring time
delays and whether it will be possible to measure time delays with just the
LSST data. The LSST cadence strategy
\citep{Marshall:2017wph} will be defined soon and the goal of this 
paper is to evaluate different cadences for our science case of 
measuring time delays in LSNe Ia. For this purpose, we have
investigated 20 different observing strategies. We used mock LSNe Ia from the Oguri and Marshall (OM10) catalog \citep{Oguri:2010} to simulate
observations, and produced the light curves for the mock
SNe images based on synthetic observables calculated with Applied Radiative Transfer
In Supernovae \citep[{\tt ARTIS};][]{Kromer:2009ce} for the spherically symmetric SN Ia model W7
\citep{1984:Nomoto}. Furthermore, we employed magnifications
maps from {\tt GERLUMPH} \citep{Vernardos:2015wta} to include the effects of
microlensing, similar to the approach followed by \cite{Goldstein:2017bny}. We then simulated
data points for the light curves following the observational sequence
from different cadences and uncertainties according to the LSST
science book \citep{2009:LSSTscience}. We used the free-knot splines estimator from Python Curve Shifting \citep[{\tt PyCS};][]{2013:Tewesb,Bonvin:2015jia} to measure the time delay from the simulated observation.

The structure of the paper is as follows. In Section
\ref{sec:Microlensing on Type Ia Supernovae} we present a theoretical
calculation of microlensing on LSNe Ia. In Section \ref{sec: LSST} we introduce relevant information about LSST and different
observing strategies investigated in this work.  In Section \ref{sec:Time-Delay
  Measurements of mock LSST LSNe Ia}, mock light curves of LSNe Ia are simulated and the time-delay measurement to quantify different LSST observing strategies is described in Section \ref{sec:Time-delay measurement}. The results are presented in Section \ref{sec:results}
before we conclude in Section \ref{sec:Summary and Future Prospects}.
Throughout this paper, magnitudes are given in the AB system.


\section{Microlensing on Type Ia Supernovae}
\label{sec:Microlensing on Type Ia Supernovae}

In this section we describe the calculation of microlensed SNe Ia
light curves combining magnifications maps and a theoretical SNe Ia
model. The relevance
of microlensing on LSNe Ia has been shown theoretically by
\citet[e.g.,][]{DoblerKeeton06}, \cite{Goldstein:2017bny} and \cite{Bonvin:2018b} and, as mentioned before, the first detected LSNe Ia shows
discrepancies between models and observation which might be partly due
to microlensing \citep{More:2016sys,Yahalomi:2017ihe,Foxley-Marrable:2018dzu}. Therefore to simulate more realistic light curves of LSNe Ia we included microlensing in our studies. In Section \ref{sec: Microlensing} magnifications maps are described and Section \ref{sec:SNeIa and the 1D projection} explains the radiative transfer code \texttt{ARTIS} used to calculate synthetic observables. In addition the projection of the 3D simulation output to 1D is discussed including the geometrical delay as described by \cite{Bonvin:2018b}. In Section \ref{sec: Microlensing on SNe Ia} a comprehensive derivation of microlensed light curves of SNe Ia is presented.

\subsection{Magnification maps for microlensing}
\label{sec: Microlensing}
Microlensing is the effect of additional magnification or
demagnification caused by stars, or other compact objects with comparable properties, of the lensing galaxy. We used
magnification maps based on {\tt GERLUMPH} \citep[J.~H.~H.~Chan et al.~in
preparation]{Vernardos:2015wta} to model the effect of microlensing on a SN Ia . These maps are created using the
inverse ray-shooting technique
\citep[e.g.,][]{Kayser:1986,Wambsganss:1992,Vernardos:2013vma} and are pixellated maps
containing magnification factors $\mu$ at the source plane. The three main parameters for the
maps are the convergence $\kappa$, the shear $\gamma$, and the smooth
matter component $s$ which is defined as the ratio of the smooth
matter convergence $\kappa_s$ to the total convergence $\kappa$. For
simplicity, we assumed $s=0.6$ in our investigation. Estimated $s$ values at image positions of galaxy-scale lenses typically vary between $0.3$ and $ 0.8$ \citep[e.g.,][]{Schechter2014,Chen2018,Bonvin:2019xvn} and therefore cover a much broader range.
Nevertheless \cite{Goldstein:2017bny} investigated a few different $s$ values and found that the effect of microlensing on LSNe Ia depends more on the spatial distribution of the radiation than on the precise $s$ value. Even though we over- or underestimate the microlensing effect slightly (depending on the mock lens system) by fixing $s$ in our work, this is done in the same way for all cadence strategies investigated in this work, thus leaving the overall message unchanged. A further investigation of different $s$ values will be presented in S. Huber et al.~(in preparation). The Einstein radius
$\Rein$ is the characteristic scale of the map at the source plane, defined as
\begin{equation}
\Rein=\sqrt{\frac{4 G \langle M \rangle}{c^2} \frac{\source \ls}{\lens}}.
\label{basics: Einstein Radius physical coordinate in cm}
\end{equation}

We assume a Salpeter initial mass function (IMF) with a mean mass of
the point mass microlenses of $\langle M \rangle = 0.35
M_\odot$. Details of the IMF are not relevant for our studies (J.~H.~H.~Chan et al.~in
preparation). The angular diameter distances $\source$, $\lens$, and $\ls$ are measured from us to the source, from us to the lens, and between the
lens and the source, respectively.  If we assume a flat $\Lambda$CDM
cosmology and neglect the contribution of radiation, we can calculate
the angular diameter distance via
\begin{equation}
\scalebox{1.2}{$\angular = \frac{c}{H_0 (1+z_2)} \int_{z_1}^{z_2} \frac{\dd z}{\sqrt{\matter (1+z)^3+ \vac }}.$}
\end{equation}

\begin{figure}
\centering
\includegraphics[scale=0.65]{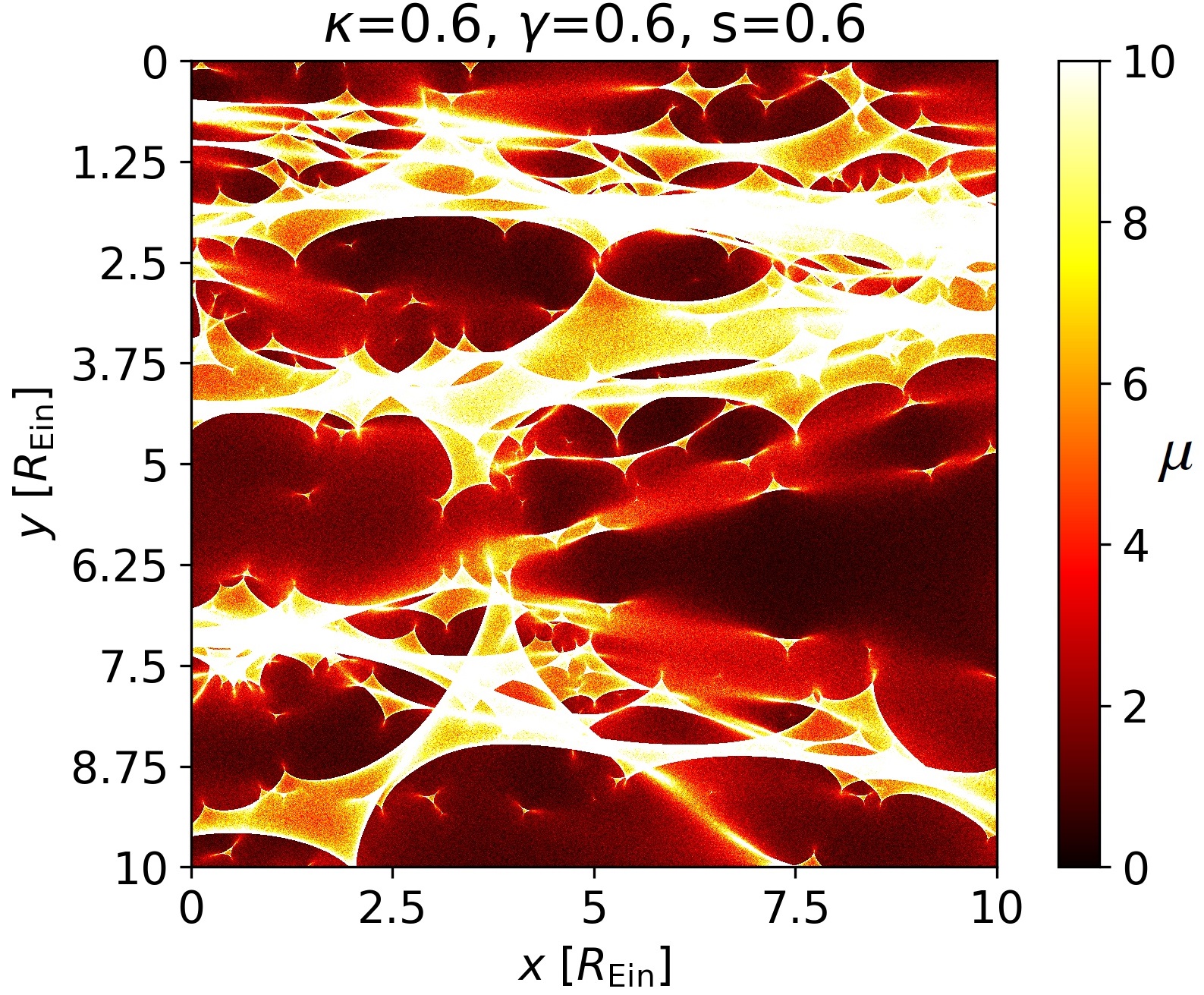}
\caption[Microlensing map for $\kappa=0.6$, $\gamma=0.6$ and
$s=0.6$.]{Example magnification map for $\kappa=0.6$,
  $\gamma=0.6$ and $s=0.6$. The color scheme illustrates the different
  magnification factors $\mu$ at the source plane depending on the $x$ and $y$
  coordinate. Many micro ``caustics" are visible separating regions of
  high and low magnification.}
\label{fig: microlensing map}
\end{figure}

Our maps have a resolution of $20000 \times 20000$ pixels and the
total size of the maps is set to $10 \Rein \times 10 \Rein$. Therefore the
size of one square pixel of the magnification map is
\begin{equation}
\scalebox{1.2}{$\dmag=\frac{10 \Rein}{20000}=\frac{1}{
1000} \sqrt{\frac{G  \langle M \rangle}{c^2} \frac{\source \ls}{\lens}}.$}
\label{eq: pixelsize micro map}
\end{equation}
For the simulated LSST LSNe Ia in Section \ref{sec:Time-Delay Measurements of mock LSST LSNe Ia}, the size of these microlensing maps ranges from $\SI{4.12e-2}{\parsec}$ to $\SI{2.70e-1}{\parsec}$ with a median of $\SI{1.02e-1}{\parsec}$.
As an example, a magnification map for $\kappa=0.6$ and $\gamma=0.6$ is shown in
Figure \ref{fig: microlensing map}, where $\Rein = \SI{7.2e-3}{\parsec} = \SI{2.2e16}{\cm} $ assuming an iPTF16geu like configuration.

\subsection{Theoretical SNe Ia model and the 1D projection}
\label{sec:SNeIa and the 1D projection}

To combine magnification maps with SNe Ia, we adopt a similar approach as \cite{Goldstein:2017bny} where the spherically symmetric W7
model \citep{1984:Nomoto} and the Monte Carlo-based radiative transfer
code SEDONA \citep{Kasen:2006ce} were used.
 
\begin{figure}
\centering
\includegraphics[scale=0.7]{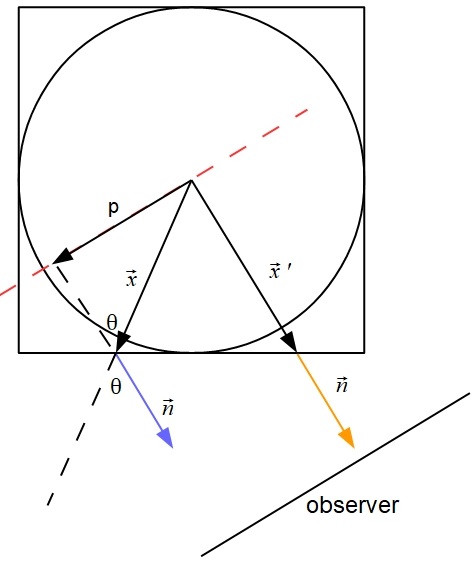}
\caption[1D projection of SNe.]{Slice through
  spherically symmetric SN enclosed in 3D simulation box in order to
  explain 1D projection of SN and definition of observer time defined in Equation (\ref{SN: observer time}).}
\label{fig: 1D projection}
\end{figure}

For our analysis, we also rely on the W7 model, but calculate synthetic
observables with the radiative transfer code {\tt ARTIS}
\citep{Kromer:2009ce}, which stands for Applied Radiative Transfer In
Supernovae and is a Monte Carlo based code to solve the frequency and
time-dependent radiative transfer problem in 3D. Thus, {\tt ARTIS} is not a
deterministic solution technique, where the radiative transfer equation is
discretized and solved numerically, but a probabilistic approach in which the radiative transfer process is simulated by a large number of Monte Carlo
packets, whose propagation is tracked based on the methods developed
by \cite{Lucy:1999,Lucy:2001ts,Lucy:2003zx,Lucy:2004fz}. In this procedure, $\gamma$-ray photon
packets from the radioactive decay of \nickel \, to \cobalt \, and the
successive decay of \cobalt \, to \iron \, are converted into UVOIR
(ultraviolet-optical-infrared radiation) packets which are then
treated with the full Monte Carlo radiative transport procedure. In
the propagation of UVOIR packets, bound-free, free-free, and especially
bound-bound processes are taken into account. Once a packet escapes
from the SN ejecta and the computational domain (which we refer to as a
simulation box), the position $\vec{x}$ where it
escapes the simulation box, the time $t_\mathrm{e}$ when it leaves and
the propagation direction $\vec{n}$ are stored in addition to the
energy and frequency. For the spherically symmetric ejecta the
  interaction of a photon packet stops after leaving the ejecta surface
  so in general before hitting the simulation box. For an
illustration of two photon-packets leaving the simulation box in the
same direction, see Figure \ref{fig: 1D projection}.

Typically one is interested in spectra and light curves, to compare observations to theoretical models. To get this information from
numerical simulations, all escaping packets have to be binned in
frequency and time, alongside the solid angle for asymmetric models. Since the microlensing effect depends on the location of the
source as shown in Figure \ref{fig: microlensing map}, spatial
information of the SN is needed as well. Therefore, we have to project
the 3D SN onto a 2D plane perpendicular to the observer and get the
specific intensity as a function of wavelength, time, and spatial
coordinates $x$ and $y$. Throughout this work, we assume that SNe Ia
can be treated with spherical symmetry and therefore no binning in solid
angle is necessary. While this is exact for an inherent 1D model like
W7 and good for multi-dimensional simulations that lead to nearly spherically
symmetric ejecta like some delayed detonations \citep{Seitenzahl:2013}
and sub-Chandrasekhar detonations \citep{Sim:2010}, this approximation
is questionable for models that lead to strongly asymmetric ejecta
like the violent merger \citep{Pakmor:2011,Pakmor:2012apjl}.

In the 1D case, the spatial dependency of the specific intensity
reduces to the dependency on the impact parameter $p$, that is, the projected
distance from the ejecta center. To construct this, we consider a
plane containing the position $\vec{x}$, where a photon-packet has
left the 3D simulation box, and the propagation direction
$\vec{n}$. This is illustrated in Figure \ref{fig: 1D projection} for
two packets leaving at different positions but propagating in the same
direction. Because of the vast distance of the SN, the observer is defined as a plane perpendicular to
$\vec{n}$. The radial
coordinate where the photon leaves the box is $r=\sqrt{\vec{x}^2}$, 
and the angle between the position vector $\vec{x}$ and the
propagation direction $\vec{n}$ is $\cos \theta = \frac{\vec{x} \cdot
  \vec{n}}{|\vec{x}| |\vec{n}|}$. Then, the impact parameter is
defined as
\begin{equation}
p =r \sin \theta = r \sqrt{1-\cos^2 \theta}, \qquad \theta \in [0,\pi].
\label{eq:impact parameter radial}
\end{equation}

From Figure \ref{fig: 1D projection} we see that different photon-packets, leaving the box at different positions but at the same time after explosion $t_\mathrm{e}$, will reach the observer at different times. If we assume that the orange packet from Figure \ref{fig: 1D projection} reaches the observer at time $t^\prime$ and the blue packet at time $t$ we can relate both times via $t = t^\prime + \frac{d^\prime-d}{c}$,
where $d = r \cos \theta $ and $d^\prime = |\vec{x}^\prime|$. The time when the orange packet reaches the observer can be expressed as $t^\prime = t_\mathrm{e} + C$, where C is a constant defining the distance from the observer to the simulation box for the orange packet. From this we can write $
t=t_\mathrm{e} + C + \frac{d^\prime - d}{c}$. Since the comparison to real observations is always performed relative to a maximum in a chosen band we are only interested in relative times. Therefore we can simplify the equation for $t$ by defining a reference plane at the center of the SN perpendicular to the propagation
direction $\vec{n}$ (red dashed line). For this reference plane $C=-\frac{d^\prime}{c}$ which leads to the observer time
\begin{equation}
t=t_\mathrm{e} - \frac{r \cos\theta}{c},
\label{SN: observer time}
\end{equation} 
as defined in \cite{Lucy:2004fz} which accounts for the geometrical delay described in \cite{Bonvin:2018b}. We will refer to the observer time $t$ as the time since explosion. With the definition of the time $t$ and the impact parameter
$p$, the energy is binned in these two quantities\footnote{Technical
  detail: Since the box expands with the SN over time, the impact
  parameter $p$ is a function of time. To eliminate this time
  dependency, one can assume that the SN is homologously expanding \citep{Ropke:2004ji} and
  therefore simply divide the impact parameter by the observer time as in \cite{Goldstein:2017bny}. The unit of this new impact parameter is
  therefore $\si{\cm\per\s}$ instead of $\si{\cm}$ and the unit of the new
  specific intensity is $\si{\erg\s\per\cubic\cm}$ instead of
  $\si{\erg\per\s\per\cubic\cm}$.} as well as in wavelength
$\lambda$. The emitted specific intensity can then be calculated via
\begin{equation}
\specificE = \frac{\dd E}{4 \pi \dd t \ds \lambda \, 2 \pi p \ds p},
\end{equation} 
where the factor $4 \pi$ is needed as a normalization over the unit sphere.
 
\subsection{Microlensed flux of SNe Ia}
\label{sec: Microlensing on SNe Ia}

To calculate microlensed light curves, one has to first determine the
observed spectral flux for a SN, which can be calculated for a source of
angular size $\Omega_0$ on the sky as
\begin{equation}
F_{\lambda,\mathrm{o}} = \int_{\Omega_0} I_{\lambda,\mathrm{o}} \cos \theta_\mathrm{p} \ds\Omega.
\end{equation}
Here $ I_{\lambda,\mathrm{o}}$ is the specific intensity at the
position of the observer. In Figure \ref{fig: SNe projected
  perpendicular to line of sight} a spherical source (gray disk) is
placed perpendicular to the line of sight at $\theta_\mathrm{p}=0$. The disk
represents the projected emitted SN specific intensity
$I_{\lambda,\mathrm{e}}$.
\begin{figure}
\centering
\includegraphics[scale=0.6]{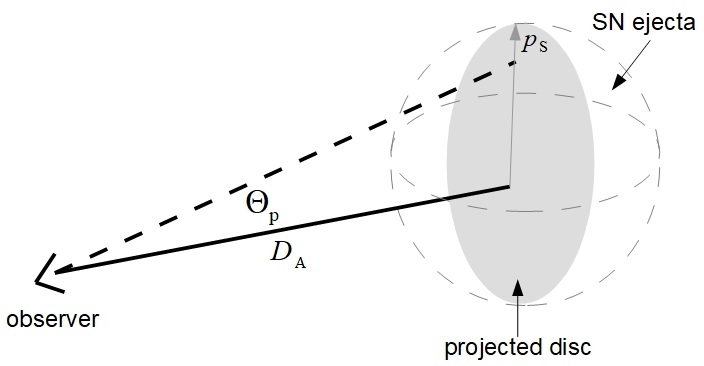}
\caption[SN projection onto a disc.]{SN projected
  onto disk perpendicular to line of sight to observer. The
  center of the disk with radius $p_\mathrm{S}$ is placed at
  $\theta_{\rm p}=0$ at an angular diameter distance of $D_\mathrm{A}$ from
  the observer. }
\label{fig: SNe projected perpendicular to line of sight}
\end{figure}
Since the source size is much smaller than
the angular diameter distance to the source, we use the approximation
for small angles and get $\theta_\mathrm{p} = \frac{p}{\angular}$ and $\cos\theta_\mathrm{p} \approx 1$, which means that we assume parallel light
rays. Therefore $\dd \Omega = \dd \phi  \ds \theta_\mathrm{p} \, \theta_\mathrm{p} = \frac{1}{\angular^2} \dd \phi \ds p \, p  $ and the spectral flux can be expressed as
\begin{equation}
\scalebox{1.1}{$F_{\lambda,\mathrm{o}}=\frac{1}{\angular^2}\int_0^{2 \pi} \dd \phi \int_0^{p_\mathrm{S}} \dd p \, p \,  \, I_{\lambda,\mathrm{o}},$}
\label{eq:non redshifted radiaion flux as integral over specific intensity in radial P coordinats without microlensing}
\end{equation}
where $p_\mathrm{s}$ is the source radius of the projected disk.
The next step is to relate the specific intensity at the observer's
position to the source position. Hereby, we have to take into account
that the specific intensity is redshift dependent. According to
Liouville’s theorem $I_\nu / \nu^3$ is Lorentz invariant \citep[page 414]{mihalas1984foundations} and therefore we have
$I_\lambda \propto \lambda^{-5}$. Since the emitted wavelength
$\lambda_\mathrm{e}$ can be related to the observed one,
$\lambda_\mathrm{o}$, via $
\lambda_\mathrm{o}=\lambda_\mathrm{e}(1+z)$ we find that
$I_{\lambda,\mathrm{o}}=I_{\lambda,\mathrm{e}}/(1+z)^5$. Therefore by
using $\lum = (1+z)^2 \angular$ the spectral flux reduces to
\begin{equation}
F_{\lambda,\mathrm{o}}=\frac{1}{\lum^2 (1+z)}\int_0^{2 \pi} \dd \phi \int_0^{p_\mathrm{S}} \dd p \, p \, I_{\lambda,\mathrm{e}}.
\label{eq:radiaion flux as integral over specific intensity in radial P coordinats without microlensing}
\end{equation}
To add the effect of microlensing $I_{\lambda,\mathrm{e}}$ has to be
replaced with $\mu I_{\lambda,\mathrm{e}}$, which is possible since
lensing conserves surface brightness. The value $\mu$ is the
microlensing magnification\footnote{We break here with
  the traditional nomenclature adopted in radiative transfer, where
  $\mu$ stands for $\cos \theta$. Instead, $\mu$ denotes the
  magnification factor throughout this work.} as a function of $\phi$ and $p$. Therefore
we get
\begin{equation}
F_{\lambda,\mathrm{o}}=\frac{1}{\lum^2(1+z)}\int_0^{2 \pi} \dd \phi \int_0^{p_\mathrm{S}} \dd p \, p \, \mu \, I_{\lambda,\mathrm{e}}.
\label{eq:radiaion flux with microlensing as integral over specific intensity in radial P coordinats}
\end{equation}
We note that this equation is in agreement with \cite{Hogg:2002yh} and \cite{Goldstein:2017bny}, where in the latter the flux is calculated in the supernova frame ($F_{\lambda,\mathrm{e}}$) instead of the observer frame ($F_{\lambda,\mathrm{o}}$).


The projected specific intensity inferred from simulations is a
discrete function in time, wavelength, and impact parameter and denoted
as $I_{\lambda_j,\mathrm{e}}(t_i,p_k)$. Because of the spherical
symmetry of W7, it has just a 1D radial dependency whereas the
magnification map is obtained on a 2D cartesian grid. To combine both quantities as needed in Equation (\ref{eq:radiaion flux with
  microlensing as integral over specific intensity in radial P
  coordinats}), it is necessary to transform one of both discrete
quantities into the other coordinate system. We choose to interpolate
the specific intensity onto a 2D cartesian grid:
\begin{equation}
 I_{\lambda_j,\mathrm{e}}(t_i,p_k) \rightarrow I_{\lambda_j,\mathrm{e}}(t_i,x_l,y_m).
\label{eq:Specific Intensity Interpolation onto cartesian grid}
\end{equation}
For this, we construct a cartesian grid with a pixel size $\Delta
x=\Delta y \equiv \dmag$. To get accurate results, $\Delta p
\gtrsim \dmag$ is required but to save computational memory we
restrict ourselves to
\begin{equation}
  \Delta p \approx \dmag.
\label{eq:criteria for bin size}
\end{equation}
As the SNe Ia ejecta expand, $\Delta p$ grows. Since $\dmag$ is a
fixed quantity defined by Equation (\ref{eq: pixelsize micro map}), we
interpolate the magnification map to a finer or coarser grid to
fulfill the criteria in Equation (\ref{eq:criteria for bin size}) using
the Python library scipy\footnote{https://www.scipy.org/}
\citep{Scipy:2001}. To get $I_{\lambda_j,\mathrm{e}}(t_i,x_l,y_m)$ for
a given time $t_i$ we interpolate $I_{\lambda_j,\mathrm{e}}(t_i,p_k)$
in $p$ and evaluate it for all grid points $(x_l,y_m$). Therefore the
spectral flux at time $t_i$ after explosion can be calculated via
\begin{equation}
\scalebox{0.9}{$F_{\lambda_j,\mathrm{o, cart}}(t_i)=\frac{1}{\lum^2(1+z)} \sum_{l=0}^{N-1} \sum_{m=0}^{N-1} I_{\lambda_j,\mathrm{e}}(t_i,x_l,y_m) \, \mu(x_l,y_m) \, \dmag^2.$}
\label{eq:flux discrete for spectra}
\end{equation}  

For the calculation of fluxes and light curves for
astronomical sources at redshift $z$ we have
\begin{equation}
 t_\mathrm{o} = t_\mathrm{e} \, (1+z) \qquad \mathrm{and} \qquad \lambda_\mathrm{o} = \lambda_\mathrm{e} \, (1+z).
\label{eq:redshift time and wavelength}
\end{equation}
To calculate microlensed light curves for the six LSST filters (details about LSST in Section \ref{sec: LSST}) we combine Equation (\ref{eq:flux discrete for spectra}) with the
transmission function $S_\mathrm{X}(\lambda)$ for LSST filter X. We
calculate AB-magnitudes as described by \cite{Bessel:2012} such that 
\begin{equation}
\scalebox{0.9}{$
m_\mathrm{AB,X}(t_i) = -2.5 \log_{10} \left( \frac{\sum_{j=0}^{N_\lambda-1} S_\mathrm{X}(\lambda_j) \, F_{\lambda_j,\mathrm{o,cart}}(t_i)  \, \Delta \lambda_j \, \lambda_j}{\sum_{j=0}^{N_\lambda-1} S_\mathrm{X}(\lambda_j) \, c \, \Delta \lambda_j / \lambda_j} \times \si{\square\cm\over\erg} \right)  - 48.6$}
\label{snmicro: light curves ab magnitudes}
\end{equation}
for the magnitude at the i-th time bin for filter X. 

\begin{figure*}[htbp]
\centering

\subfigure{\includegraphics[scale=0.42]{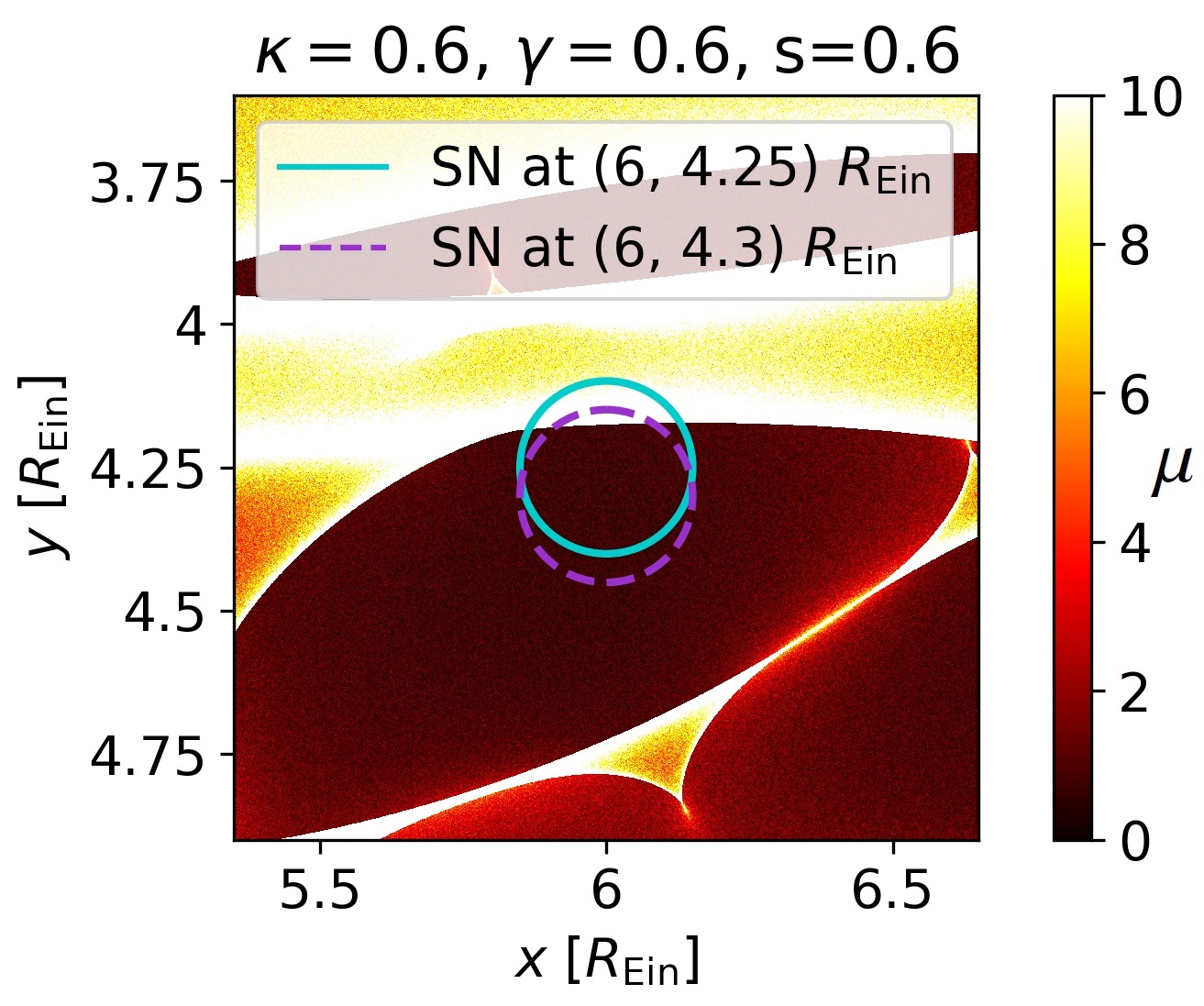}}
\subfigure{\includegraphics[scale=0.42]{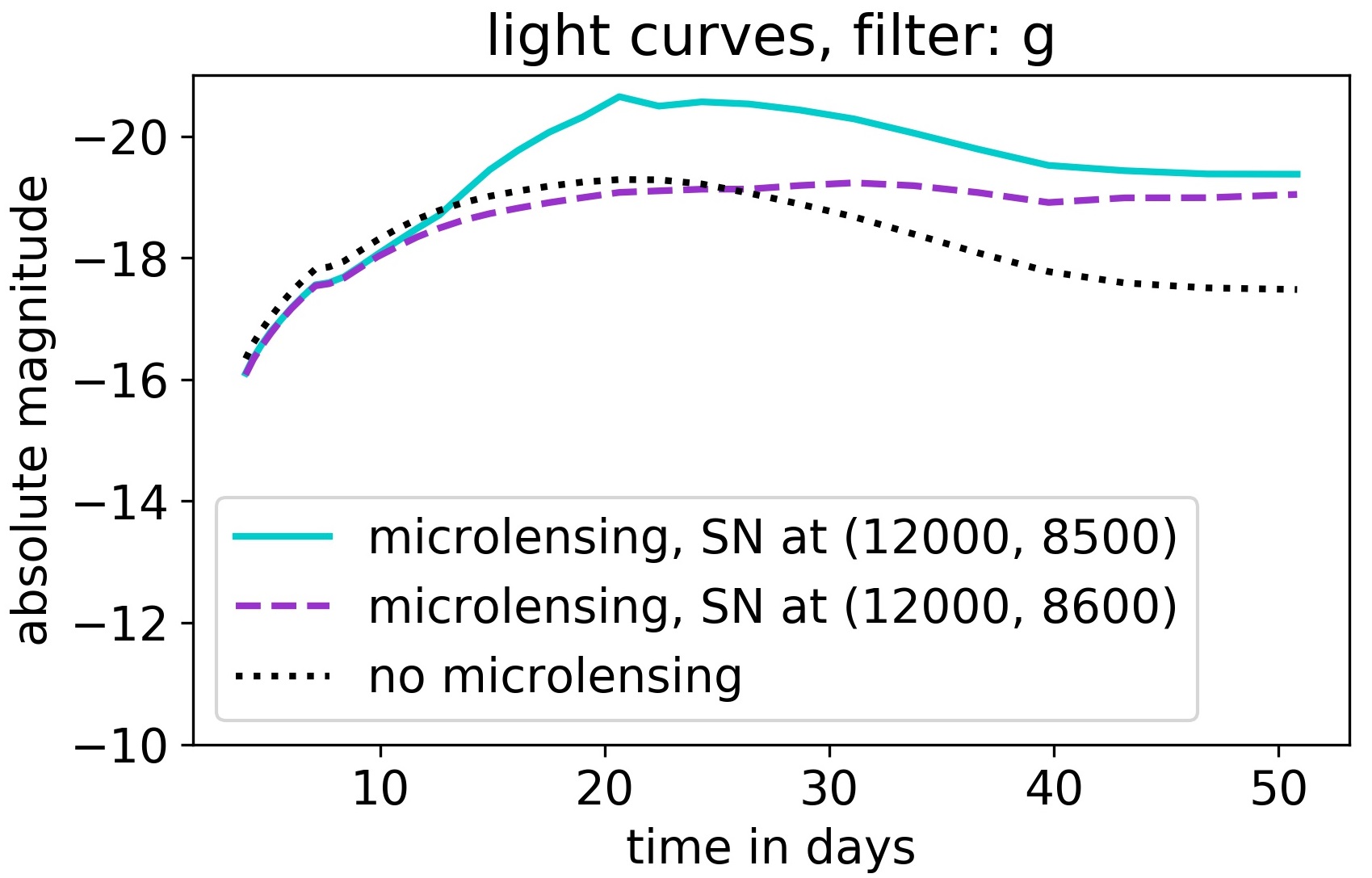}}
\subfigure{\includegraphics[scale=0.42]{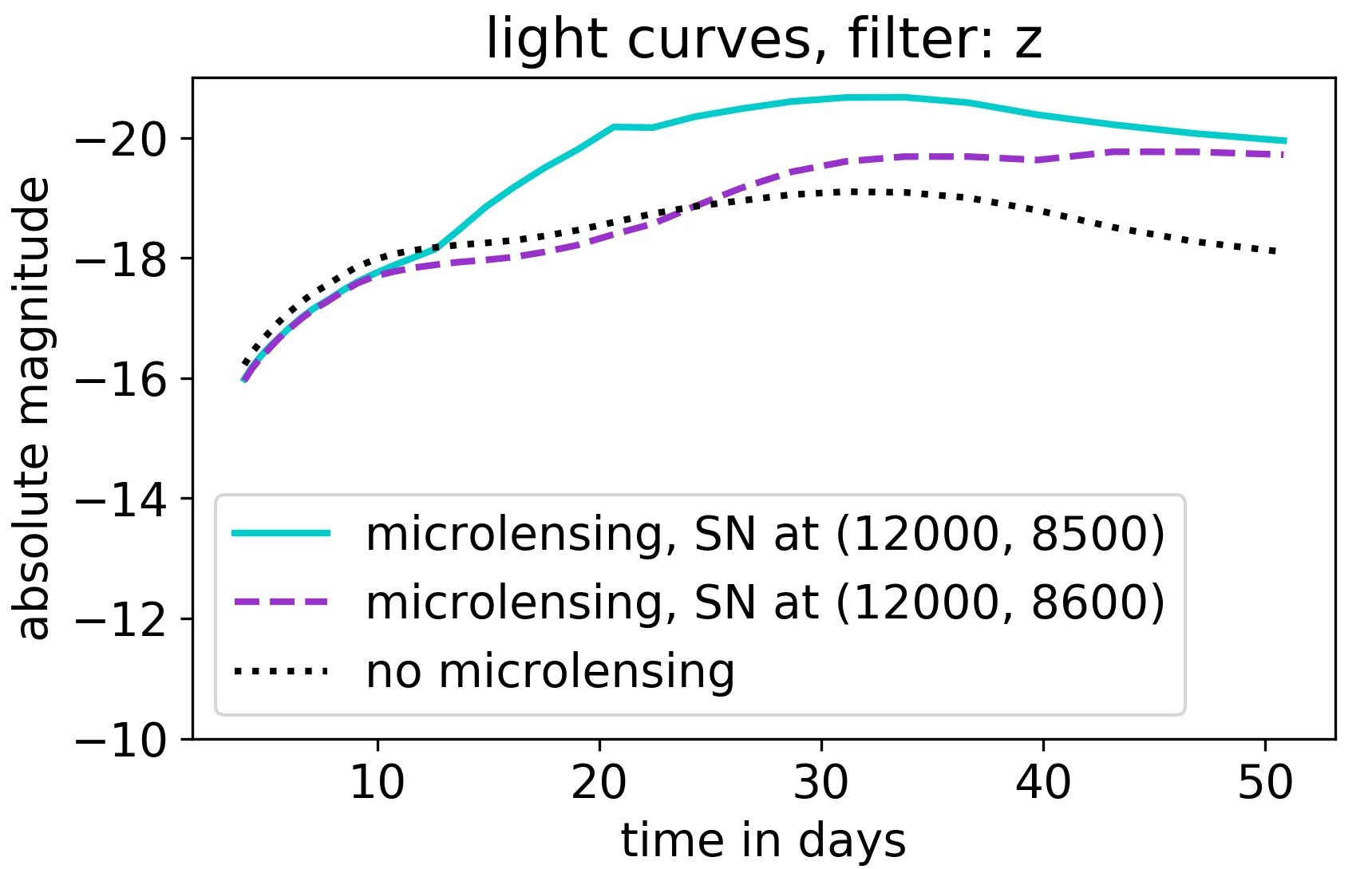}}
\caption{Influence of microlensing on light curves \textit{g} and \textit{z} for two different positions (solid cyan and violet dashed) as shown in left panel at 21 days after explosion for magnification map of Figure \ref{fig: microlensing map} where $\Rein = \SI{7.2e-3}{\parsec}$. The case of no microlensing is shown as black dotted line 
in the middle and right panels. 
We see that microlensing can cause distortion of light curves, shift the peaks and therefore add uncertainties to time-delay measurements between images undergoing different microlensing.}
\label{fig: micro influence on light curves}
\end{figure*}

Light curves in absolute magnitudes are shown in Figure \ref{fig: micro influence on light curves} for the \textit{g} and \textit{z} bands. It is important to catch the light curve peaks of different images of a LSNe Ia to measure time delays. While we have a single peak for rest-frame light curves \textit{u} and \textit{g} we find a secondary peak in the redder bands where we could ideally catch both peaks for delay measurements. In addition to the non microlensed case (dotted black), light curves with microlensing (solid cyan and dashed violet) for two different positions (see left panel) in the magnification map from Figure \ref{fig: microlensing map} are shown. 
The microlensed light curves are highly distorted and peaks are shifted, which adds large uncertainty to the time-delay measurement between different images based on light curves that undergo different
microlensing.

A more detailed investigation of microlensing is presented in Appendix \ref{sec: Case study}, where also spectra and color curves are discussed. We find from the investigated magnification map (Figure \ref{fig: microlensing map}) 
an achromatic phase for some color curves up to approximately $25-30$ days, as reported in \cite{Goldstein:2017bny}; however, other color curves show a shorter or non-existent achromatic phase. Our investigation also indicates that
the achromatic phase depends highly on the specific intensity
profiles and therefore the investigation of different explosion models is necessary
to explore this further (S. Huber et al., in preparation). Furthermore, some color curves from \texttt{ARTIS} are different in shape from the ones of \texttt{SEDONA}, which is important since features like peaks are necessary to measure time delays.
Even though color curves seem to be more promising for measuring time delays \citep[as suggested by][and discussed in Appendix \ref{sec: Case study}]{Goldstein:2017bny}, 
we use light curves instead for our further investigation because the 
sparse sampling of LSST does not provide directly color 
curves.  
Since color information is more easy to obtain with triggered
follow-up observations, it is promising to develop color
curve fitting methods in the future.

\section{Large Synoptic Survey Telescope (LSST)}
\label{sec: LSST}

The LSST will target about $\SI{20000}{\square\deg}$ of the
southern hemisphere with a field of view of $\SI{9.6}{\square\deg}$. Observations will be taken in six broad photometric bands \textit{ugrizy} and each position in the survey area will be repeatedly observed over
time, where each visit is composed of one or two back-to-back exposures
in the observing strategies currently under consideration. About $90 \%$ of the observing time will be spent on the
$\SI{18000}{\square\deg}$ wide-fast-deep survey (WFD), where
the inter-night gap between visits in any filter is about three days \citep{2009:LSSTscience}. The
rest of the 
time will be used for other regions like the northern Ecliptic, the south Celestial Pole, the Galactic Center, and a few ``deep drilling fields" (DDFs) where single fields ($\SI{9.6}{\square\deg}$)
will be observed to a greater depth in
individual visits. 

The scientific goals of LSST include exploring the nature of dark energy
and dark matter, exploring the outer regions of the solar system, and
completing the inventory of small bodies in the solar system. These science goals
restrict the cadence strategy 
but still leave a certain amount of freedom. For example, to detect
fast-moving transients like asteroids, a revisit of an observed field
within an hour is usually necessary. Such a revisit is planned if the first
observation was taken in one of the bands \textit{g, r, i,} or \textit{z} and is done in the same filter as the first observation for most of the cadence strategies under investigation in this work. For more
details, see \cite{2009:LSSTscience}.

As the LSST Project is in the process of finalizing the cadence
strategy, this paper 
investigates how different cadence strategies will influence the
possibility of measuring time delays for LSNe Ia. We specifically look at what is termed as a ``rolling
cadence", where the overall idea is to subdivide the WFD and focus on different subdivided parts in different years, with the final ten-year static survey performance being the same as the nominal ten-year survey. This strategy is one way to provide a better sampling but it will reduce the number of seasons. 
A specific case for a rolling cadence is the one with two declination bands, which subdivides the WFD (with a declination from 0 to $\SI{-60}{\deg}$) into a northern
region covering declination from 0 to $\SI{-30}{\deg}$ and a southern
one with declination in $-30$ to $\SI{-60}{\deg}$. The idea is then to
visit the northern part only in odd years (year one, three, five, seven, and nine) and the
southern part in even years (year two, four, six, eight, and ten) or vice versa. 

We investigate 20 different observing strategies which are potential LSST cadences or of special interest for our science case. In Section \ref{sec:specifications of observing strategies} we present the different observing strategies. Readers who are more interested in the overall conclusions instead of specific details about the cadence strategies might directly jump to Section \ref{sec:categorization of observing strategies}.

\subsection{Specifications of observing strategies}
\label{sec:specifications of observing strategies}
Sixteen out of the 20 investigated cadence strategies are implemented with the {\tt OpSim} scheduler\footnote{\url{https://cadence-hackathon.readthedocs.io/en/latest/current_runs.html} and in addition \texttt{pontus\_2506} from Tiago Ribeiro.} and the remaining four are produced by {\tt alt\_sched}\footnote{\url{http://altsched.rothchild.me:8080/}} and the {\tt feature-based} scheduler\footnote{\url{https://github.com/yoachim/SLAIR\_runs}}.  
Both the {\tt OpSim} and {\tt feature-based} schedulers use a greedy
algorithm, where the sky location of the next visit is determined by
optimizing different parameters such as seeing, time lapsed since the
last visit at the location, etc. In contrast, {\tt alt\_sched} employs
a non-greedy algorithm by observing at minimum air mass and only
relaxing on that to increase season length.
The
following key points describe the different observing strategies very
briefly, where strategies with a capital letter have a larger than nominal $\SI{18000}{\square\deg}$ WFD
footprint (the color scheme is explained in Section \ref{sec:categorization of observing strategies})\footnote{A discussion within the Dark Energy Science Collaboration revealed that the three rolling cadences \texttt{kraken\_2036}, \texttt{mothra\_2045}, and \texttt{pontus\_2502} seem to lack some observations. Nevertheless, we investigate those cadences as all others, because we are mainly interested in the dependency on different parameters. Our statement about rolling cadences would stay the same even if we remove these three strategies from our investigation.}:

\begin{itemize}
\item \altsched: Non-greedy algorithm;
  revisits in the same night in different filter; visits distributed in
  \textit{ugrizy} as $\sim(8.2, 11.0, 27.6, 18.1, 25.6, 9.5)\%$.
\item \altschedrolling: Same as \texttt{alt\_sched} but as a rolling cadence with two declination bands.
\item \baseline: Greedy algorithm  like all following cadences;
  official baseline; $2 \times \SI{15}{\s}$ exposure; revisit within
  an hour in the same filter and scattered visits over WFD, four DDFs,
  northern Ecliptic, south Celestial Pole, and Galactic
  Center; distribution of visits in WFD over \textit{ugrizy} as $\sim (6.8, 9.4,
  21.5, 21.6, 20.2, 20.4)\%$. For all following cadences up to \texttt{pontus\_2506} just the main differences with respect to \texttt{baseline2018a} are listed.
\item \colossusfour: WFD cadence over Galactic Plane.
\item \colossusfive: Slightly expanded WFD.
\item \colossusseven: Single visits instead of pair visits each night.
\item \krakentwosix: Unofficial baseline with improved slew time.
\item \krakenfive: Nine DDFs instead of four.
\item \krakenthreesix: Standard WFD cadence in year one, two, nine, and ten and a rolling
  cadence with three declination bands in between.
\item \krakentwo: Single 30\,s exposure instead of $2 \times 15{\s}$
  exposure.
\item \krakenfour: Very large WFD footprint of
  \SI{24700}{\square\deg}; five DDFs; single visits instead of visits in
  pairs each night.
\item \mothrafive: A rolling cadence 
in WFD (two dec. bands).
\item \mothranine: Similar to \texttt{mothra\_2045} but on a very large WFD footprint
  (\SI{24700}{\square\deg}).
\item \nexusseven: Similar to \texttt{kraken\_2036} but on a WFD footprint of
  \SI{24700}{\square\deg}.
\item \pontuszerozerotwo: Very large WFD footprint
  (\SI{24700}{\square\deg}) and five DDFs.
\item \pontusnine: $2 \times \SI{15}{\s}$ visits replaced by $1 \times
  \SI{20}{\s}$ in \textit{grizy} and $1 \times \SI{40}{\s}$ in \textit{u} band.
\item \pontusfivezerotwo: A rolling cadence (two dec. bands) in WFD where the baseline cadence stays on at a reward level of $25\%$.
\item \pontusfivezerosix: Revisits in the same night in different filter.
\item \rollingopsim: A rolling cadence (two dec. bands) in WFD where the de-emphasized band is set to reach $25\%$ of it’s usual number of visits in a year; paired visits in \textit{g, r,} and \textit{i}.
\item \rollingmixopsim: A rolling cadence similar to rolling\_10yrs\_opsim but with revisits in different
  filters.
\end{itemize}

\subsection{Categorization of observing strategies}
\label{sec:categorization of observing strategies}
From our investigation (in Section \ref{sec:results}), we find that the main relevant parameters for
measuring time delays in LSNe Ia are the cumulative season length ($t_\mathrm{eff}$),
mostly in terms of the total number of LSNe Ia, and the mean
inter-night gap ($t_\mathrm{gap}$; also referred as sampling frequency or sampling) concerning the quality of the light curves.
These two parameters are defined later in this section. For categorizing different observing strategies $t_\mathrm{gap}$ and $t_\mathrm{eff}$ are shown
in Figure \ref{fig:WFD cumulative season length and inter night
  gap} for 20 LSST observing strategies and from this we can separate them into three different categories with respect to the current LSST baseline cadence strategy (\texttt{baseline2018a}): 
%
\begin{itemize}
\item \textcolor{orange}{$``\mathrm{baseline \, like}"$: baseline-like cadence strategies in terms of sampling respectively cadence ($t_\mathrm{gap}$) and
  cumulative season length ($t_\mathrm{eff}$)}
\item \textcolor{blue}{$``\mathrm{higher \, cadence \, \& \, fewer \, seasons}"$: higher cadence but shorter cumulative season length}
\item \textcolor{magenta}{$``\mathrm{higher \, cadence}"$: higher cadence and baseline-like cumulative season}
\end{itemize}
Readers interested in general properties of the strategies should focus on these three categories which are highlighted by the category names and their corresponding colors.
Observing strategies in blue $\rolling$ are all rolling cadences. The
alternating observation pattern for different years leads to a shorter
cumulative season length and hence an improved sampling. Magenta strategies $\better$ provide a better mean inter-night gap
than the baseline cadence by reducing the exposure time, doing the
revisits of the same field within an hour in different filters or by
just doing single visits of a field within a night. For this reason,
these strategies provide  sampling similar to rolling cadences 
but they leave the cumulative season length close to the baseline
cadence. Rolling cadences which keep the WFD on a $25\%$ reward level
have a cumulative seasons length similar to the baseline cadence but
do not provide a better mean inter-night gap and are therefore listed in category $\baselinelike$\footnote{except for \texttt{rolling\_mix\_10yrs\_opsim} where the revisit in different filters
  improves the sampling frequency.}.

The mean cumulative season length and mean inter-night gap from a simulation of a given observing strategy are calculated by taking the mean of all fields under
consideration. We look at two different cases. The first case
considers 719 LSST fields from the WFD survey\footnote{The 719 WFD fields contain all fields with $\mathrm{Dec}
\in [-58,-2] \, \si{\deg}$ and $\mathrm{RA} \in [0,120] \cup [330,360]
\, \si{\deg}$, where all DDFs are excluded.}, which is shown as black solid line
in Figure \ref{fig:WFD cumulative season length and inter night
  gap}, with the shaded region marking $99\%$ of the fields. In the
second case we consider for comparison all 5292 LSST fields covering 
the entire sky. We only take into account those fields
where observations are taken, which is shown as blue dashed line. In the upper panel, cadences with the black
  solid line below the black dot-dashed line are those with a
  significantly better inter-night gap than the baseline cadence (i.e., magenta $\better$ and blue $\rolling$ strategies), whereas the others are baseline-like (orange $\baselinelike$). From the lower panel we distinguish between strategies with a cumulative season length
  similar to the baseline cadence (magenta $\better$ and orange $\baselinelike$) and a significantly worse
  cumulative season length (blue $\rolling$). The area of the WFD footprint is not plotted explicitly because relative differences in the area are smaller than those in the cumulative season length. 
Nevertheless cadence strategies with a capital (small) letter have a nominal WFD footprint of $24700 \, (18000) \, \si{\square\deg}$.

\begin{figure*}
\centering
\includegraphics[width=0.6\textwidth]{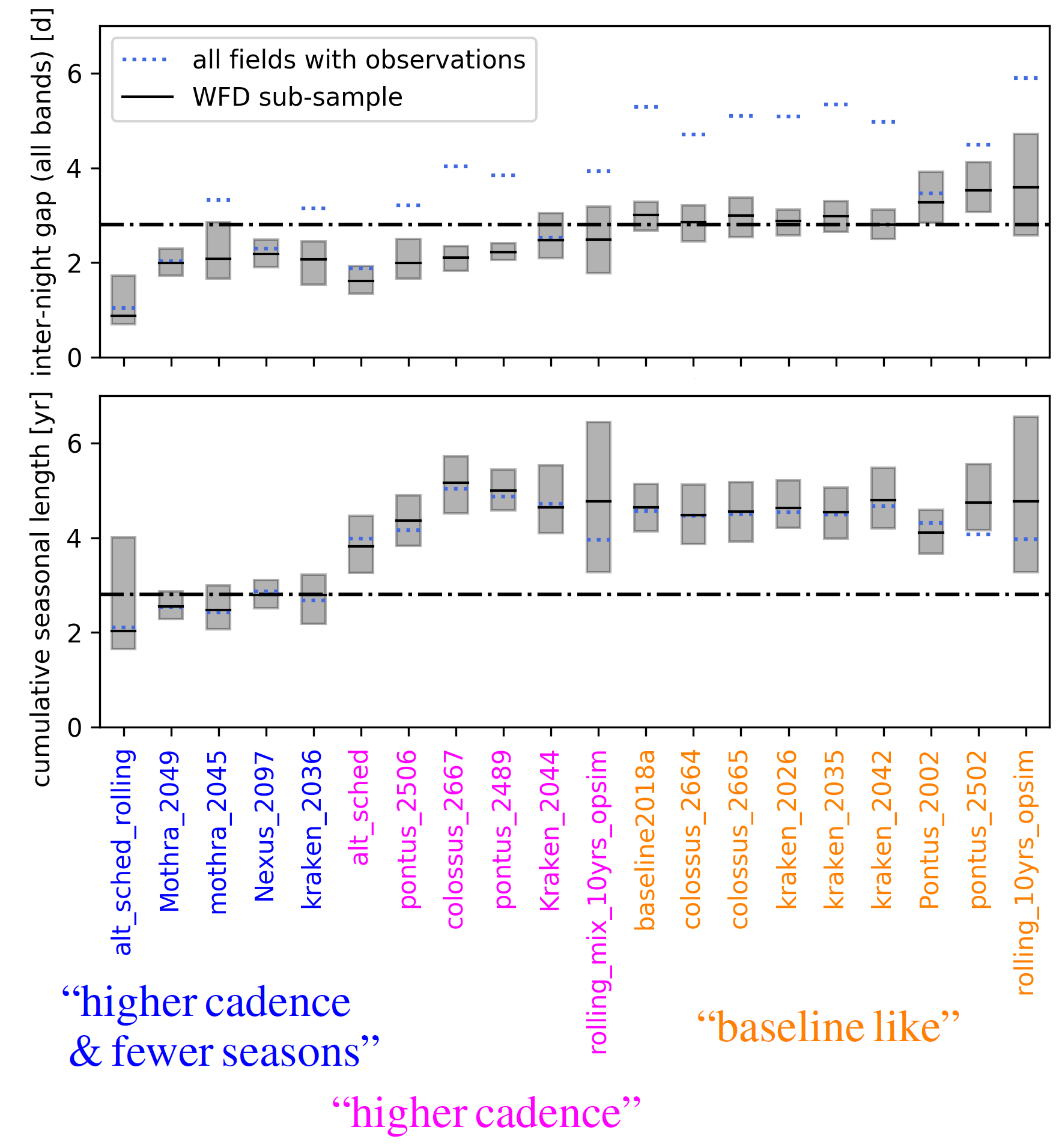}
\caption{Mean inter-night gap (upper panel) and mean cumulative
  season length (lower panel) for 20 different observing strategies
 to define the three categories $\rolling$, $\better$, and $\baselinelike$ as described in Section \ref{sec:categorization of observing strategies}. }
\label{fig:WFD cumulative season length and inter night gap}
\end{figure*}

The cumulative season length is the summed up season length over all
seasons. A season gap for an LSST field is defined if no observation
in any filter is taken for 85 days\footnote{To avoid unrealistically
  long seasons, we split a season if the season length is longer than 320
  days at the biggest gap. Seasons with a season length shorter than 10
  days are removed from the simulations.}. The mean cumulative season length
of all fields under consideration is shown in the lower panel of
Figure \ref{fig:WFD cumulative season length and inter night gap}. For
the inter-night gap, shown in the upper panel of Figure \ref{fig:WFD
  cumulative season length and inter night gap}, the revisits of a
field within hours in the same filter are summarized into a single
visit. Since SNe do not typically change over such a short time scale, the data
points are combined into a single detection with reduced
uncertainty. For some of the observing strategies, the mean
inter-night gap between the picked WFD fields deviates significantly
from the consideration of all fields, which is due to time spent on other surveys like northern
hemisphere, the southern Celestial Pole, and the Galactic Center.

\section{Generating realistic LSST mock light curves of LSNe Ia}
\label{sec:Time-Delay Measurements of mock LSST LSNe Ia}

The goal of this section is to describe how mock LSST light curves
for LSNe Ia are obtained for different cadence strategies. We used mock LSNe Ia from
the OM10 catalog \citep{Oguri:2010}, where we assumed
the spherically symmetric SN Ia W7 model
\citep{1984:Nomoto} for each image to simulate observations randomly. Synthetic light curves were produced with the
radiative transfer code {\tt ARTIS} \citep{Kromer:2009ce} where we included
the effect of microlensing via magnifications maps from {\tt GERLUMPH}
\citep[J.~H.~H.~Chan in preparation]{Vernardos:2015wta} following
Section \ref{sec: Microlensing on SNe Ia}. We then simulated data
points for the light curves, following the observation pattern from
different cadences and uncertainties according to the LSST science
book \citep{2009:LSSTscience}.
In Section
\ref{sec:OM10} we describe the OM10 mock catalog for strong lenses and
Section \ref{sec:Light curves for various observing strategies}
illustrates how we simulated mock light curves for mock LSNe Ia from OM10.

\subsection{Mock LSNe Ia from the OM10 catalog}
\label{sec:OM10}
The OM10 catalog \citep{Oguri:2010} is a mock lens catalog for
strongly lensed quasars and supernovae for LSST. For our purpose, we
focus on the LSNe Ia in the catalog. We expect about 45 spatially resolved LSNe Ia for
the ten-year LSST survey, under the assumption of OM10, namely a survey
area of $\Omega_\mathrm{OM10}=\SI{20000}{\square\deg}$ and a season
length of three months. Additionally, the 10$\sigma$ point source limiting magnitude in the \textit{i} band for a single visit is assumed to be $23.3$. The catalog contains LSNe Ia with two images (doubles) and four images (quads), but includes only those systems where
the multiple images are resolved (minimum image separation of
$\SI{0.5}{\arcsec}$) and the peak of the \textit{i}-band magnitude (of the fainter image for a double or the 3rd brightest image for a quad) falls in an observing season and is 0.7 mag brighter than the 10$\sigma$ point source limiting magnitude. Since we used the W7 model for our mock light curves and we got random microlensing magnification, we allowed automatically for fainter systems up to 25 mag in \textit{i}-band\footnote{98\% brighter than 24.0 mag and 41\% brighter than 22.6.}, instead of the sharp OM10 cut of 22.6 mag. Applying the cut as in OM10 is not necessary, because we used the 5$\sigma$ depth from simulations of the LSST observing strategies to create realistic light curves with uncertainties. Therefore, systems which are too faint will provide overall worse time-delay measurements than bright ones, making it unnecessary to exclude them in advance. Furthermore, applying no cut in magnitude allows us to draw conclusions about fainter systems not in the OM10 catalog, which are also relevant for time-delay measurements.

The mock catalog assumes as a lens mass model a Singular
Isothermal Ellipsoid \citep[SIE;][]{Kormann:1994} and the convergence for the
SIE is given in \cite{Oguri:2010} via
\begin{equation}
\kappa(\theta_1,\theta_2)= \frac{\thetaein \sqrt{1-e}}{2} \frac{\lambda(e)}{\sqrt{\theta_1^2+(1-e)^2 \theta_2^2}},
\label{MockCurves:konvergence OM}
\end{equation}
where $(\theta_1,\theta_2)$ are the lens coordinates, $\thetaein$ is the Einstein radius in arcsec, $e$ is the
ellipticity and $\lambda (e)$ the dynamical 
normalization defined in \cite{Oguri:2012}.  The lens mass
distribution is then rotated by its position angle.

The OM10 catalog is composed of two parts. The first part is the input
for the SIE model containing properties of the source and the lens,
such as redshift, velocity dispersion, source positions, and so
on. This first part is used to calculate mock images using GLAFIC
\citep{Oguri:2010GLAFIC} and therefore predict image positions,
magnifications, and time delays, which is the second part of the OM10
catalog. Furthermore, a microlensing map like the one in Figure \ref{fig:
  microlensing map} is needed to get the macro and microlensing magnification for different images, and therefore $\kappa$ and $\gamma$ have
to be known for each of the mock images\footnote{In principle also the
  smooth matter fraction $s$ but for simplicity we assumed as before $s=0.6$.}. We
calculated these parameters analytically for the SIE model following
equations from \cite{Kormann:1994}, \cite{Oguri:2010} and
\cite{Oguri:2012}, and checked
the consistency by comparing to magnification factors predicted by
GLAFIC.

The distribution of the source redshift and the time-delay of all OM10
mock systems is shown in Figure \ref{fig:OM10 source redshift
  distribution}. For quad systems, the maximum of the six possible time
delays (between pair of images) is shown. All 417 LSNe Ia from OM10
correspond to the blue line. To reduce the computational effort for the investigations in Section \ref{sec:results} we restrict ourselves to a subsample of 202 mock LSNe Ia (101 mock quads and 101 mock doubles) which is represented by the orange line. We find LSNe Ia for
a source redshift of 0.2 to 1.4 where most of them are around 0.8. In
terms of time delays, most of the systems have a maximum delay shorter
than 20 days. There are only a few systems with very long time delays (greater than 80 days).

\begin{figure}[h!]
\centering
\subfigure{\includegraphics[width=0.45\textwidth]{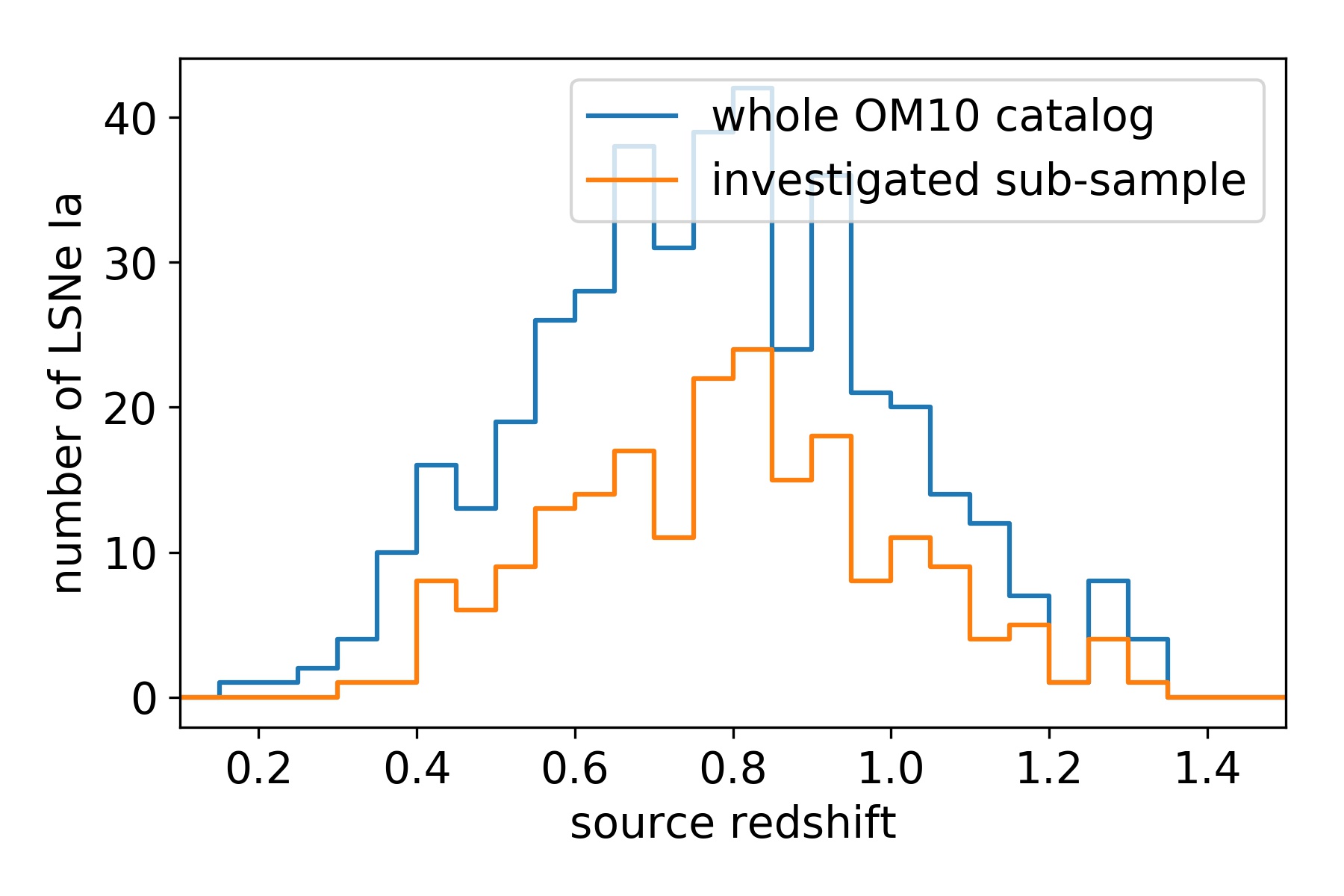}}
\subfigure{\includegraphics[width=0.469\textwidth]{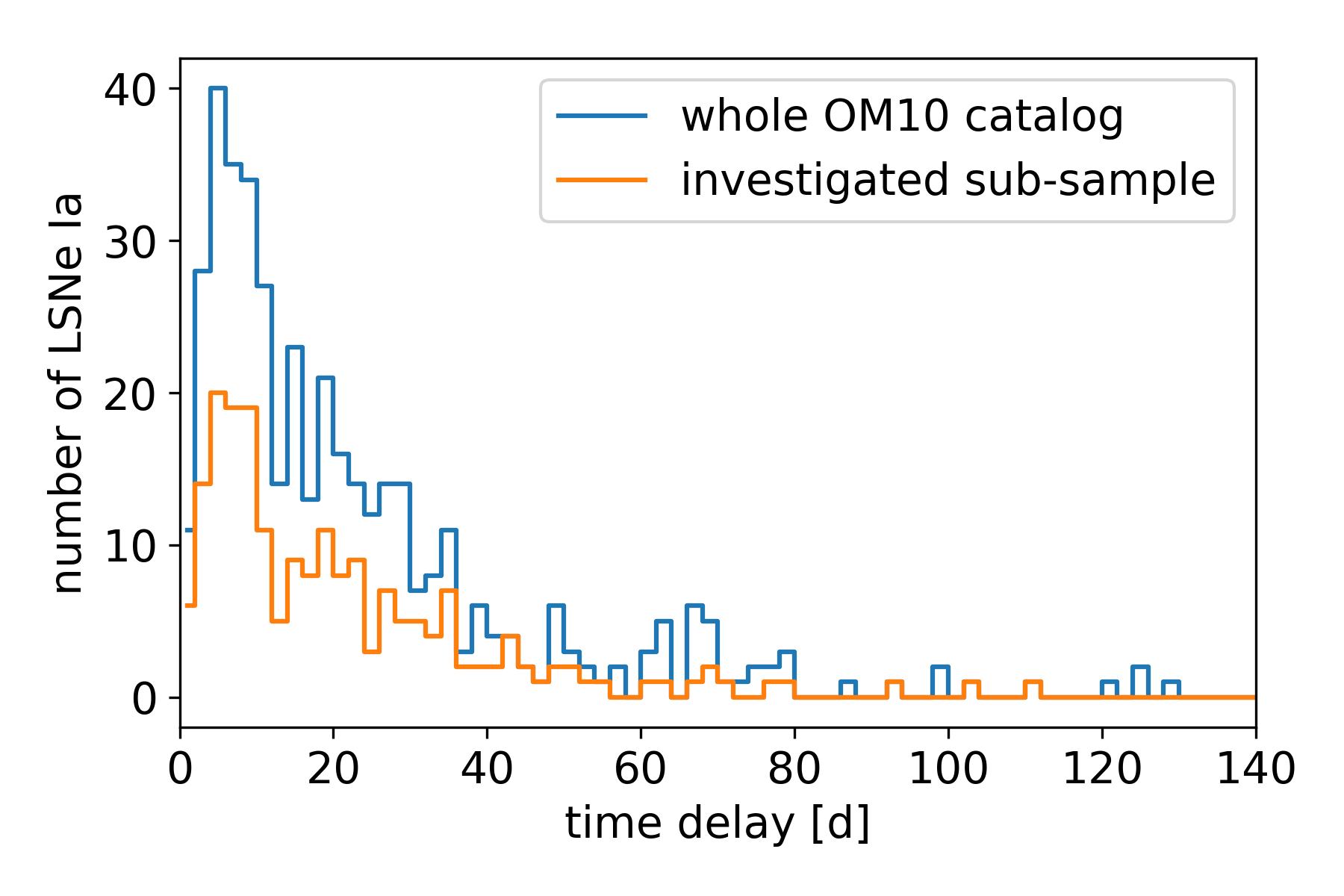}}
\caption{Source redshift (upper panel) and time-delay (lower panel)
  distribution of LSNe Ia from the OM10 catalog. The blue line
  shows the whole catalog (417 mock systems). The orange line
  shows the subsample of 202 mock systems (101 randomly picked quads and 101 randomly picked doubles) under investigations in
  Section \ref{sec:results}. For the time-delay distribution, the maximum time delay is shown (just relevant for quads) and there are
  three systems not in the plot with time delays greater than
  140 days. The highest delay of a LSNe Ia in the OM10 catalog is 290
  days.}
\label{fig:OM10 source redshift distribution}
\end{figure}

\subsection{Sampling of the light curves for various LSST observing strategies}
\label{sec:Light curves for various observing strategies}
\begin{figure}[h!]
\centering
\subfigure{\includegraphics[width=0.44\textwidth]{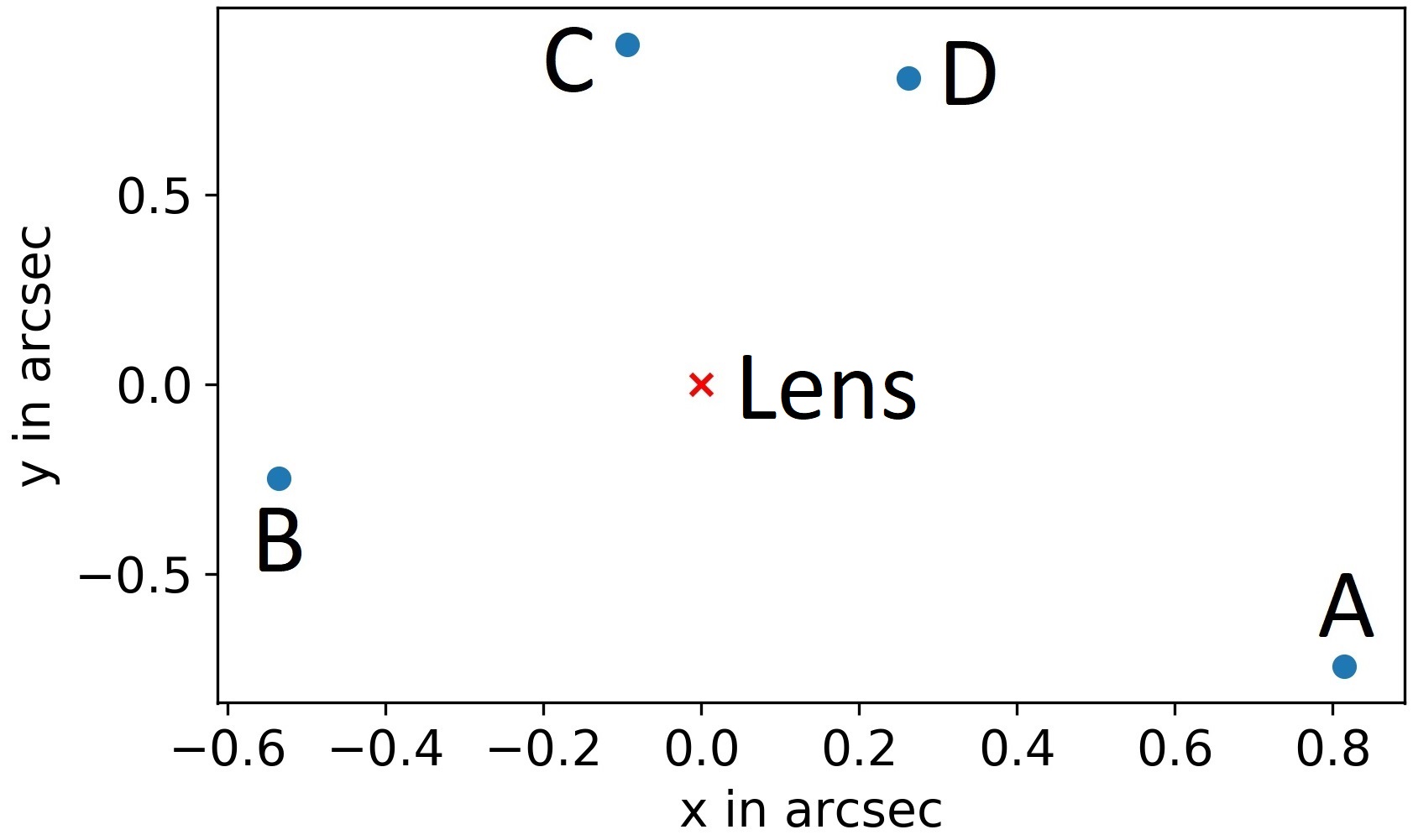}}
\subfigure{\includegraphics[width=0.5\textwidth]{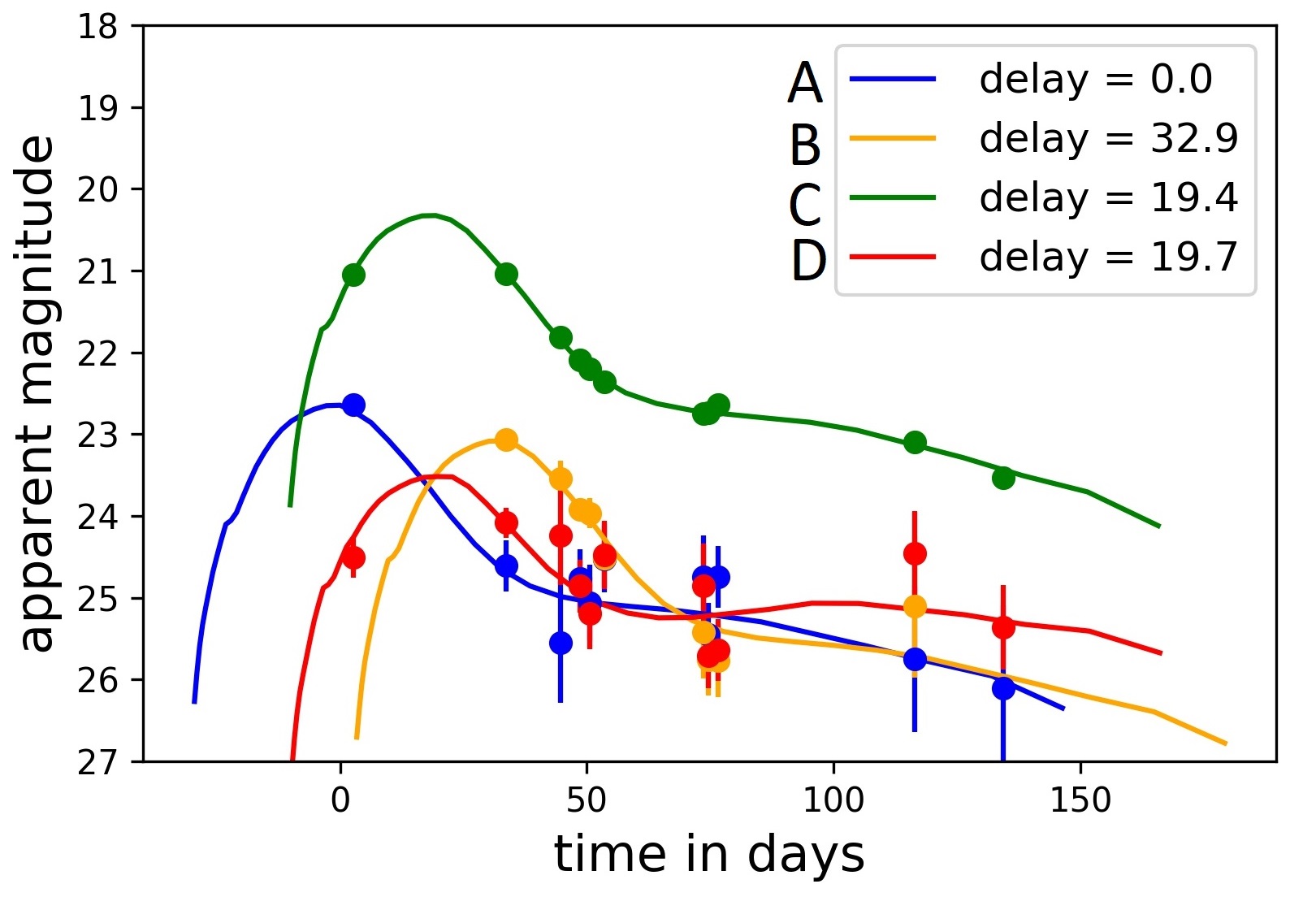}}
\caption{Synthetic \textit{i}-band light curves (lower panel) of a mock quad
  LSNe Ia (upper panel) to illustrate simulated observations. The redshift of the source is $0.71$ and is taken into account. The observation sequence is for a random
  field in the WFD survey for the \texttt{baseline2018a}
  cadence.}
\label{fig: simulated observation}
\end{figure}

To simulate observations, we randomly picked 202 mock LSNe Ia from the OM10
catalog (see orange curves in Figure \ref{fig:OM10 source redshift distribution}) and produced synthetic microlensed light curves for the mock
SNe images following Section \ref{sec: Microlensing on SNe Ia}. As an
example a mock quad system and the corresponding light curves (each
image in a random position in its corresponding microlensing map) is
shown in Figure \ref{fig: simulated observation}. Image A arrives
first followed by C, D, and B. In the simulated light curves of image D
(red solid line), an ongoing microlensing event is visible as
additional brightening about $\SI{80}{\day}$ after the peak, which is not visible
in the other three images.

\begin{figure*}
\centering
\includegraphics[width=\textwidth]{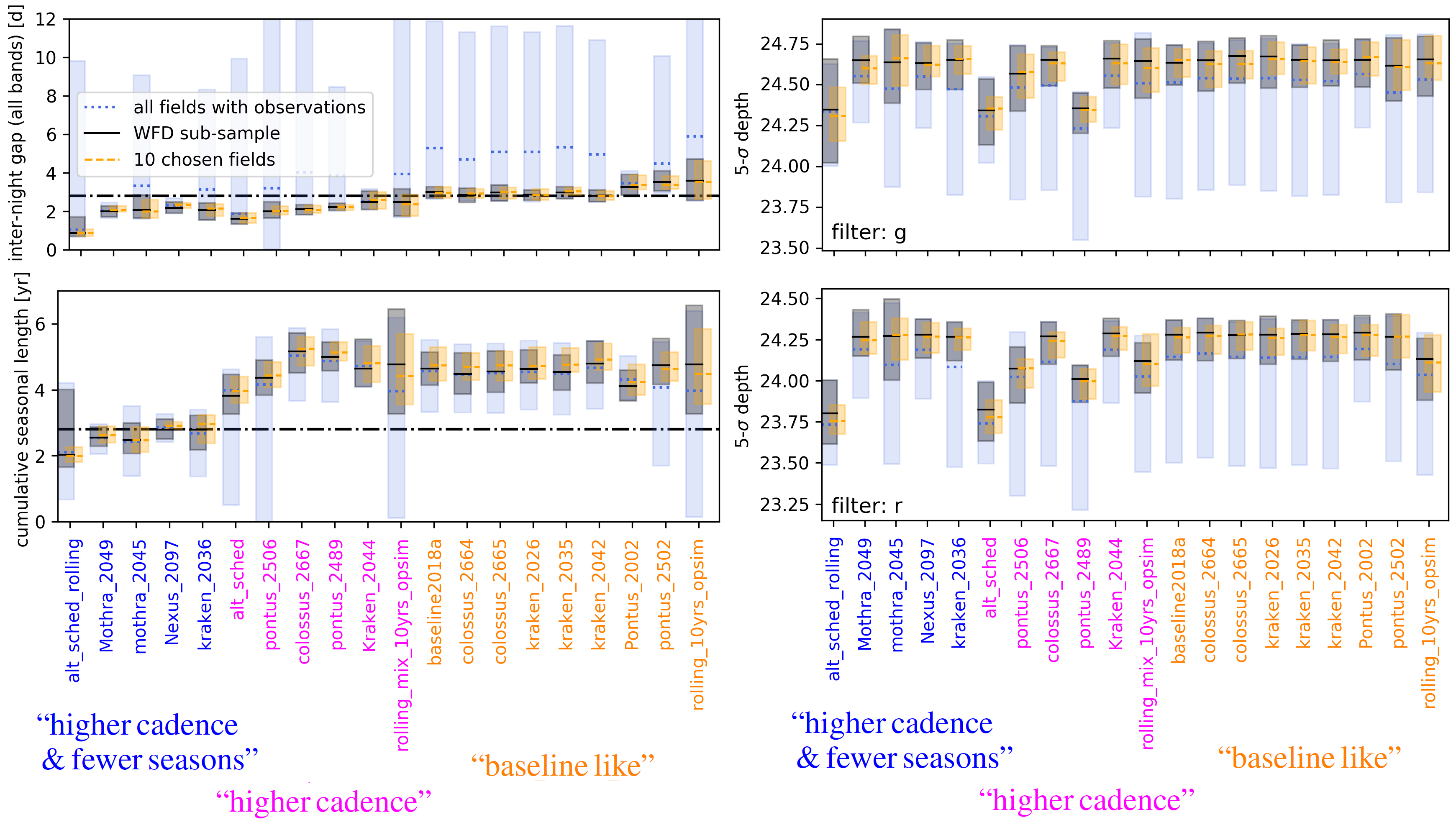}
\caption[Comparison of inter-night gap, cumulative season length, and
5$\sigma$ depth of ten fields under investigation to the
subsample of 719 WFD fields.]{Comparison of inter-night
  gap, cumulative season length, and 5$\sigma$ depth of ten fields
  under investigation (orange) to sample of 719 WFD (black)
  fields. In addition, all 5292 LSST fields where
  observations are taken (blue) are shown. The lines indicate the mean and the shaded
  area includes everything up to the 99th percentile. We see that the
  ten chosen fields are representative for the WFD survey but not for
  the whole survey.}
\label{fig:Comparison 10 fields to WFD fields}
\end{figure*}

To get simulated data points from the theoretical light curves as shown
in Figure \ref{fig: simulated observation}, we combined the
light curves with an observing sequence of visits. This is illustrated
for the \texttt{baseline2018a} cadence in Figure \ref{fig:observation patter
  LSST 10 year survey} where for one field in the WFD, all observations
within the 10-year survey are shown. For this purpose, we picked 10
fields in the WFD survey which are listed in Table \ref{tab: 10 wfd
  fields}\footnote{We have not added dithering to observing strategies simulated with the OpSim scheduler, which means that we underestimated the number of visits slightly.}. That these ten fields are representative for the WFD survey
is shown in Figure \ref{fig:Comparison 10 fields to WFD fields}. Here
the mean inter-night gap (top left panel), mean cumulative season
length (bottom left panel) and mean 5$\sigma$ depth for bands \textit{g} (top
right panel) and \textit{r} (bottom right panel) for our ten fields (orange),
WFD fields (black) and all fields (blue) are shown, while the shaded region
encloses the 99th percentile.

For each of the ten fields for a given cadence, we considered the following
for each visit of the field: date (mjd), filter(s) observed, and
5$\sigma$ point-source depth $m_5$. The depth is needed to calculate the
photometric uncertainties $\sigma_1$ according to the \cite{2009:LSSTscience} (see Appendix \ref{sec:Appendix LSST uncertainty}).
The magnitude for each data point can then be calculated via
\begin{equation}
m_\mathrm{data} = m_{\mathrm{W7}} + r_\mathrm{norm} \sigma_1,
\label{eq:noise realization random mag including error LSST science book}
\end{equation}
where $r_\mathrm{norm}$ is a random number following the normal
distribution and $ m_{\mathrm{W7}}$ is the magnitude of the data point
from the theoretical W7 model.  By placing the synthetic light curves
(shown as solid lines in Figure \ref{fig: simulated observation})
randomly in one of the fields in Table \ref{tab: 10 wfd fields}, 
randomly in time following the detection criteria from the OM10
catalog, and using Equation (\ref{eq:noise
  realization random mag including error LSST science book}), we created
simulated data points as illustrated in Figure \ref{fig: simulated
  observation}. If two or more data points are taken within one hour
in the same filter we combined them into a single measurement, because SNe typically do not change on such time scales. Specifically,
two data points $m_\mathrm{data,1} + \sigma_1$ and $m_\mathrm{data,2}
+ \sigma_2$ observed at time $t_1$ and $t_2$, where $ t_1 \le t_2 \le
t_1+ \SI{1}{\hour}$, were combined into a single one as
\begin{equation}
 m_\mathrm{combined} + \sigma_\mathrm{combined},
\end{equation}
where
\begin{equation}
\scalebox{1.1}{$m_\mathrm{combined} = \frac{m_1/\sigma_1^2 + m_2/\sigma_2^2}{1/\sigma_1^2 + 1/\sigma_2^2}, \quad \sigma_\mathrm{combined} = \sqrt{\frac{1}{1/\sigma_1^2+1/\sigma_2^2}}. $}
\end{equation}
We assigned to the combined data point the time $t_\mathrm{combined} = (t_1 + t_2)/2.$
\begin{figure}
\centering
\includegraphics[width=0.5\textwidth]{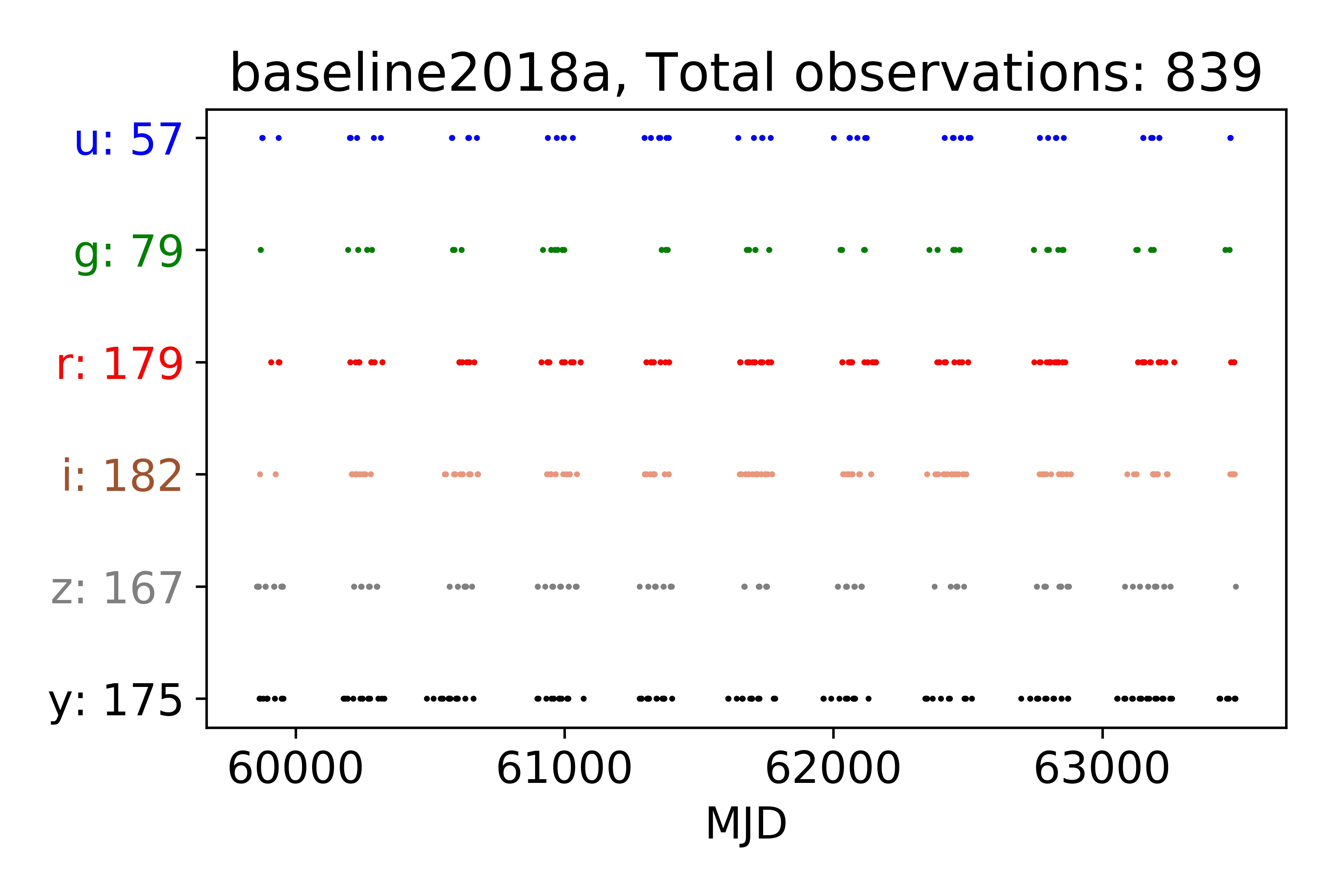}
\caption[Observation sequence for a WFD field in LSST.]{Illustration of Modified Julian Date (MJD) and
  filters when observations are taken over 10-year survey for field
  number four from Table \ref{tab: 10 wfd fields} for observing strategy \texttt{baseline2018a}. The y axis shows the six LSST filters and the number of observations taken in that filter.}
\label{fig:observation patter LSST 10 year survey}
\end{figure}
\begin{table*}[htbp]
\centering
\begin{tabular}{c|c|c|c|c|c|c|c|c|c|c}
field number & 1 & 2 & 3 & 4 & 5& 6 & 7 & 8 & 9 & 10  \\
\hline
RA in deg& 0.0 & 32.1 & 65.8 & 50.9 &44.9& 125.6 & 155.0 & 207.7 & 304.3 & 327.5  \\
\hline
DEC in deg& -7.4 & -44.2 & -7.2 & -30.0 & -50.9& -11.4 & -25.6 & -45.3 & -55.2 & -35.9  \\
\end{tabular}
\caption[Ten wide fast deep fields of LSST, where observation sequence is considered.]{Ten fields of WFD survey, where observational sequence for different cadences is considered, which is used to determine fraction of systems with measured time delay as discussed in Section \ref{sec:results}. We investigate the observing sequence at the centers of the listed fields.}
\label{tab: 10 wfd fields}
\end{table*}

\section{Time-delay measurements}
\label{sec:Time-delay measurement}
In this section we describe how we estimate time-delays from the simulated observations to quantify different observing strategies. We investigate 202 mock LSNe Ia (already mentioned in Section \ref{sec:Time-Delay Measurements of mock LSST LSNe Ia}) for each cadence strategy to have sufficient statistics, where we pick $50\%$ doubles and $50\%$
quads. We define a system with ``good" time delay measurement as a systems where the accuracy is below $1\%$ and the precision is below $5\%$. To estimate accuracy and precision we investigate for each of the mock systems, 100 random starting configurations. A starting configuration corresponds to a random
position in the microlensing map and a random field from Table
\ref{tab: 10 wfd fields}, where it is placed randomly in one of the
observing seasons such that the peak of the \textit{i}-band magnitude of the fainter image for a double or the 3rd brightest image for a quad falls in the observing season. We used the same random positions in the microlensing map for each mock image for all observing strategies investigated here, to avoid uncertainties due to different microlensing
patterns. For each of these starting
configurations, we then draw 1000 different noise realizations of light
curves following Equation (\ref{eq:noise realization random mag including error LSST science book}). For each of these realizations we have to estimate the time delay and compare it to the true value. 

\begin{figure}
\centering
\includegraphics[width=0.5\textwidth]{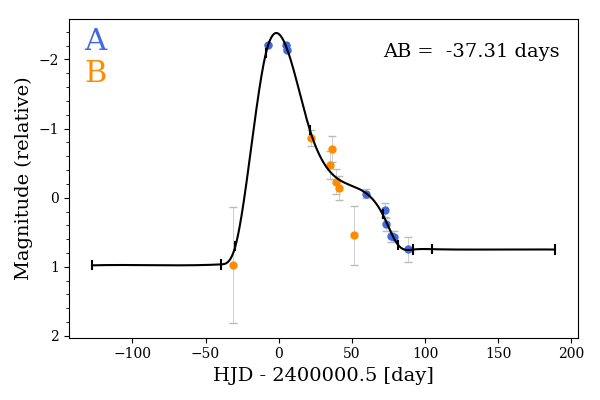}
\caption{Illustration of spline fitting technique for a double mock LSNe Ia at redshift $0.27$ for the \textit{i}-band light curve. The black line corresponds to the spline fit of the data (blue and orange), where the knots positions (small vertical ticks on the black lines) as well as the magnitude and time shifts have been iteratively optimized to minimize a chi-square term, resulting in the measured delay indicated in the top-right.}
\label{fig:PyCS illustration}
\end{figure}

To get a measured time delay from the mock data we used the
free-knot splines estimator from {\tt PyCS} 
\citep[Python Curve Shifting;][]{2013:Tewesb,Bonvin:2015jia}. As a spline, a piecewise polynomial
function of degree three is used. The polynomial pieces are connected by
knots, where for the optimization process, the initial number of knots
has to be specified. The polynomial coefficients and the knot
positions are free variables to optimize. To avoid clustering of the
knots a minimum knot separation is also defined in advance \citep{Molinari2004}. The basic
idea of the optimizer is to fit a single intrinsic spline to two light
curves from different images and shift the data iteratively in time and
magnitude, and modify the spline parameters, to get a time-delay
measurement.  
We show in Figure \ref{fig:PyCS illustration} an example of
the fitting of the spline to two light curves, with one light curve
time-shifted by the time delay to increase overlap with the other.
Both the spline parameters and the time delay between the two curves
are optimized by reducing the residuals in the fit of the spline to
the two light curves.  Even with noiseless data, we would get a spread
of delays from PyCS due to the range of splines that could fit to the
data equally well.  Densely sampled light curves with little
microlensing would restrict the range of delays.
We do not explicitly include
additional spline components to model the microlensing
variation. An analysis that models separately the intrinsic and microlensing variability is deferred to future work.

{\tt PyCS} was initially developed to measure time delays in strongly
lensed quasars, and is not yet optimized for LSNe Ia, such as fitting
simultaneously multiple filters and using SN template light curves.
Nonetheless, \citet{Rodney:2015uyq} used the tools of {\tt PyCS} to
measure the time delays between the multiple images of SN Refsdal as
one of the approaches, and also fit SN templates to the light
curves as another approach.  The resulting delays from both
approaches were consistent with each other. While both methods did not
explicitly include the effects of microlensing, the residuals of the
light curves of SN Refsdal suggested that no major microlensing event
occurred in the case of SN Refsdal \citep{Rodney:2015uyq}.  The
template-fitting approach was also used by \citet{Goldstein:2017bny}
to fit to mock light curves and color curves, although in an idealized
scenario without noise and high-cadence sampling.
\citet{Goldstein:2017bny} found the fitting of templates to light
curves yielded time-delay uncertainties of approximately $4\%$, limited by
microlensing distortion of light curves, whereas the fitting to color
curves in the achromatic phase provided approximately $1\%$ uncertainties in the
delays.  For our LSST light curves, we opt to use {\tt PyCS} on light
curves given that (1) color curves are not available from LSST data
given the sampling cadence, and (2) there is currently no publicly
available template-fitting software accounting for microlensing, an
effect that can significantly distort the light curves as shown in
Section \ref{sec:Microlensing on Type Ia Supernovae}.

Applying {\tt PyCS} to individual filter's light curves, we get a single
independent time delay for each filter. This means that we have for the given LSST filter $f$, the $j$-th starting configuration and the $k$-th noise realization a deviation from the true time delay:
%
\begin{equation}
\tau_{\mathrm{d, }f,j,k} = \frac{\Delta t_{\mathrm{measured, }f,j,k} - \Delta t_{\mathrm{true, }f,j,k}}{\Delta t_{\mathrm{true, }f,j,k}}.
\label{eq: deviation from true time delay}
\end{equation}
For each observing strategy and double LSNe Ia, we have thus 1 (delay for the one pair of images) $\times \, 6$ (filters) $\times \, 100$ (starting configurations) $\times \, 1000$ (noise realisations)
time-delay deviations as in Equation (\ref{eq: deviation from true time delay}).
For the six pairs of images for a quad system, we have a sample of $6
\times 6 \times 100 \times 1000$. 

To exclude starting configurations which are completely wrong in comparison to most of the investigated systems we calculated separately for each starting configuration
the median $\tau_{\mathrm{d,50, }f,j}$ and the error as $\delta_{f,j} =
(\tau_{\mathrm{d,84, }f,j}-\tau_{\mathrm{d,16 },f,j})/2$, where
$\tau_{\mathrm{d,50, }f,j}$, $\tau_{\mathrm{d,84, }f,j}$ and
$\tau_{\mathrm{d,16, }f,j}$ are the 50th, 84th, and 16th percentile
from the 1000 noise realizations. Furthermore, we combined the six filters via
the weighted mean into a single time-delay deviation
$\tau_{\mathrm{d,50},j} \pm \delta_j$, where 
\begin{equation}
\tau_\mathrm{d,50,j} = \frac{\sum_{f=\mathrm{ugrizy}} \tau_{\mathrm{d,50, }f,j} / \delta_{f,j}^2 }{\sum_{f=\mathrm{ugrizy}} 1/\delta_{f,j}^2}, \qquad \delta_j = \sqrt{\frac{1}{\sum_{f=\mathrm{ugrizy}} 1/\delta_{f,j}^2}}.
\end{equation} 
%
This is possible since the distribution of the time-delay deviation for each filter is
approximately Gaussian.
From this we exclude
``catastrophic failures" which are starting configurations with
$\delta_j \ge 2 \bar{\delta}_j$ or
$|\tau_{\mathrm{d,50},j}-\bar{\tau}_{\mathrm{d,50},j}| \ge 5
\delta_j$, which occur for about $10\%$ of the starting
configurations independent of the observing strategy. The bar indicates the mean, that is,
\begin{equation}
\bar{\delta}_j = \frac{1}{100} \sum_{j=1}^{100} \delta_j \qquad \mathrm{and} \qquad \bar{\tau}_{\mathrm{d,50},j} = \frac{1}{100} \sum_{j=1}^{100} \tau_{\mathrm{d,50},j}.
\end{equation}
The failures are likely due to a bad starting time of the supernova
in the season (such as at the beginning or end of season, where some
of the light curves of the multiple images would be incomplete due to
seasonal gap) and strong microlensing distortions.  These effects
could be easily identified in real lens systems, and provide advance
warning of potentially problematic delay inference.  In addition,
simulations of light curves mimicking those of real lens systems could
be used to identify catastrophic failures of problematic systems and
avoid the use of their time delays for further analysis such as
cosmography.

After excluding catastrophic failures we are left with about 90 of the 100 initial starting configurations leading to approximately $ 90 \times 1000 \approx 90000$ time-delay
deviations $\tau_{\mathrm{d, }f,j,k}$ for each filter $f$. From these we define accuracy as the median $\tau_{\mathrm{d,50, }f}$
and precision as $\delta_f = (\tau_{\mathrm{d,84,
  }f}-\tau_{\mathrm{d,16 },f})/2$, where $\tau_{\mathrm{d,84, }f}$ is
the 84th and $\tau_{\mathrm{d,16, }f}$ the 16th percentile of the 90000 starting configuration and noise realizations, that is, over the $j$ and $k$ indexes. Since the time-delay deviations from the
six filters are independent, we combined them into a single time-delay
deviation. This means that in the end, we have
for one strategy and a mock LSNe Ia a single $\tau_\mathrm{d,50} \pm \delta$ per pair of images, where
\begin{equation}
\tau_\mathrm{d,50} = \frac{\sum_{f=\mathrm{ugrizy}} \tau_{\mathrm{d,50, }f} / \delta_f^2 }{\sum_{f=\mathrm{ugrizy}} 1/\delta_f^2}, \qquad \delta = \sqrt{\frac{1}{\sum_{f=\mathrm{ugrizy}} 1/\delta_f^2}}.
\label{eq: accuracy and precission}
\end{equation}
To use the weighted mean here is possible since the time-delay distributions for different filters are approximately Gaussian.
\section{Results: cadence strategies for LSNe}
\label{sec:results}
In this section, we present the results of the investigation of the
different cadence strategies presented in Section \ref{sec: LSST}. We
distinguish between two different cases: (1) using LSST data only for
measuring time delays, and (2) using LSST just as a discovery machine
for LSNe Ia and getting the time delay(s) from follow-up observations.

Given that $H_0 \propto {\Delta t_\mathrm{true}^{-1}}$, where
$\Delta t_\mathrm{true}$ is the time delay between two images, we aim
for accuracy ($\tau_\mathrm{d,50}$ in Equation (\ref{eq: accuracy and
  precission})) smaller than $1 \%$ and precision ($\delta$ in Equation (\ref{eq: accuracy and
  precission})) smaller
than $5\%$. We refer to systems fulfilling these requirements as
systems with good time delays. A quad system is counted as
successful if at least one of the six delays fulfills these demands.  The
accuracy requirement is needed for measuring $H_0$ with $1\%$
uncertainty, and the precision requirement ensures that the delay
uncertainty does not dominate the overall uncertainty on $H_0$ given
typical mass modeling uncertainties of about $5\%$
\citep[e.g.,][]{Suyu2018}.

\subsection{Number of LSNe Ia}
Before comparing cadence strategies based on the time-delay
measurements, we first estimate the total number of LSNe Ia for
different observing strategies. Since different observing strategies
have different survey areas and different cumulative season lengths, 
the number of LSNe Ia deviates from the predicted number from OM10. We
approximate the total number of LSNe Ia as
\begin{equation}
\label{eq: total number of LSNe Ia from modified OM 10}
N_\mathrm{LSNe Ia, cad} \approx N_\mathrm{LSNe Ia, OM 10} \frac{\Omega_\mathrm{cad}}{\Omega_\mathrm{OM10}} \frac{\bar{t}_\mathrm{eff,cad}}{t_\mathrm{eff, OM10}},
\end{equation}
where $N_\mathrm{LSNe Ia, OM10} = 45.7$, $\Omega_\mathrm{OM10} =
\SI{20000}{\square\deg}$ and $t_\mathrm{eff, OM10}=\SI{2.5}{\year}$
from \cite{Oguri:2010}. The
effective respectively cumulative season length for a given cadence strategy is given via $\bar{t}_\mathrm{eff,cad}$,
where we have averaged over the sample of 719 WFD
fields. The survey area for a given observing
strategy is $\Omega_\mathrm{cad}$. Instead of taking the nominal values
($\SI{24700}{\square\deg}$ for large footprint strategies and
$\SI{18000}{\square\deg}$ for rest) we calculated the area from fields
represented by our study, which are the fields with a mean cumulative
season length and inter-night gap similar or even better than the 719
WFD fields, that means, cumulative season length ($t_\mathrm{eff}$)
longer than the lower 99th percentile and inter-night gap
($t_\mathrm{gap}$) shorter than the upper 99th percentile. Additionally we  take into account the 5$\sigma$ depth ($m_5$), where we
  consider only the main relevant bands \textit{g, r, i,} and \textit{z}. Here we
consider all fields with ($m_5+0.2 \mathrm{mag}$) greater than the
lower 99th percentile of the 719 WFD fields. The relaxed 5$\sigma$
depth is necessary in order to represent the wider areas as suggested
by the nominal values\footnote{This leads to a few percent
  overestimation of the total number of LSNe Ia with good time
  delays for large footprints in comparison to the
  $\SI{18000}{\square\deg}$. Nonetheless, since we find that the improvement
  due to wider area is too small this is not a problem and does not
  affect the overall conclusions of our work.}.  The area can
then be calculated from the number of fields fulfilling the above
defined criteria ($N_\mathrm{cad,criteria})$, multiplied with the field of view of
$\SI{9.6}{\square\deg}$, taking into account the overlap factor of the
fields: 
\begin{equation}
\Omega_\mathrm{cad} = f_\mathrm{overlap} \cdot N_\mathrm{cad,criteria} \cdot \SI{9.6}{\square\deg},
\end{equation}
where
\begin{equation}
f_\mathrm{overlap}=\frac{4 \pi \cdot (\SI{180}{\deg}/\pi)^2}{5292 \cdot \SI{9.6}{\square\deg}} \approx 0.812.
\end{equation} 
The total number of fields is 5292, which cover the entire sky, as
noted in Section \ref{sec: LSST} and the numerator corresponds to the surface area of a sphere in $\mathrm{deg}^2$. Therefore,  $\Omega_\mathrm{cad}$ is equivalent to $4 \pi N_\mathrm{cad,criteria}/5292$ in units of $\mathrm{rad}^2$.  The results from Equation (\ref{eq: total number of LSNe Ia from
  modified OM 10}) for the 20 investigated cadences are shown in Table
\ref{tab: total number of LSNe Ia from OM 10}. We find that mainly the
cumulative season length sets the order of the table and therefore for rolling cadences with a lower number of observing
seasons (blue $\rolling$ strategies) many LSNe Ia will not be detected, because
of the alternating observation scheme.
\begin{table}
\tabcolsep=0.15cm
\centering
\begin{tabular}{c|c|cc} & $N_\mathrm{LSNe Ia,cad}$ & $\bar{t}_\mathrm{eff,cad}$ in yr & $\Omega_\mathrm{cad}$ in $\si{\square\deg}$ \\
\hline
\krakenfour        &  101.9 &  4.64 &  24010 \\
\pontuszerozerotwo &   86.0 &  4.11 &  22926 \\
\colossusseven     &   84.0 &  5.16 &  17797 \\
\pontusnine        &   81.1 &  5.00 &  17758 \\
\rollingopsim      &   79.1 &  4.77 &  18148 \\
\rollingmixopsim   &   78.9 &  4.76 &  18132 \\
\krakentwo         &   78.0 &  4.79 &  17828 \\
\colossusfive      &   76.8 &  4.55 &  18475 \\
\pontusfivezerotwo &   76.3 &  4.74 &  17602 \\
\colossusfour      &   74.6 &  4.48 &  18202 \\
\baseline          &   73.4 &  4.64 &  17306 \\
\krakenfive        &   73.4 &  4.54 &  17680 \\
\krakentwosix      &   72.4 &  4.63 &  17119 \\
\pontusfivezerosix &   72.2 &  4.36 &  18132 \\
\altsched          &   61.7 &  3.81 &  17703 \\
\nexusseven        &   52.2 &  2.79 &  20471 \\
\mothranine        &   50.9 &  2.55 &  21874 \\
\krakenthreesix    &   45.2 &  2.79 &  17719 \\
\altschedrolling   &   37.9 &  2.03 &  20463 \\
\mothrafive        &   37.2 &  2.48 &  16417 \\ 
\end{tabular}  
\caption[Total number of LSNe Ia for different observing strategies based on OM10.]{Total number of LSNe Ia over 10-year survey calculated via equation (\ref{eq: total number of LSNe Ia from modified OM 10}) where $69\%$ are doubles and $31\%$ are quads. To understand the differences between the multiple strategies also the cumulative season length $\bar{t}_\mathrm{eff,cad}$ and the survey area $\Omega_\mathrm{cad}$ are shown. The total number depends on the selection criteria assumed in \cite{Oguri:2010}. If we relax the criteria like the image separation these numbers will be higher, but the order will be unchanged. Since differences in $\bar{t}_\mathrm{eff,cad}$ are much larger than in $\Omega_\mathrm{cad}$ the cumulative season length mostly sets the order of the table.}
\label{tab: total number of LSNe Ia from OM 10}
\end{table}

\subsection{LSST data only}
\label{sec:LSST Data only}
Here, we quantify the 20 investigated cadences for the case of
using LSST data only for measuring time delays. We have investigated
101 randomly picked quads and 101 randomly picked doubles. The
distribution of the source redshifts and time delays are shown as
orange lines in Figure \ref{fig:OM10 source redshift
  distribution}. The 202 systems are used to determine the fraction
$f_{a}$ of systems with good time delays:
\begin{equation}
f_{a} = \frac{N_{\Delta t,a}}{N_{a}} \qquad a = \mathrm{double, quad},
\label{eq:fraction of sytems with good delays}
\end{equation}
where $N_{\Delta t,a}$ is the number of systems with good time
delays and $N_{a}=101$ for $a=\mathrm{double, quad}$. Since we have
picked the same amounts of 
doubles and quads, whereas the real ratio between doubles and quads in
the OM10 catalog is 
$69:31$, the total fraction can be calculated as
\begin{equation}
f_\mathrm{total} = 0.69 f_\mathrm{double} + 0.31 f_\mathrm{quad}. 
\label{eq:res:LSST DATA only}
\end{equation}
The fractions of doubles $f_\mathrm{double}$ and quads
$f_\mathrm{quad}$ as well as the total fraction
$f_\mathrm{total}$ are shown in Table \ref{tab:LSST only
  fraction of systems}. It becomes clear that the fraction of systems
with good delays depends mostly on the inter-night gap, where
strategies with better sampling (blue $\rolling$ and magenta $\better$ strategies) provide
higher fractions.

\begin{table}[h!]
\centering
\begin{tabular}{c|c|cc}
&$f_\mathrm{total}$& $f_\mathrm{double}$ & $f_\mathrm{quad}$ \\
\hline
\altschedrolling   &  17.2 &  21.8 &  6.9 \\
\altsched          &  13.5 &  17.8 &  4.0 \\
\rollingmixopsim   &  10.2 &  13.9 &  2.0 \\
\pontusfivezerosix &   9.1 &  11.9 &  3.0 \\
\colossusseven     &   9.1 &  11.9 &  3.0 \\
\pontusnine        &   7.4 &   9.9 &  2.0 \\
\rollingopsim      &   6.8 &   8.9 &  2.0 \\
\mothrafive        &   6.1 &   7.9 &  2.0 \\
\krakenfour        &   5.8 &   7.9 &  1.0 \\
\krakentwo         &   5.8 &   7.9 &  1.0 \\
\nexusseven        &   4.8 &   6.9 &  0.0 \\
\krakentwosix      &   4.8 &   6.9 &  0.0 \\
\mothranine        &   4.7 &   5.9 &  2.0 \\
\krakenthreesix    &   4.7 &   5.9 &  2.0 \\
\colossusfive      &   3.7 &   5.0 &  1.0 \\
\baseline          &   3.7 &   5.0 &  1.0 \\
\colossusfour      &   3.4 &   5.0 &  0.0 \\
\krakenfive        &   2.0 &   3.0 &  0.0 \\
\pontusfivezerotwo &   1.4 &   2.0 &  0.0 \\
\pontuszerozerotwo &   1.4 &   2.0 &  0.0 \\
\end{tabular}
\caption[Fraction of systems with good delays for using LSST data only.]{Fraction of systems (in \%) of 202 investigated mock systems (101 doubles and 101 quads) where time delay has been measured with accuracy smaller than $1 \%$ and precision smaller than $5 \%$ for using LSST data only. The total fraction $f_\mathrm{total}$ accounts for the expected 69:31 ratio of doubles and quads from OM10 (see Equation (\ref{eq:res:LSST DATA only})). The investigation has been done for the ten fields listed in Table \ref{tab: 10 wfd fields}. These are not the final results as the total number of detected LSNe Ia is not taken into account.}
\label{tab:LSST only fraction of systems}
\end{table}

We determined the value of a given cadence strategy for our science
case, by combining Table \ref{tab: total number of LSNe Ia from
  OM 10} and \ref{tab:LSST only fraction of systems}. The results for
the 10-year survey are shown in Figure \ref{fig:LSST only, total
  number with good delays}. One sees that the key for obtaining a high
number of LSNe Ia with good delays is short inter-night gap while
keeping the cumulative season length baseline-like (magenta $\better$ strategies). Only for the strategy \texttt{alt\_sched\_rolling}, the much better
sampling can compensate for the short cumulative season length.

From the upper panel of Figure \ref{fig:distribution,LSST only, altsched}, it becomes
clear that only nearby systems ($z \lesssim 0.9$) with long time
delays ($\Delta t \gtrsim \SI{25}{\day}$) are measured
successfully. High redshift systems are overall fainter and the larger photometric errors make delay measurements more uncertain. Shorter time delays are not accessible because of the
sparse sampling and microlensing uncertainties. Looking at the total
number in Figure \ref{fig:LSST only, total number with good delays}, we
find that even the best strategies provide just a handful of systems
and therefore using just LSST data for measuring time delays is not
ideal. Therefore we investigate the prospects of
using follow-up observations in combination with LSST data.

\begin{figure}[h!]
\centering
\includegraphics[width=0.5\textwidth]{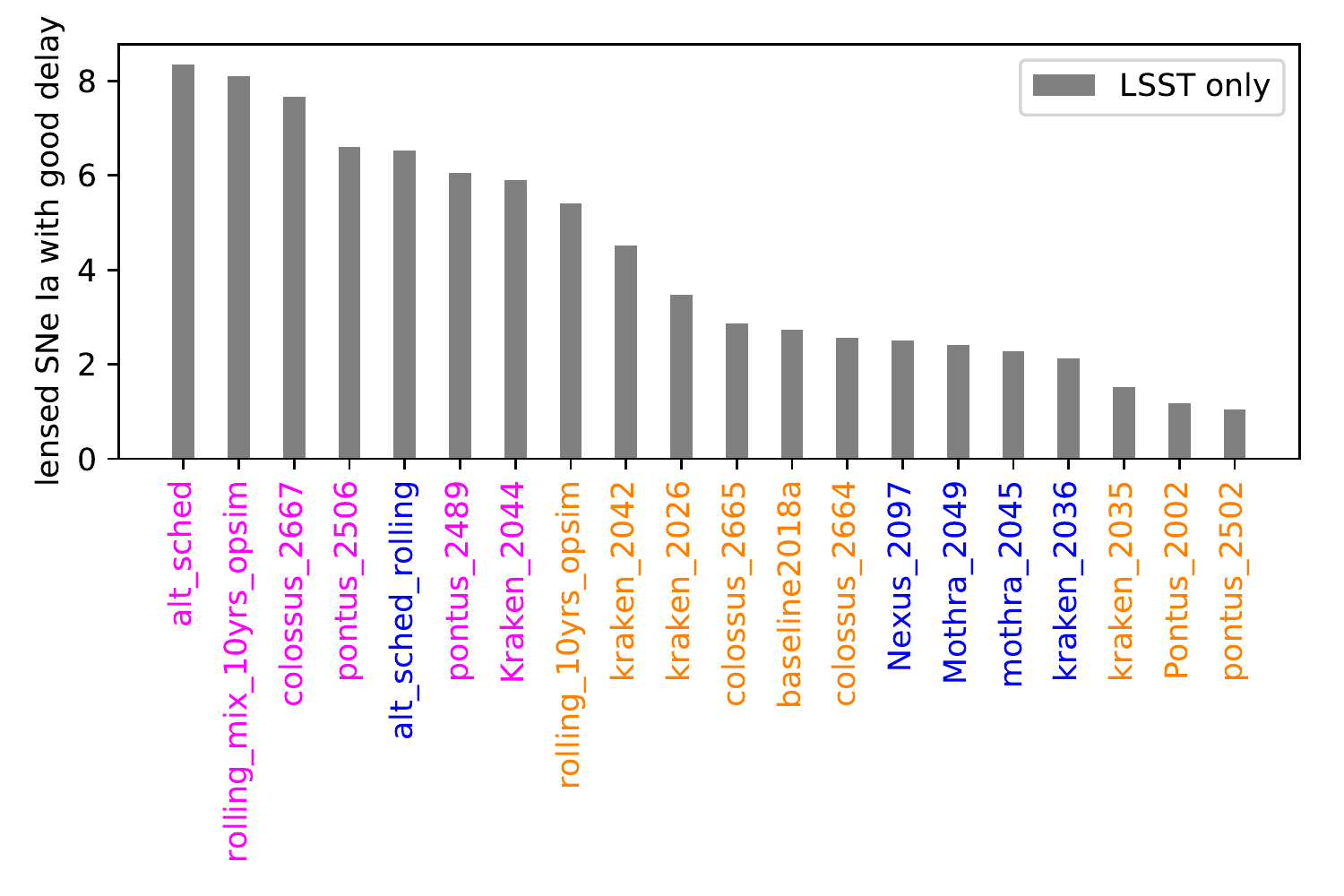}
\caption{Number of LSNe Ia for 10-year survey where time delay has been measured with accuracy $<1\%$ and precision $<5\%$ for using only LSST data.}
\label{fig:LSST only, total number with good delays}
\end{figure}

\begin{figure}
\centering
\includegraphics[width=0.45\textwidth]{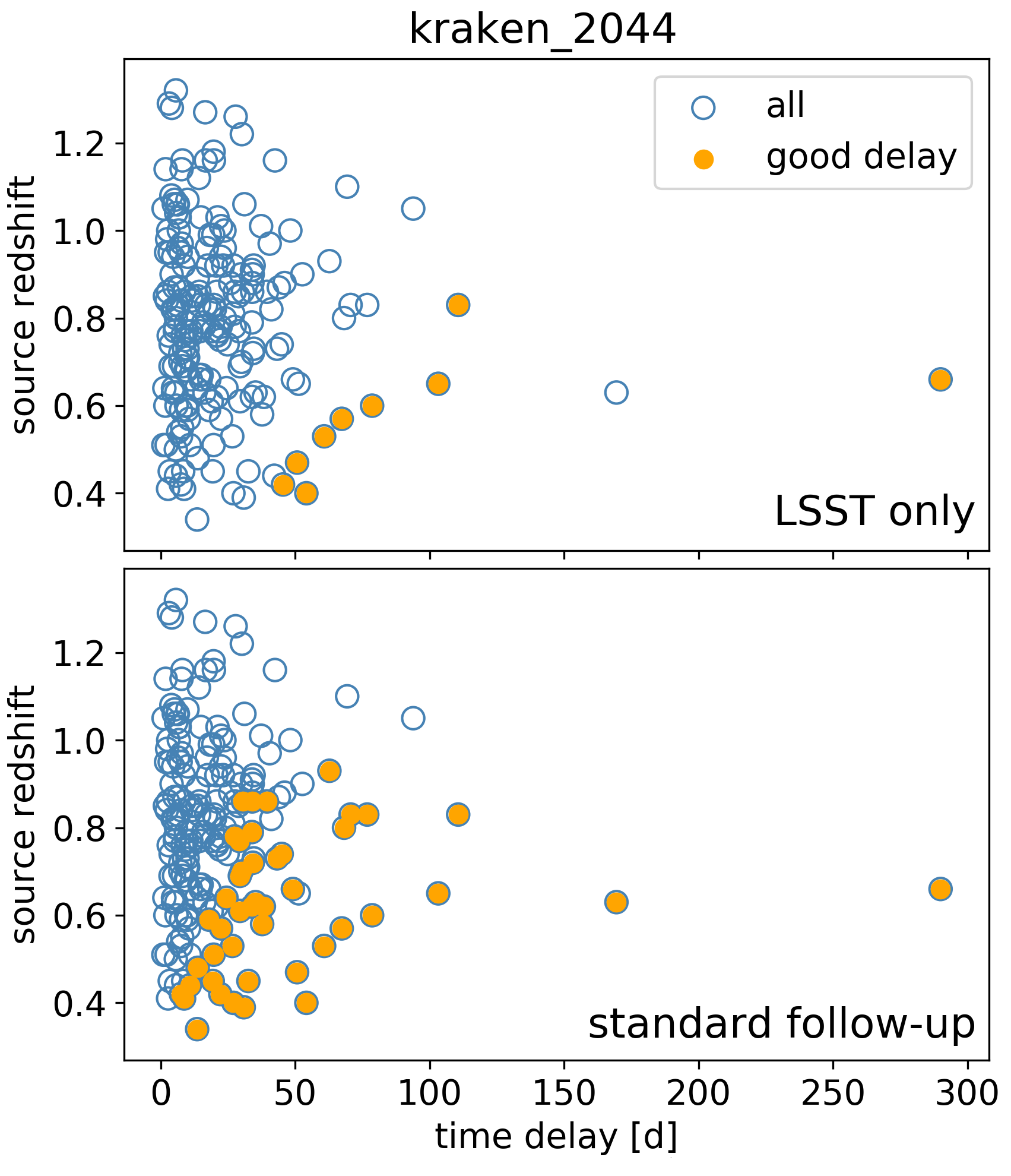}
\caption{Time-delay and source-redshift distribution for 202
  investigated mock LSNe Ia for ``LSST only" (upper panel) and ``LSST + follow-up" (lower panel) for observing
  strategy \texttt{kraken\_2044}. For a quad system, just a single delay is
  shown, either the first successful measured time-delay or the
  maximum of the six possible time delays. The blue circles show all 202
  investigated systems and the orange filled dots correspond to
  systems where the time delay has been measured with accuracy better
  than $1 \%$ and precision better than $5 \%$. Comparing the two panels we see significant improvement going from ``LSST only" to ``LSST + follow-up", which we find for most of the observing strategies as suggested by Table \ref{tab:LSST+follow-up fraction of systems}.}
\label{fig:distribution,LSST only, altsched}
\end{figure}

\subsection{LSST and follow-up observation}
\label{sec:LSST+follow-up}
Here, we investigate 20 different LSST observing strategies for
using LSST just as a discovery machine. For the time delay measurement we
assumed follow-up observation in the three filters \textit{g, r,} and \textit{i}, going to a
depth of $m_{5,\mathrm{g}}=\SI{24.6}{\mag}$,
$m_{5,\mathrm{r}}=\SI{24.2}{\mag}$ and
$m_{5,\mathrm{i}}=\SI{23.7}{\mag}$, which are similar to the depth of
the baseline cadence. These depths correspond to an observing time of approximately $\SI{6}{\minute}$ per filter and night on a 2 m telescope, which is despite diameter assumed to be identical to LSST (e.g., detector sensitivity). We adopt a realistic scenario where follow-up
starts two days after the third data point exceeds the 5$\sigma$
depth in any filter\footnote{\cite{Goldstein:2018bue} suggests that follow-up after three data points might be optimistic, but we would like to point out that this relies on the applied classification scheme \citep{Goldstein:2017bny}  
that does not make use of all available lensing information which would help with identification.}.
The follow-up is assumed to take place every
second night in all three filters. Alternative follow-up scenarios are investigated in Section \ref{sec:number of LSNe Ia with delays}.

Assuming a 2-meter telescope is a conservative assessment of the follow-up resources. Observing with larger telescopes would be quite reasonable, which would significantly reduce the exposure time or enable greater depth. The prospects of deeper follow-up will be discussed in Section \ref{sec:number of LSNe Ia with delays}.

The fraction of systems with well measured time delays is calculated
similar to Section \ref{sec:LSST Data only} and summarized in Table
\ref{tab:LSST+follow-up fraction of systems} for the 20 investigated
observing strategies. Applying only the accuracy requirement ($\tau_\mathrm{d,50} < 1 \%$) would yield for all cadence strategies about $30 \%$ less systems from the 202 investigated ones with a slight trend for more accurate systems for cadence strategies with improved sampling. 
Since for the case of ``LSST + follow-up" accuracy is only weakly dependent on the cadence strategy, the precision requirement ($\delta < 5 \%$) sets mostly the order of Table \ref{tab:LSST+follow-up fraction of systems}. Since blue ($\rolling$) and magenta ($\better$) strategies perform better than orange ($\baselinelike$) strategies in Tables \ref{tab:LSST only fraction of systems} and \ref{tab:LSST+follow-up fraction of systems}, we see that for a good precision a short inter-night gap is important. Even though the light curves for Table \ref{tab:LSST+follow-up fraction of systems} are created via follow-up resources, the better inter-night gap is still important to detect systems earlier and get better sampled light curves, although it is less important as for ``LSST only" where the ratio between the best and worst cadence
strategy is about $12$ instead of approximately $2$ for LSST +
follow-up. This makes clear that in terms of the fraction of systems
with good delays, the sampling of the LSST observing strategy is still important but far less than if
we would rely on LSST data only.
From Table \ref{tab:LSST+follow-up fraction of systems} we see that we can increase the fraction and therefore the number of LSNe Ia with good
delays for ``LSST + follow-up" in comparison to using only LSST data by a factor of 2 to 16, depending on the cadence strategy. For a
strategy like \texttt{alt\_sched\_rolling}, the effort of triggering the above
defined follow-up observation is questionable, but for most other
strategies the improvement is significant.

In practice it is important to pick systems with good accuracy for a final cosmological sample in order to determine $H_0$. We find that the reduction due to our accuracy requirement is partly due to microlensing but also the quality of the light curve plays a role since follow-up with greater depth provide more systems with accurate time delays. The prospects of greater depth are investigated in Section \ref{sec:number of LSNe Ia with delays} and one way to mitigate the effect of microlensing is the use of the color information as discussed in Appendix \ref{sec: Case study}. From Figure \ref{fig:Accuracy as function of time delay} we see that for ``LSST + follow-up" nearly all time delays greater than 20 days yield an accuracy within one percent, whereas going for short delays is dangerous in terms of adding bias to a final cosmological sample. 

\begin{figure}
\centering
\includegraphics[scale=0.6]{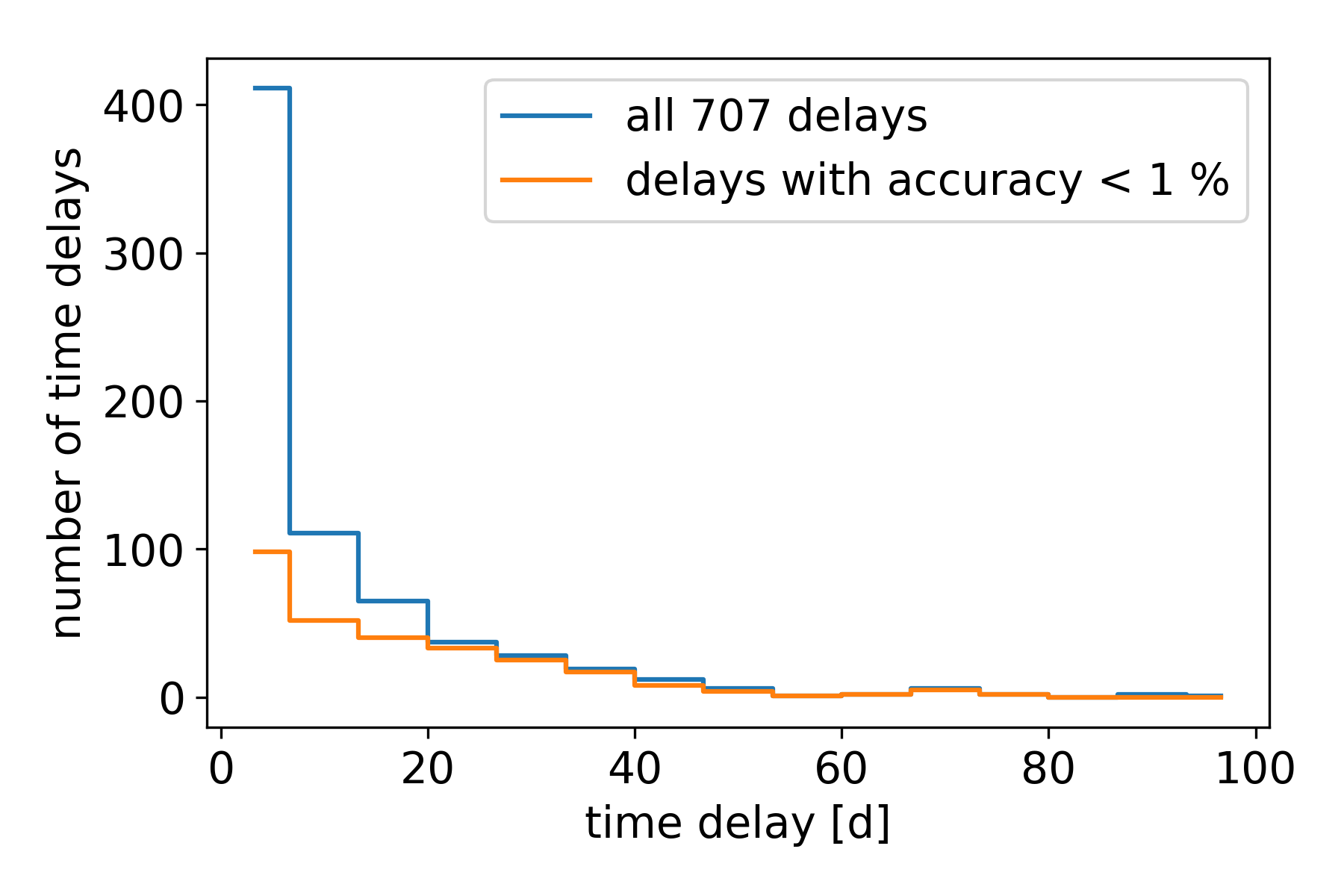}
\caption{Duration distribution for all 707 possible time delays (blue) and time delays with accuracy better than $1 \%$ (orange) from 202 investigated systems for ``LSST + follow-up" and observing strategy \texttt{colossus\_2667}. Nearly all time delays are accurate for pairs of images which yield a time delay greater than 20 days.}
\label{fig:Accuracy as function of time delay}
\end{figure}

In the lower panel of Figure \ref{fig:distribution,LSST only, altsched}, we
see that similar to the case of using only LSST data, we are limited to
nearby systems ($z \lesssim 0.9$). In terms of time delays, we can
reach lower values due to the much better quality of the
light curve, but still, most of the short time delays are not
accessible because of microlensing and our cut on precision.
\begin{table}[h!]
\tabcolsep=0.15cm
\centering
\begin{tabular}{c|c|cc|c}
&$f_\mathrm{total}$& $f_\mathrm{double}$ & $f_\mathrm{quad}$ & $\frac{f_\mathrm{total,LSST+follow-up}}{f_\mathrm{total,LSST only}}$ \\
\hline
\altschedrolling   &  34.4 &  43.6 &  13.9 &   2.0 \\
\altsched          &  32.1 &  41.6 &  10.9 &   2.4 \\
\colossusseven     &  31.1 &  40.6 &   9.9 &   3.4 \\
\pontusfivezerosix &  27.0 &  34.7 &   9.9 &   3.0 \\
\mothrafive        &  26.7 &  35.6 &   6.9 &   4.4 \\
\krakenfour        &  26.7 &  34.7 &   8.9 &   4.6 \\
\krakentwo         &  25.0 &  32.7 &   7.9 &   4.3 \\
\krakentwosix      &  24.3 &  31.7 &   7.9 &   5.1 \\
\krakenthreesix    &  24.0 &  31.7 &   6.9 &   5.1 \\
\pontusnine        &  23.6 &  30.7 &   7.9 &   3.2 \\
\mothranine        &  23.6 &  30.7 &   7.9 &   5.0 \\
\rollingmixopsim   &  23.3 &  30.7 &   6.9 &   2.3 \\
\nexusseven        &  23.3 &  30.7 &   6.9 &   4.9 \\
\baseline          &  23.3 &  30.7 &   6.9 &   6.3 \\
\pontuszerozerotwo &  22.0 &  28.7 &   6.9 &  16.1 \\
\krakenfive        &  22.0 &  28.7 &   6.9 &  10.7 \\
\colossusfive      &  22.0 &  28.7 &   6.9 &   5.9 \\
\colossusfour      &  22.0 &  28.7 &   6.9 &   6.4 \\
\pontusfivezerotwo &  20.3 &  26.7 &   5.9 &  14.8 \\
\rollingopsim      &  18.2 &  23.8 &   5.9 &   2.7 \\ 
\end{tabular}
\caption[Fraction of systems with good delays for using LSST as discovery machine in combination with follow-up observation to get time delays.]{Fraction of systems (column two, three \& four in \%) of 202 investigated mock systems (101 doubles and 101 quads) where time delay has been measured with accuracy smaller than $1\%$ and precision smaller than $5\%$ for using LSST as a discovery machine and getting time delays from follow-up observations. The investigation has been done for the ten fields listed in Table \ref{tab: 10 wfd fields}. The 5th column shows how much better a cadence performs in comparison to using LSST data only. This table is insufficient to rank different cadence strategies because the total number of detected LSNe Ia is not taken into account.}
\label{tab:LSST+follow-up fraction of systems}
\end{table}

By combining Tables \ref{tab: total number of LSNe Ia from OM 10} and
\ref{tab:LSST+follow-up fraction of systems}, we get the total amount
of LSNe Ia with good time delays as shown in Figure \ref{fig:LSST only
  and LSST+follow-up, total number with good delays}. We note that the presented results have errors within $10 \%$ due to uncertainties in the calculated area and sampling. Another point is that we do not apply the sharp OM10 cut of 22.6 mag as mentioned in Section \ref{sec:OM10}. We find that we are also able to get good time delays for fainter systems ($> \SI{22.6}{\mag}$) although in number they are a factor of at least 1.7 fewer than for bright ones ($\le \SI{22.6}{mag}$). This means that the numbers presented in Table \ref{tab: total number of LSNe Ia from OM 10} and therefore also the numbers in Figure \ref{fig:LSST only and LSST+follow-up, total number with good delays} are a conservative estimate and in reality we can expect even more systems with well measured time delays. An overly optimistic version of Figure \ref{fig:LSST only and LSST+follow-up, total number with good delays} is presented in Appendix \ref{sec:Appendix optimistic estimate of the number of LSNe Ia with well measured delay}. While these sources of uncertainties might change the ordering presented in Figure \ref{fig:LSST only and LSST+follow-up, total number with good delays} slightly, it does not influence our overall conclusions which will be presented in the following. 
  
We see that for
the current baseline strategy we would expect about $17$ LSNe Ia with
good delays over the 10-year survey. To increase this number, the most
promising strategies are those with a baseline-like cumulative season
length $\bar{t}_\mathrm{eff,cad}$ and an enhanced sampling (magenta $\better$
strategies). To achieve this, the most efficient way would be to get
rid of the revisit within the same night (compare \texttt{colossus\_2667} to
\texttt{baseline2018a}). Because this would make the science case of fast
moving objects impossible, we think a reasonable compromise is to do
the revisit within the same night in a different filter \citep{Lochner:2018}. This performs
worse than doing single visits but still better than doing the revisit
in the same filter (compare \texttt{pontus\_2506} to \texttt{colossus\_2667} and
\texttt{baseline2018a}). In terms of the cumulative season length, it seems
appropriate to stay with a baseline-like season length of about 170
days and ten seasons. Further improvement can be achieved by the replacement of the $2 \times
\SI{15}{\s}$ exposure by $1 \times \SI{30}{\s}$ to improve efficiency
(compare \texttt{kraken\_2042} to \texttt{baseline2018a}).

Although our numbers for an extended WFD area by $\SI{6700}{\square\deg}$ (compare \texttt{Kraken\_2044} and
\texttt{colossus\_2667}, and \texttt{Pontus\_2002} and \texttt{baseline2018a}) are increased, we only find this for ``LSST + follow-up".  For ``LSST only", strategies with a smaller WFD footprint perform better. Therefore we suggest to stick with the WFD footprint of $\SI{18000}{\square\deg}$, as used for 16 of the 20 investigated observing strategies, but we are also fine with $\SI{24700}{\square\deg}$. 
Concerning the depth of the observing strategy most of the investigated
strategies provide a similar 5$\sigma$ depth as the baseline cadence (see right panels of Figure \ref{fig:Comparison 10 fields to WFD fields}). Those strategies with a slightly lower 5$\sigma$ depth (\texttt{alt\_sched}, \texttt{alt\_sched\_rolling} and \texttt{pontus\_2489}) show no significant deviations in the results, which is related to their enhanced sampling in comparison to the baseline cadence.
Another
interesting scenario to investigate is the redistribution from visits
in \textit{y} band to more useful bands for LSNe Ia as done in \texttt{alt\_sched}. This
means going from a distribution of visits in \textit{ugrizy}: (6.8, 9.4, 21.5,
21.6, 20.2, 20.4)\% to (8.2, 11.0, 27.6, 18.1, 25.6, 9.5)\%. Because
of the many differences between \texttt{alt\_sched} and \texttt{baseline2018a}, a direct
comparison is impossible but we expect some improvement. A simulation implementing the redistribution with the greedy
algorithm used for \texttt{baseline2018a} would be helpful to quantify
this.

Furthermore, a very important result: most rolling cadence strategies are disfavored for our LSNe Ia science case. For
these cadence strategies, the shortened cumulative season lengths
$\bar{t}_\mathrm{eff,cad}$ lead to an overall more negative impact on
the number of LSNe Ia with delays, compared to the gain from the
increased sampling frequency.

\begin{figure*}[h!]
\centering
\includegraphics[width=0.85\textwidth]{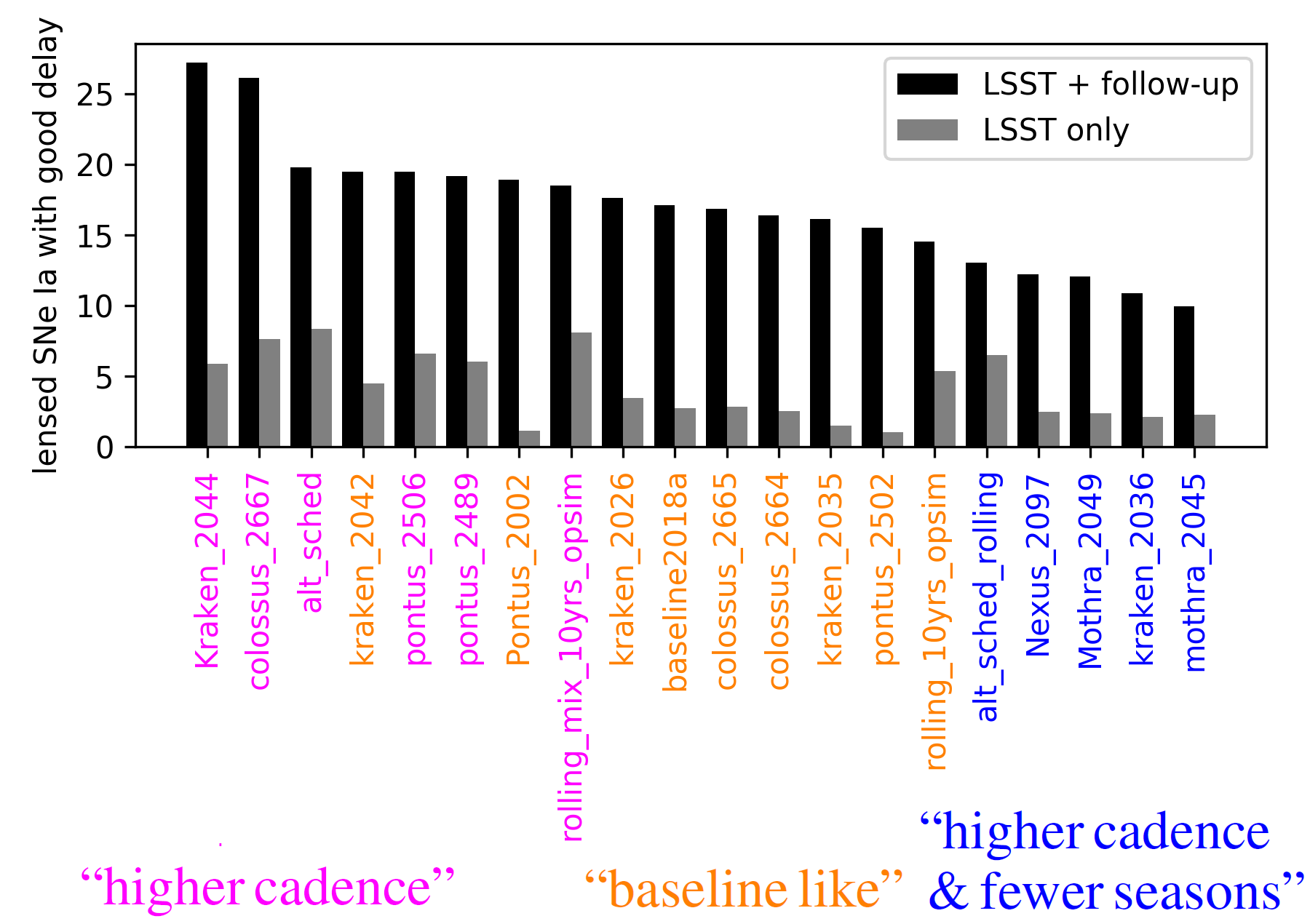}
\caption{Number of LSNe Ia for 10-year survey where time delay has been measured
  with accuracy $< 1\%$ and precision $< 5\%$ by using
  LSST as discovery machine in combination with follow-up
  observations for measuring time delays (black bars) and using only
  LSST data (gray bars, see also Figure \ref{fig:LSST only, total
    number with good delays}).  Follow-up is
  every second night in filters \textit{g, r,} and \textit{i}, starting two nights after
  third LSST detection (with brightness exceeding 5$\sigma$ depth in
  any filter). With follow-up observations, we get a substantial
  increase in the number of LSNe Ia systems with good measured delays. The numbers shown in this figure are a conservative estimate. An optimistic approach is discussed in Appendix \ref{sec:Appendix optimistic estimate of the number of LSNe Ia with well measured delay}, leading to the same overall conclusion about the categories of cadence strategies (magenta, orange, and blue) but providing about 3.5 times more LSNe Ia with well-measured delays.} 
\label{fig:LSST only and LSST+follow-up, total number with good delays}
\end{figure*}

\subsection{Different follow-up scenarios}
\label{sec:number of LSNe Ia with delays}
In this section, the prospects of increasing the number of LSNe Ia by
assuming different follow-up scenarios are discussed. For this purpose,
we have investigated a sample of 100 mock LSNe Ia (50 mock quads and
50 mock doubles). The result for 
the standard follow-up case is shown in Table \ref{tab:follow-up
  alternative scenarios and no microlensing} first row  for the two cadence strategies \texttt{baseline2018a} and \texttt{alt\_sched}. To clarify,
the standard follow-up scenario assumes observations in the three filters
\textit{g, r,} and \textit{i}, going to a depth of $m_{5,\mathrm{g}}=\SI{24.6}{\mag}$,
$m_{5,\mathrm{r}}=\SI{24.2}{\mag}$ and
$m_{5,\mathrm{i}}=\SI{23.7}{\mag}$. Follow-up is assumed every second
night in all three filters two days after the third data point exceeds the
5$\sigma$ depth in any filter.

\begin{table}
\centering
\tabcolsep=0.11cm
\begin{tabular}{cc|c|c}
& row& \texttt{baseline2018a} & \texttt{alt\_sched} \\ 
\hline
LSST + follow-up& 1 & 16.5 $(22.4 \%)$ & 21.0 $(33.9 \%)$ \\
\hline
follow-up in bands $riz$& 2 & 15.0 $(20.4 \%)$ & 20.2 $(32.7 \%)$ \\
follow-up after 2 data points& 3 & 20.0 $(27.2 \%)$ & 23.0 $(37.3 \%)$ \\
daily follow-up& 4 & 19.4 $(26.4 \%)$ & 23.3 $(37.8 \%)$ \\
follow-up every third day& 5 & 13.5 $(18.4 \%)$ & 18.0 $(29.2 \%)$  \\
deeper follow-up (1 mag)& 6 & 28.2 $(38.4 \%)$ & 27.0 $(43.8 \%)$ \\
deeper follow-up (2 mag)& 7 & 37.1 $(50.6 \%)$  & 34.0 $(55.0 \%)$ \\
deeper follow-up (4 mag)& 8 & 39.4 $(53.7 \%)$ & 37.6 $(60.9 \%)$ \\
\hline
no microlensing& 9 & 35.7 $(48.6 \%)$ & 33.3 $(53.9 \%)$\\
no microl., 1 mag deeper& 10  & 48.4 $(65.9 \%)$ & 43.2 $(69.9 \%)$  \\
\end{tabular}
\caption{Summary of different
  follow-up strategies and prospects of an
  improved analysis technique concerning modeling of
  microlensing. For the two strategies \texttt{baseline2018a} and \texttt{alt\_sched},
  the number of LSNe Ia with good quality time-delay measurements over the 10-year survey
  are shown for each considered scenario, where 100 mock LSNe Ia have been
  investigated. The percentages in the
  brackets show how many of the total numbers of LSNe Ia (73.4 for baseline2018a and 61.7 for alt\_sched from Table \ref{tab: total number of LSNe Ia from OM 10}) have well measured
  time delays. The exact definition of ``LSST + follow-up" (row 1) is
  described in the text and the scenarios from rows two to eight are
  alternative follow-up scenarios detailed in the text.  Rows nine and
  ten are hypothetical numbers interesting for future improved analysis
  techniques of microlensing.} 
\label{tab:follow-up alternative scenarios and no microlensing}
\end{table}

An alternative follow-up scenario would be to observe in bands \textit{r, i,}
and \textit{z}. The numbers in the second row are slightly worse than those for
following up in bands \textit{g, r,} and \textit{i}, even though high redshift SNe are well visible in the \textit{z} band. The reason for this is that we have assumed a baseline like 5$\sigma$ depth for the follow-up observations, with $m_{5,\mathrm{z}}=\SI{22.8}{\mag}$ which is $\SI{1.8}{\mag}$ lower than the 5$\sigma$ depth in the \textit{g} band. 

The more aggressive approach is to trigger follow-up after the
second data point exceeds the 5$\sigma$ depth (see row 3). The
improvement of $10$ to $21\%$ might look promising, but also many more
false positives will be detected and therefore some observing
time would likely be wasted on false positives.

Of further interest is also the cadence of the follow-up
observation. Therefore we consider two additional cases where we
follow-up daily (see row 4) and every third day (see row 5), instead of the standard follow-up
of every second day. While going down to observations every three days
decreases the number of LSNe Ia with good delays by about $18 \%$, daily visits improve on a level of 11 to $18 \%$. Going
from a two-days to a single day cadence increases the effort of
follow-up significantly by increasing the numbers of LSNe Ia only
slightly.

A more promising approach is to keep the follow-up observations every
two days but increase the depth. To go one magnitude deeper (see row 6) than the
average baseline depth a total
observing time of approximately $\SI{45}{\min}$ per night is needed for a 2 m telescope as in Section \ref{sec:LSST+follow-up}, which is feasible. For \texttt{alt\_sched}, this
leads to an improvement of $29\%$ in comparison to the standard follow-up
scenario and therefore slightly better than the daily follow-up
case. For \texttt{baseline2018a}, the improvement is $71\%$ and therefore
definitely worth considering the effort (compare upper two panels in
Figure \ref{fig:distribution,LSST + follow-up, baseline2018a}).

Another possibility is to go two magnitudes deeper but therefore we have
to observe approximately $\SI{2}{\hour}$ per night to get observations in 3
filters. This seems only feasible for a two-meter-telescope which can
observe simultaneously in three filters or by a telescope with a larger
diameter. For \texttt{alt\_sched}, this means an improvement in comparison to
the standard follow-up scenario of $62\%$ and for \texttt{baseline2018a} an
improvement of $125\%$. Going another two magnitudes
deeper does not increase the number of LSNe Ia significantly and
therefore going beyond two magnitudes is in our investigation not worth the effort
(compare rows seven and eight in Table \ref{tab:follow-up alternative
  scenarios and no microlensing}). 

A limiting factor of our analysis is the microlensing effect which is
not taken into account in our time-delay measurement with {\tt PyCS} and
therefore we are not able to accurately measure short time delays (see Figure \ref{fig:distribution,LSST only, altsched} and the upper
two panels of Figure \ref{fig:distribution,LSST + follow-up,
  baseline2018a}) because we do not model the bias due to microlensing magnification, which is an absolute bias in time, whereas the accuracy is relative to the length of the delay. In rows nine and ten of Table \ref{tab:follow-up
  alternative scenarios and no microlensing}, we see that we could
increase the number of LSNe Ia with good delays by a factor of $60\%$
to $120\%$ in the best case scenario, where we imagine a perfect correction for microlensing deviations. This would give us access to short time-delays
as visible in the comparison of the upper two panels and the lower two
panels of Figure \ref{fig:distribution,LSST + follow-up,
  baseline2018a} and
 therefore encourages the use of 
color curves instead of light curves to reduce the impact of microlensing on the delay measurement as suggested by
\cite{Goldstein:2017bny} and discussed in Appendix \ref{sec: Case
  study}. Also, the approach of using SNe Ia templates to fit the intrinsic light curve shape including effects of microlensing might be
reasonable and produce higher fraction of good delays.  Some of these are currently being explored
(\citeauthor{PierelRodney19} \citeyear{PierelRodney19}; Collett et al., in prep, T.~Collett, priv.~comm.).

\begin{figure}
\centering
\subfigure{\includegraphics[width=0.425\textwidth]{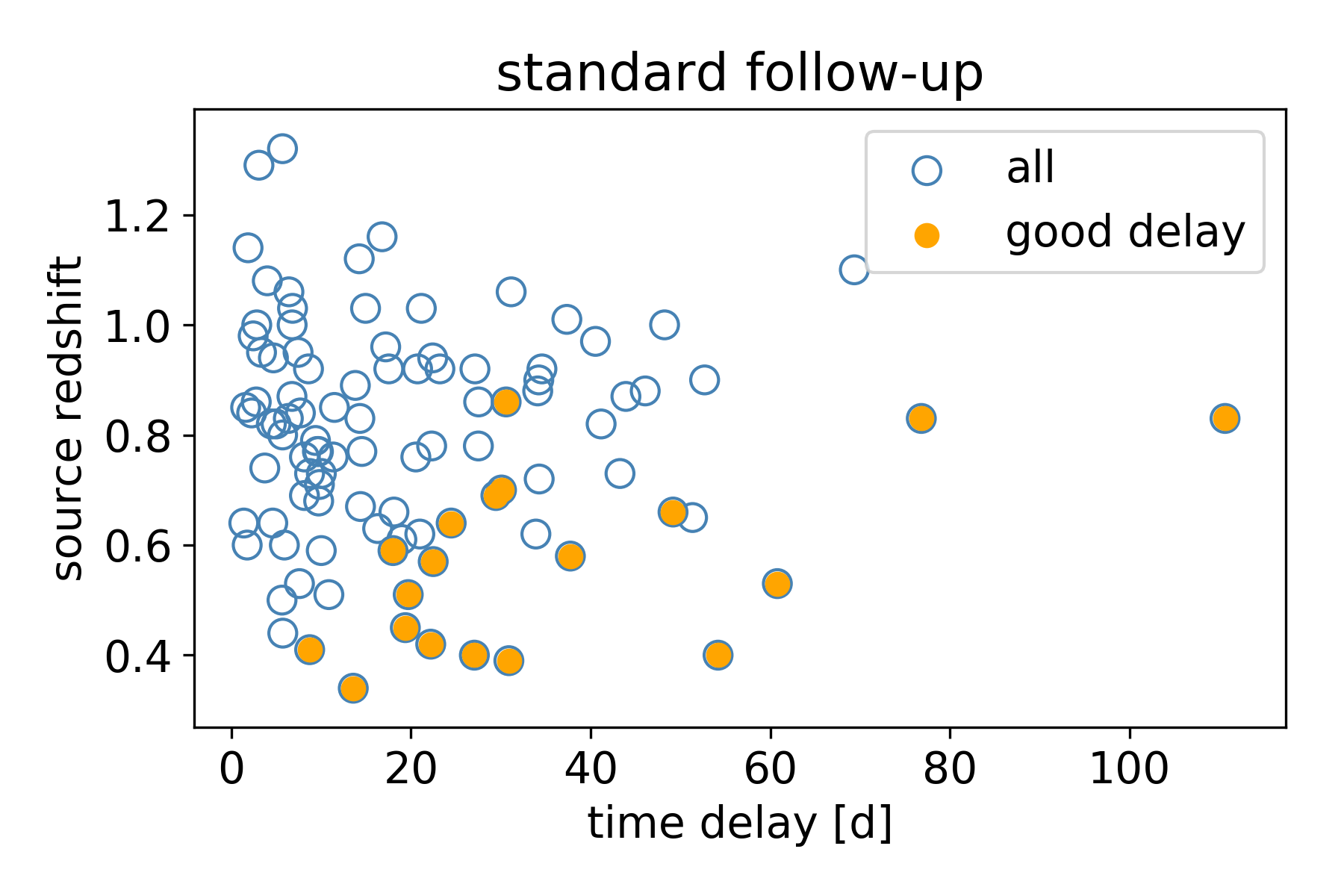}}
\subfigure{\includegraphics[width=0.425\textwidth]{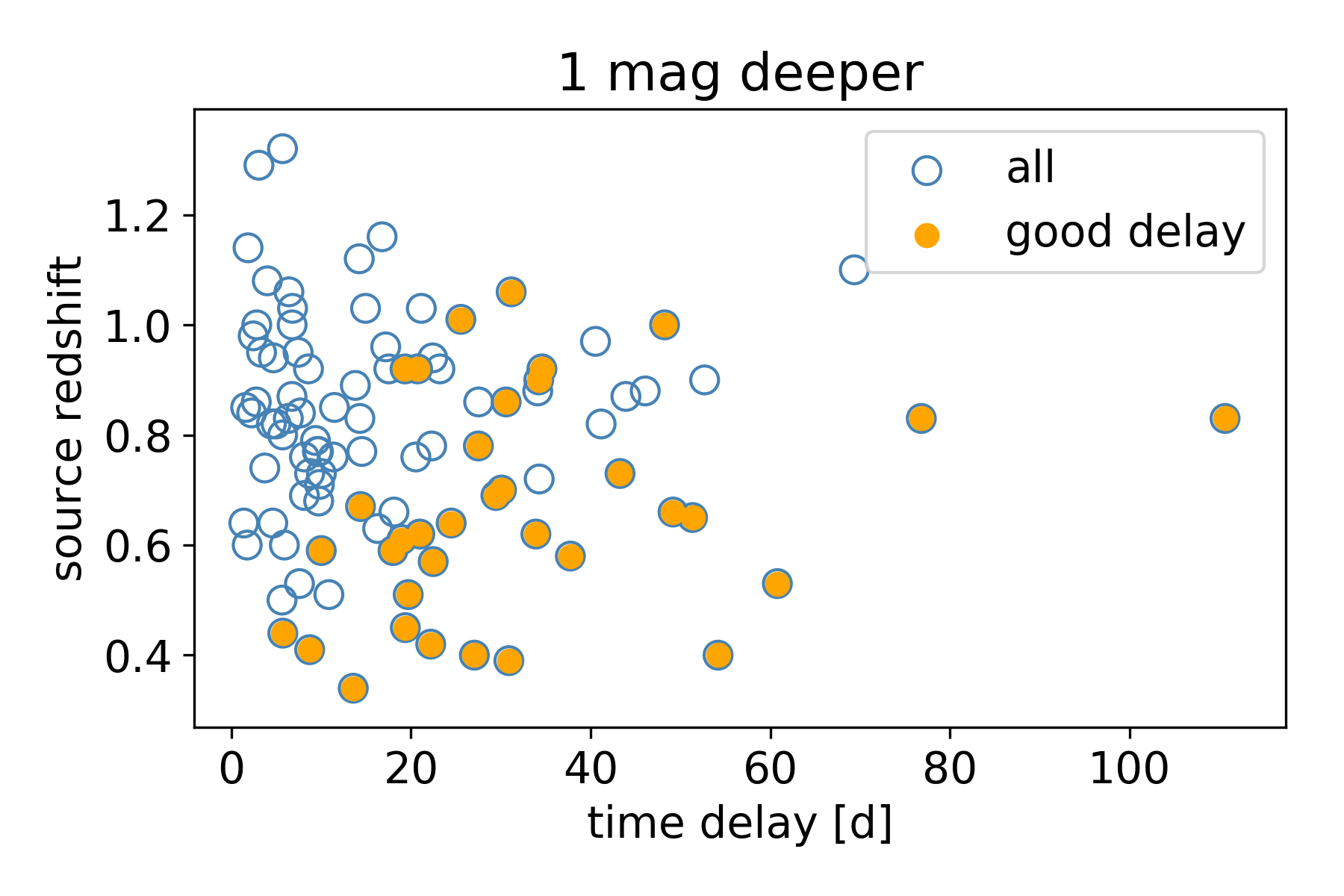}}
\subfigure{\includegraphics[width=0.425\textwidth]{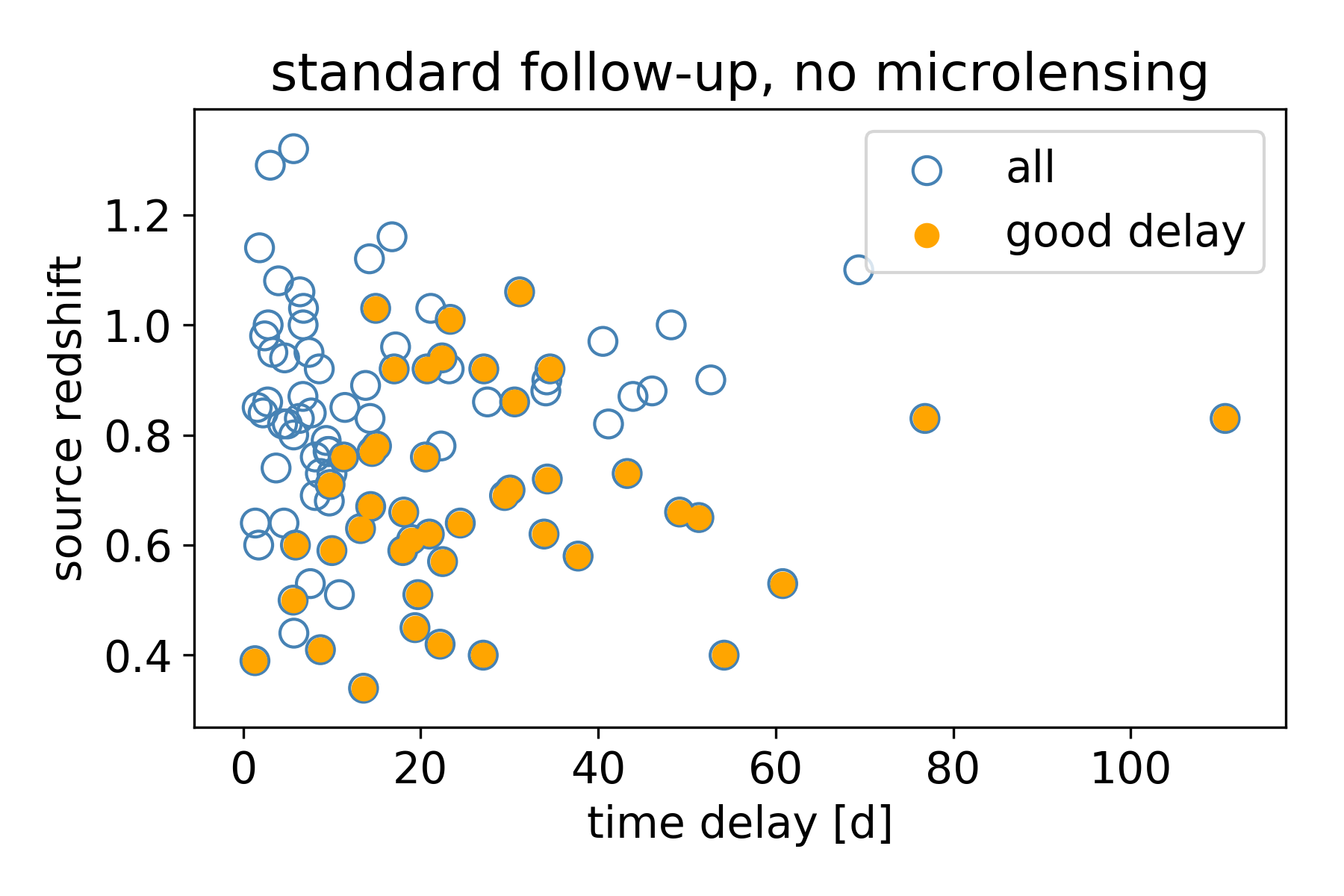}}
\subfigure{\includegraphics[width=0.425\textwidth]{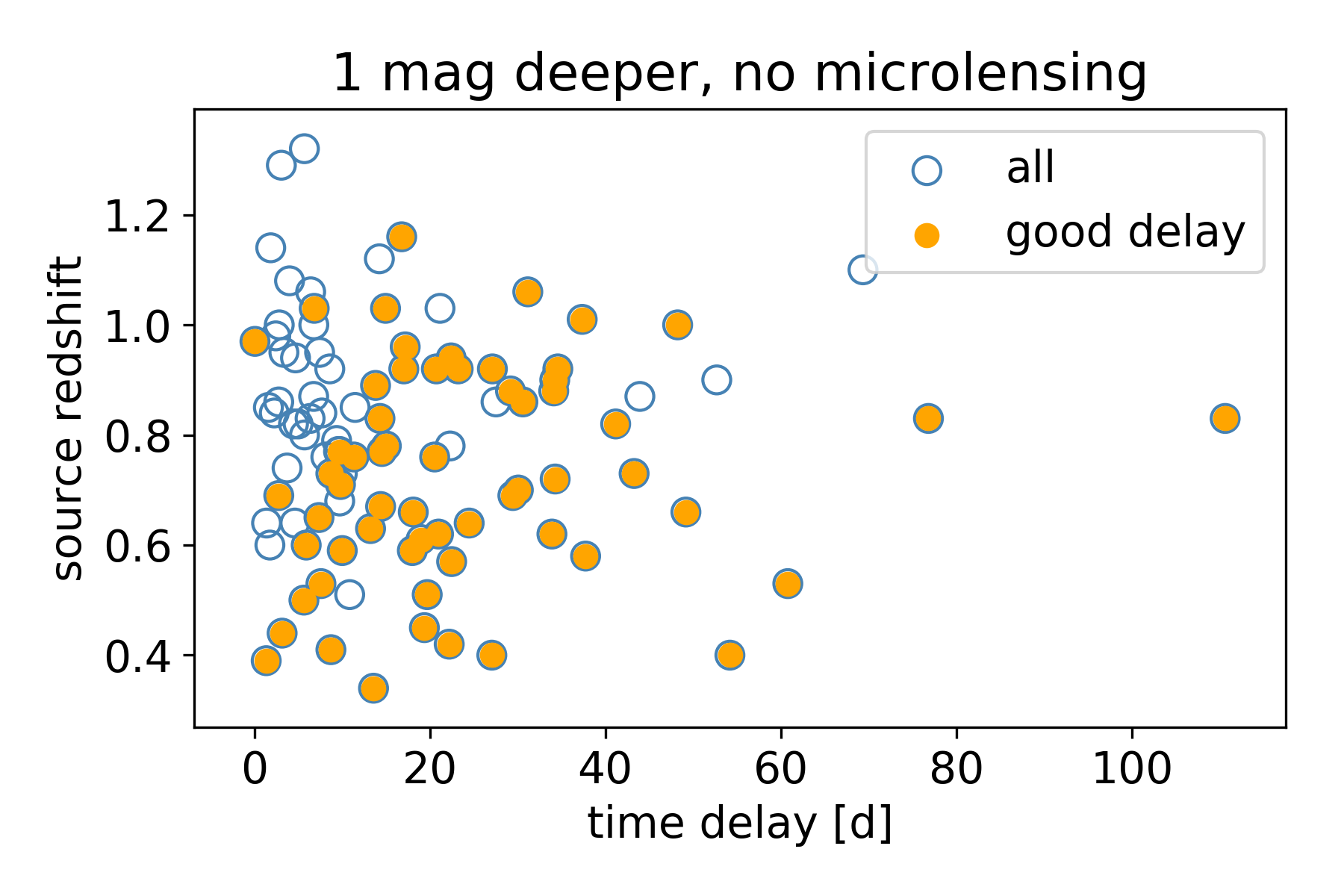}}

\caption{Time-delay and
  source-redshift distribution for 100 investigated mock LSNe for
  \texttt{baseline2018a}, similar to Figure \ref{fig:distribution,LSST only,
    altsched}. The upper two panels show the standard follow-up
  observation (first panel) and the option going one magnitude deeper
  (second panel). The lower two panels show the same follow-up
  scenarios hypothetically without microlensing. The distributions
  vary slightly because for a quad system just a single time delay is
  shown, either the first successfully measured delay or the maximum
  of the six possible delays.}
\label{fig:distribution,LSST + follow-up, baseline2018a}
\end{figure}

\section{Discussion and summary}
\label{sec:Summary and Future Prospects}
In this work, we explored different LSST cadence strategies for
measuring time delays in strongly lensed SNe Ia. As illustrated in
Figure \ref{fig:LSST only and LSST+follow-up, total number with good
  delays}, we have found that using LSST just as discovery machine in
combination with high cadence follow-up observation for the delay measurement is
the best way to increase the number of LSNe Ia with good time delays.
In contrast, using only LSST data is not ideal.

To estimate the resulting $H_0$ constraint from a sample of LSST LSNe
Ia, we assume that each LSNe Ia system with good delays yields
typically an $H_0$ measurement with approximately $5\%$ uncertainty in flat $\Lambda$CDM  (including
all sources of uncertainties such as the time-delay uncertainty
investigated in this paper, and lens mass mass modeling
uncertainties).  This is currently achieved with the best lensed
quasar systems of the H0LiCOW sample, and serves as a reference given
that we expect LSNe Ia to yield similar or better constraints than
that of lensed quasars.  While focussing only on LSNe Ia with good
delays could potentially introduce selection bias, we suspect such
biases to be small and, if present, could be corrected
\citep[e.g.,][]{Collett:16}.  Thus, for a sample of $N$ lenses, the
uncertainty on $H_0$ would scale approximately as $5\%$/sqrt$(N)$, 
assuming Gaussian uncertainties.  With LSST data only, 
the number of lensed SNe Ia from our investigation (Figure
\ref{fig:LSST only and LSST+follow-up, total number with good delays})
ranges from approximately $1-8$, depending on the strategy.  This
would yield an $H_0$ constraint with about $2-5\%$ uncertainty from the
sample. In the case of LSST with follow-up, the number of lensed SNe
increase substantially, varying from approximately $10-28$,
translating to an $H_0$ constraint with about $1-2\%$ uncertainty.
Therefore, with optimal LSST observing strategy and fast-response
follow-up, we would reach percent-level constraint on $H_0$, which is
a factor of two to five lower in uncertainty compared to the case of
LSST-only scenario.

From the investigated cadence strategies for the follow-up scenario,
we have found that observing strategies with an improved sampling by
keeping everything else baseline-like is, in general, the best
observing strategy for our science case. An ideal strategy is
presented in the following key points:
\begin{itemize}
\item Ten seasons with a season length of 170 days or
 longer
\item WFD footprint of $\SI{18000}{\square\deg}$ up to $\SI{24700}{\square\deg}$
 \item One revisit within a night in a different filter than the first visit
 \item Replacement of $2 \times \SI{15}{\s}$ exposure by $1 \times \SI{30}{\s}$
\item Distribution of visits like \texttt{alt\_sched} [\textit{ugrizy} as $\sim(8.2, 11.0, 27.6, 18.1, 25.6, 9.5)$\%]. 
\end{itemize}

Another very important point is that most of the suggested rolling
cadences are clearly disfavored for our science case
because many LSNe Ia will not even be detected due to the reduced
cumulative season length. The only rolling cadence which performed
well is \texttt{rolling\_mix\_10yrs\_opsim}, but this is most likely because the WFD
stays on in the background and additionally revisits are done in
different filters, which can partly compensate for the not ideal
``rolling" feature.

We have assumed that follow-up observations starts two days after the
third LSST data point exceeds the 5$\sigma$ depth. The follow up is done
every second night in three filters \textit{g, r,} and \textit{i} to a depth of
$m_{5,\mathrm{g}}=\SI{24.6}{\mag}$,
$m_{5,\mathrm{r}}=\SI{24.2}{\mag}$ and
$m_{5,\mathrm{i}}=\SI{23.7}{\mag}$, which is feasible with a 2-meter telescope.  To improve on that mainly a
greater depth is of interest. Follow-up observations going one magnitude
deeper than the baseline 5$\sigma$ depth, or even two magnitude deeper,
if feasible, will increase the number of LSNe Ia with good time-delays
significantly. Going beyond two magnitude deeper is not worth the
effort.

We would like to point out that we have only investigated LSNe
Ia. Although a single lensed Core-Collapse (CC) SN is less valuable
than a LSNe Ia
(given the standardizable light curves of SNe Ia)
, the larger sample of lensed CC SNe, primarily type IIn \citep{Goldstein:2018bue,Wojtak:2019hsc}, which will be detected by
LSST makes them as well relevant for time-delay cosmography. Due to
the different light curve shapes and luminosities the optimal cadence
strategy for measuring time delays in CC SNe might be different from
the one for LSNe Ia. At least in terms of total number of lensed CC
SNe the strategies will be ordered in the same way as in Table
\ref{tab: total number of LSNe Ia from OM 10} but the numbers will be
a factor of 1.8 higher \citep{Oguri:2010}. In terms of measuring time delays the improved sampling requested from our investigation of LSNe Ia will be also helpful for the case of CC SNe. To investigate the prospects
of measuring time delays in lensed CC SNe similar to the case of LSNe
Ia the specific intensity from a theoretical model is required.

In terms of analyzing the data it seems promising to find ways to
reduce the impact of microlensing. One possibility will be the use of
color curves instead of light curves. To do this, it might be worth to
implement SNe template fitting instead of splines into {\tt PyCS}. With the 
recent discovery of the very first LSNe system and the expected
sample from LSST, our work demonstrates that time-delay cosmography as
envisioned by \cite{Refsdal:1964} has bright prospects in the LSST era.

\begin{acknowledgements}
We thank W.~Hillebrandt, S.~Blondin, D. A.~Goldstein for useful discussions, 
and the internal LSST DESC reviewers S.~Rodney, A.~Goobar, and T.~E.~Collett for their feedback that improved the presentation of our paper. We also thank the anonymous referee for helpful comments.
SH and SHS thank the Max Planck Society for support through the Max
Planck Research Group for SHS. This project has received funding from
the European Research Council (ERC) under the European Union’s Horizon
2020 research and innovation programme (grant agreement No
771776). This research was supported in part by Perimeter Institute
for Theoretical Physics. Research at Perimeter Institute is supported
by the Government of Canada through the Department of Innovation,
Science and Economic Development and by the Province of Ontario
through the Ministry of Research, Innovation and Science. 
UMN has been supported by the Transregional Collaborative Research
Center TRR33 ``The Dark Universe" of the Deutsche
Forschungsgemeinschaft.
VB, JHHC and FC acknowledge support from the Swiss National Science
Foundation and through European Research Council (ERC) under the European
Union's Horizon 2020 research and innovation programme (COSMICLENS:
grant agreement No 787866).
DR acknowledges support from the
National Science Foundation Graduate Research Fellowship under Grant
No.~DGE 1752814.
HA has been supported by the Rutgers Discovery Informatics
Institute Fellowship of Excellence in Computational and Data Science
for academic years 2017-2018, 2018-2019.  
MK acknowledges support from
the Klaus Tschira Foundation. 
This work was supported in part by World Premier International
Research Center Initiative (WPI Initiative), MEXT, Japan, and JSPS
KAKENHI Grant Number JP15H05892 and JP18K03693.
The DESC acknowledges ongoing support from the Institut National de Physique Nucl\'eaire et de Physique des Particules in France; the Science \& Technology Facilities Council in the United Kingdom; and the Department of Energy, the National Science Foundation, and the LSST Corporation in the United States.  DESC uses resources of the IN2P3 Computing Center (CC-IN2P3--Lyon/Villeurbanne - France) funded by the Centre National de la Recherche Scientifique; the National Energy Research Scientific Computing Center, a DOE Office of Science User Facility supported by the Office of Science of the U.S.\ Department of Energy under Contract No.\ DE-AC02-05CH11231; STFC DiRAC HPC Facilities, funded by UK BIS National E-infrastructure capital grants; and the UK particle physics grid, supported by the GridPP Collaboration.  This work was performed in part under DOE Contract DE-AC02-76SF00515.

\end{acknowledgements}

\FloatBarrier

\bibliographystyle{aa}
\bibliography{Lit}

\FloatBarrier

\appendix
\section{Case study}
\label{sec: Case study}
To illustrate the effect of microlensing on SNe Ia in detail we calculated
microlensed spectra and light curves. For this we assumed an
iPTF16geu-like \citep{Goobar:2016uuf} configuration, which means the source redshift $\sourcez=0.409$ and the lens redshift
$\lensz=0.216$. The redshifts are needed to calculate the size of a
pixel $\dmag=\SI{3.6e-6}{\parsec}=\SI{1.1e13}{\cm}$ of the
magnification map which corresponds to an Einstein radius of $\Rein=\SI{7.2e-3}{\parsec}=\SI{2.2e16}{\cm}$ and an angular scale of
  $\Rein/\source=\SI{6.5e-12}{\rad}=\SI{1.3e-6}{\arcsec}$. Since
we only determined absolute magnitudes and rest-frame fluxes in this section, we set
$z=0$ and $\lum=\angular=\SI{10}{\parsec}$ in Equations (\ref{eq:flux
  discrete for spectra})\footnote{$\dmag=\SI{1.1e13}{\cm}$ or the interpolated
  value to fulfill Equation (\ref{eq:criteria for bin size}) is used
  for $\Delta x_l$ and $\Delta y_m$.} and (\ref{eq:redshift time and
  wavelength}).

For this case study, we looked at two specific example realizations where
we placed a SNe Ia in two different positions of the magnification map
from Figure \ref{fig: microlensing map}, corresponding to image A of
iPTF16geu \citep{More:2016sys}.

\begin{figure*}[htbp]
\centering
\includegraphics[scale=0.4]{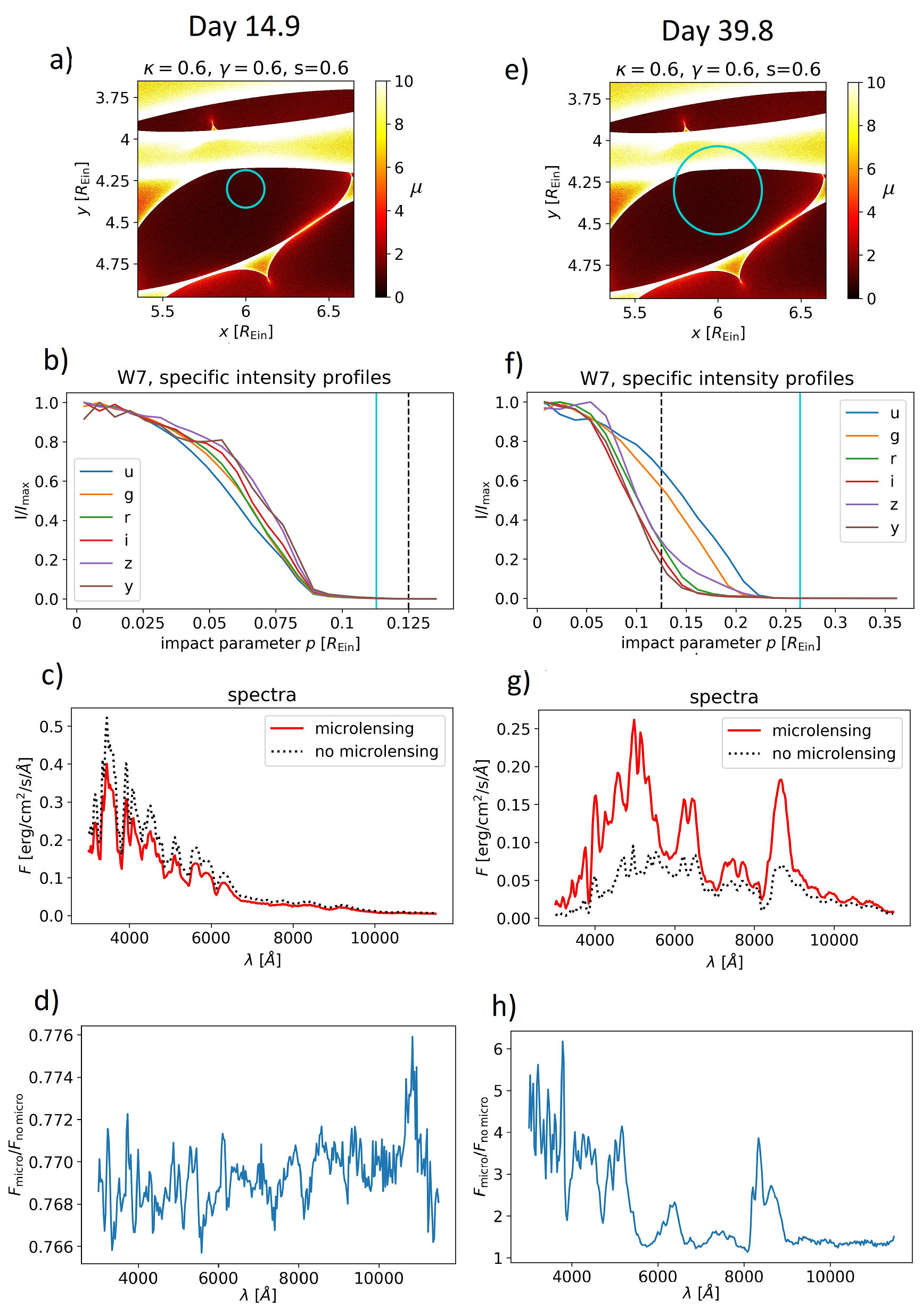}
\caption{Effect of
  microlensing on rest-frame spectrum of SN Ia centered at
  $(x,y)=(6.0, 4.3) \, \Rein$ in microlensing map from Figure \ref{fig:
    microlensing map} for two different rest-frame times since
  explosion $t$. The Einstein Radius defined in Equation \ref{basics: Einstein Radius physical coordinate in cm} is $\Rein = \SI{7.2e-3}{\parsec} = \SI{2.2e16}{\cm}$. For a discussion see Appendix \ref{sec: Case study}.}
\label{fig: micro non micro comparison non extreme case}
\end{figure*}
\begin{figure*}[htbp]
\centering

\subfigure{\includegraphics[scale=0.42]{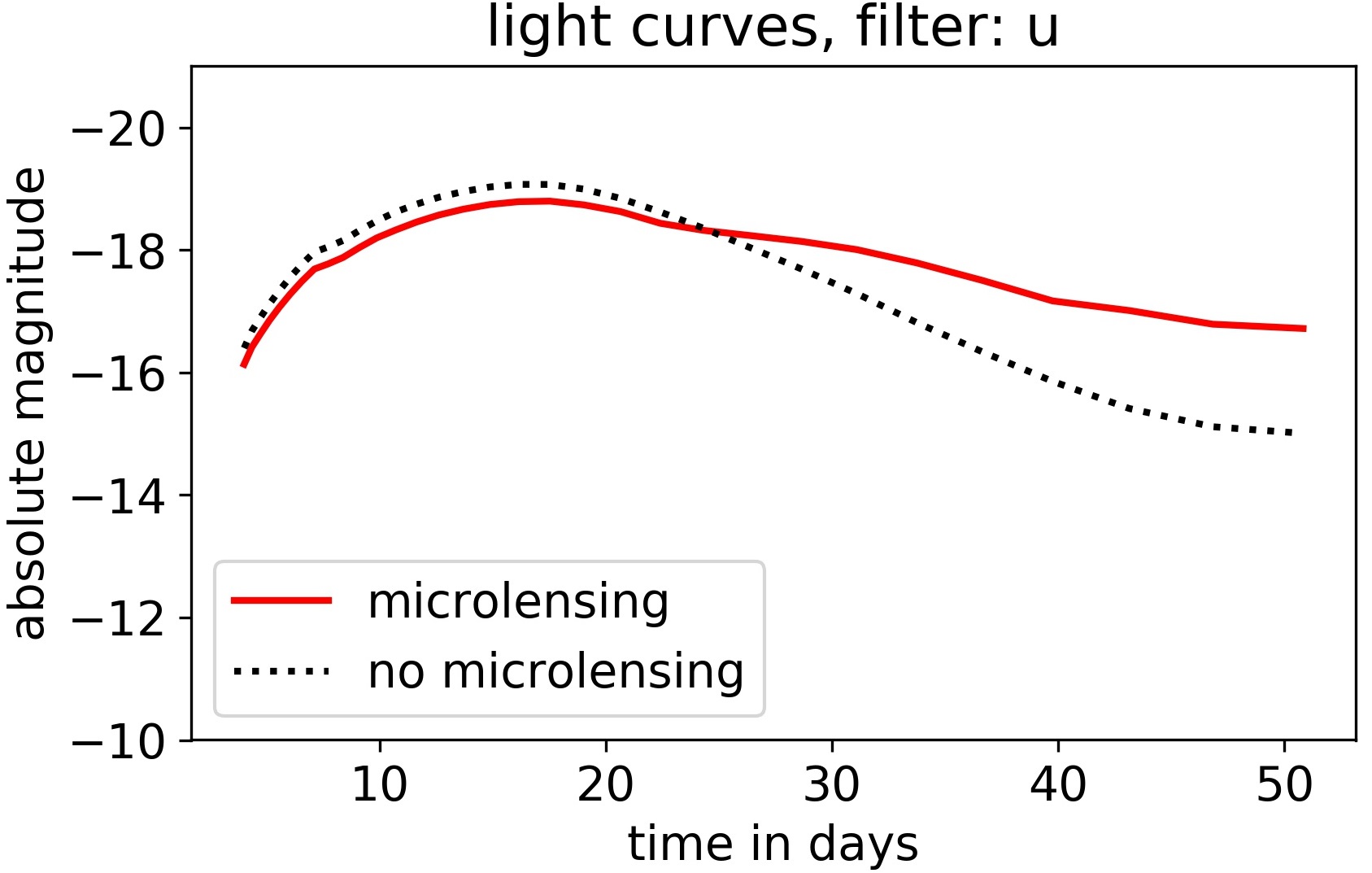}}
\subfigure{\includegraphics[scale=0.42]{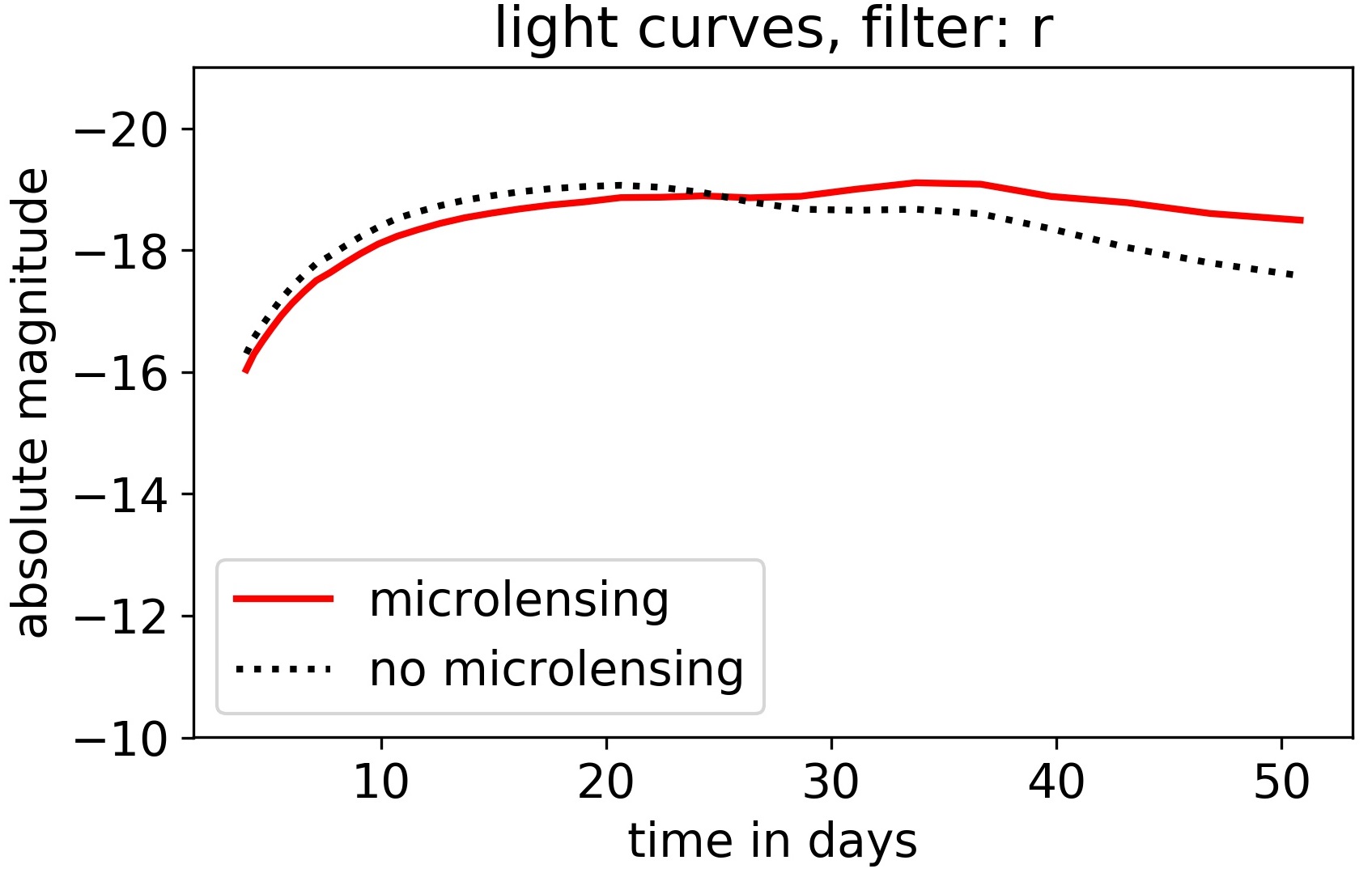}}
\subfigure{\includegraphics[scale=0.42]{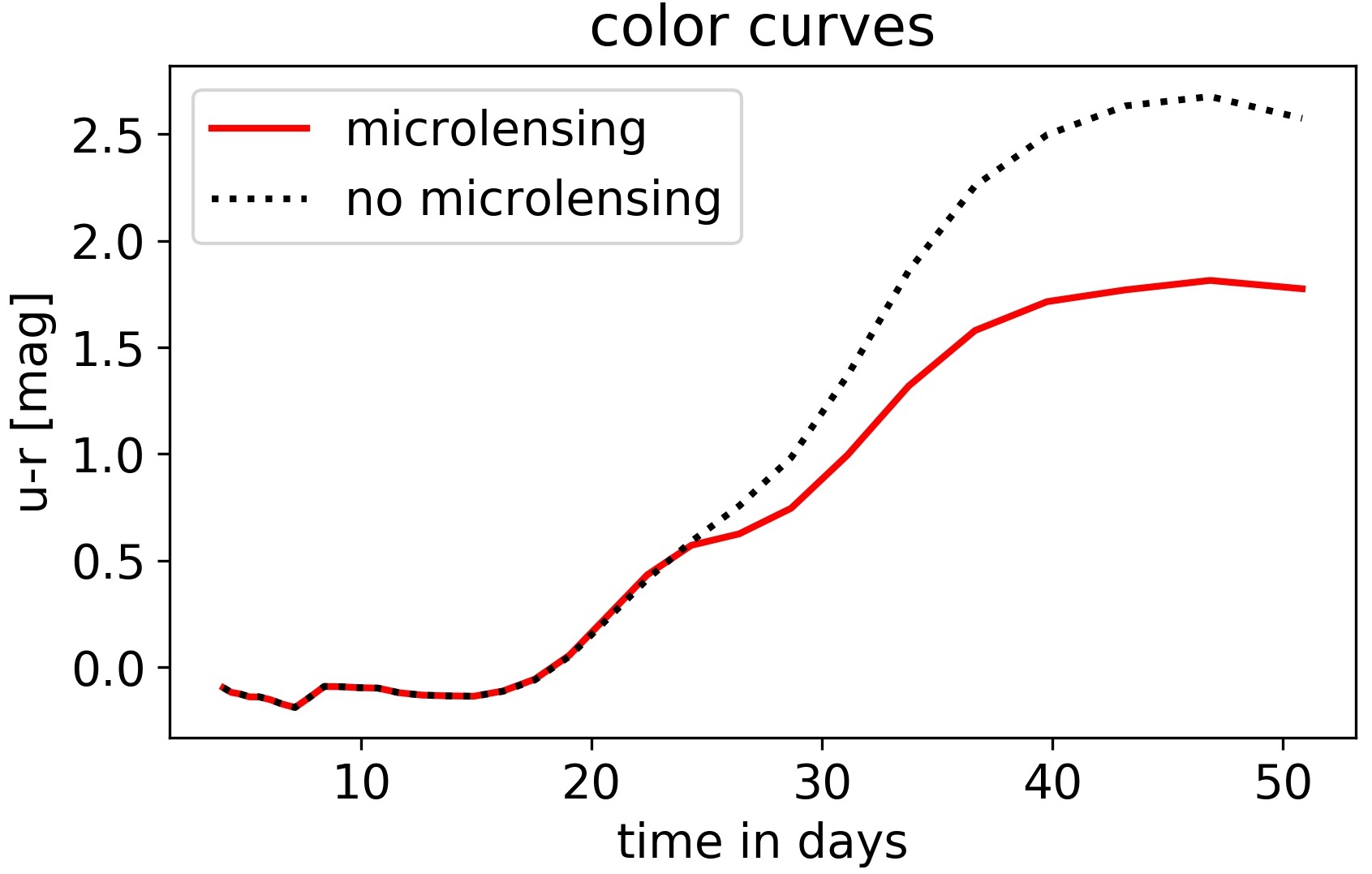}}
\caption{Influence of microlensing on two light curves and corresponding
  color curve for the SN shown in Figure \ref{fig: micro non micro comparison non extreme case}. As
  long as demagnification due to microlensing is similar in \textit{u} and
  \textit{r} bands, it cancels out in the color curve.}
\label{fig: micro influence on light curves and color curve non extreme case}
\end{figure*}
\begin{figure*}[htbp]
\centering
\includegraphics[scale=0.48]{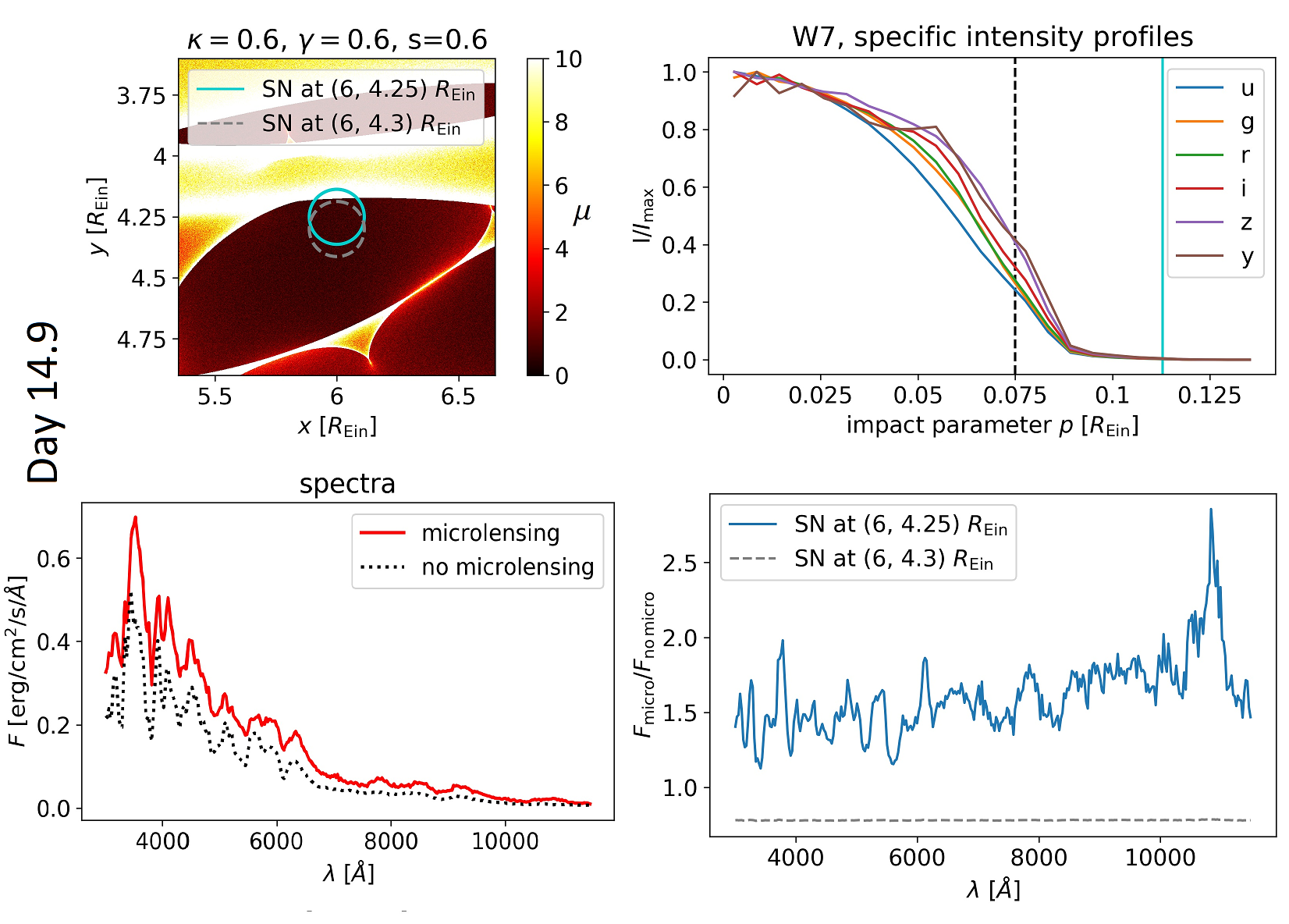}
\caption{Effect of
  microlensing on SN Ia spectrum at rest-frame time
  $t=\SI{14.9}{\day}$ after explosion, similar as in Figure \ref{fig:
    micro non micro comparison non extreme case} but at slightly
  different position of SN: $(x,y)=(6, 4.25) \, \Rein$, where $\Rein = \SI{7.2e-3}{\parsec} = \SI{2.2e16}{\cm}$. For a better comparison, the case from Figure \ref{fig: micro non micro comparison non extreme case} is shown as gray dashed line in the upper-left and lower-right panels.}
\label{fig: micro non micro comparison extreme case}
\end{figure*}

\begin{figure*}[htbp]
\centering
\subfigure{\includegraphics[scale=0.4]{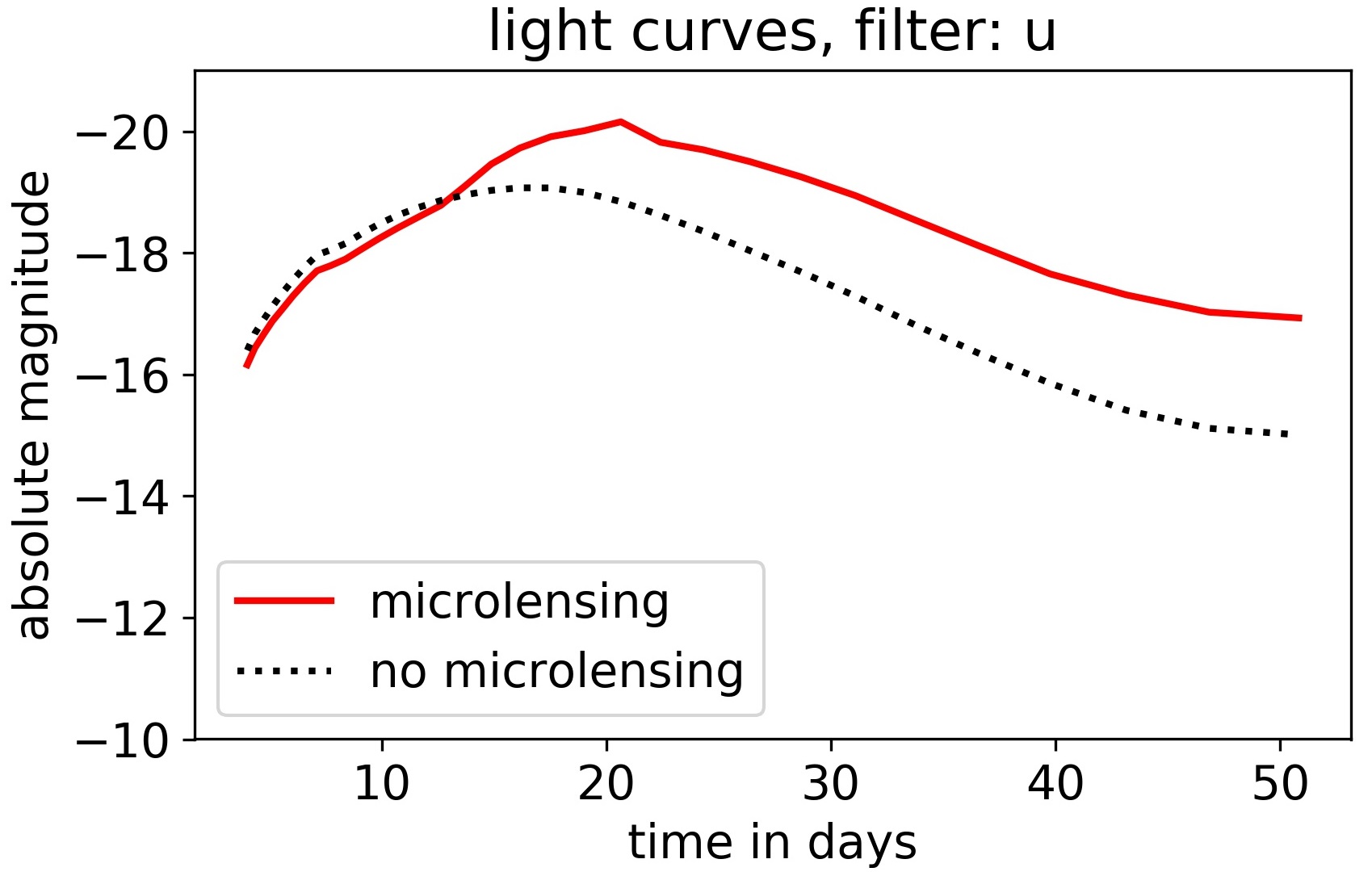}}
\subfigure{\includegraphics[scale=0.4]{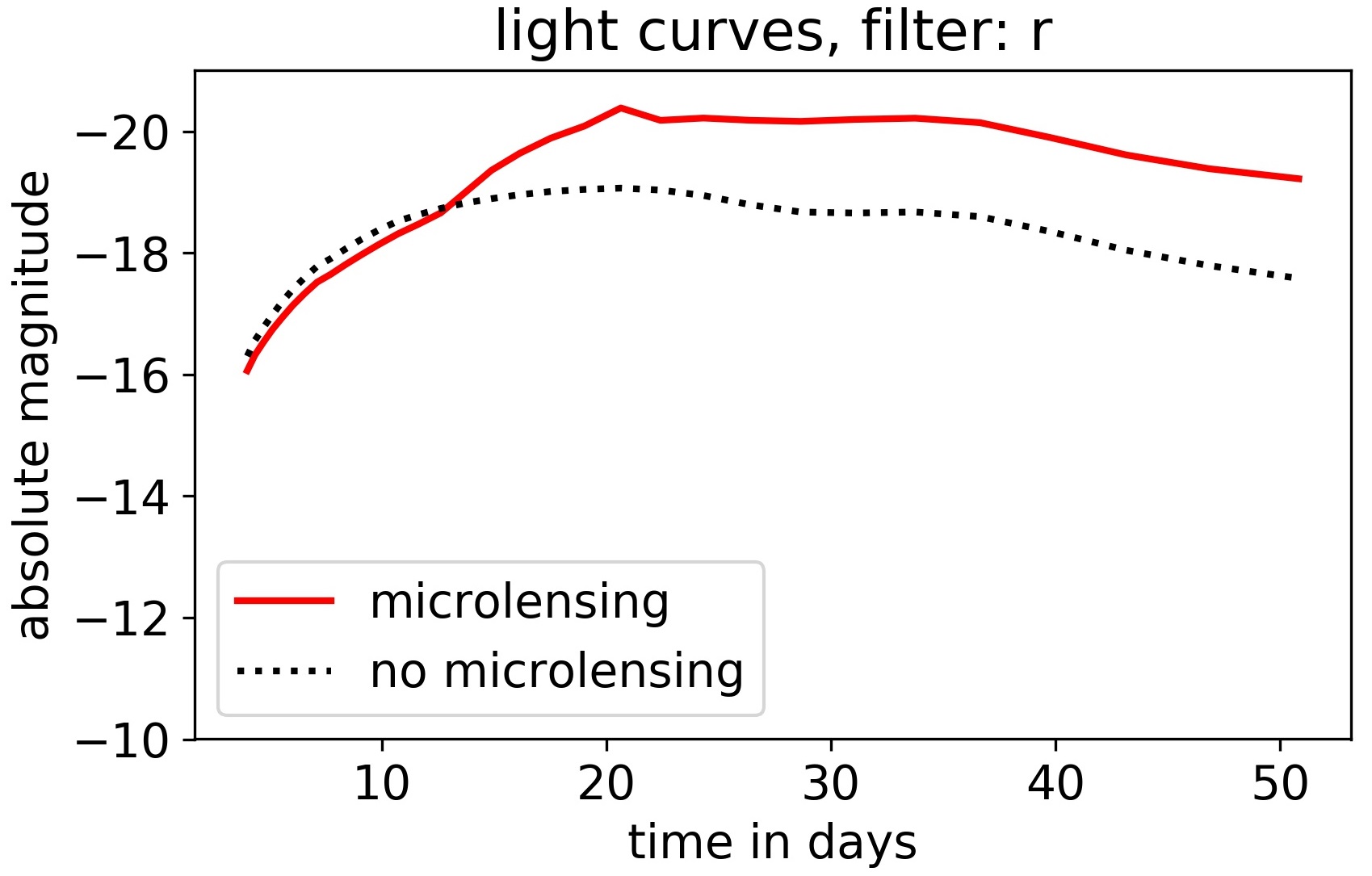}}
\subfigure{\includegraphics[scale=0.4]{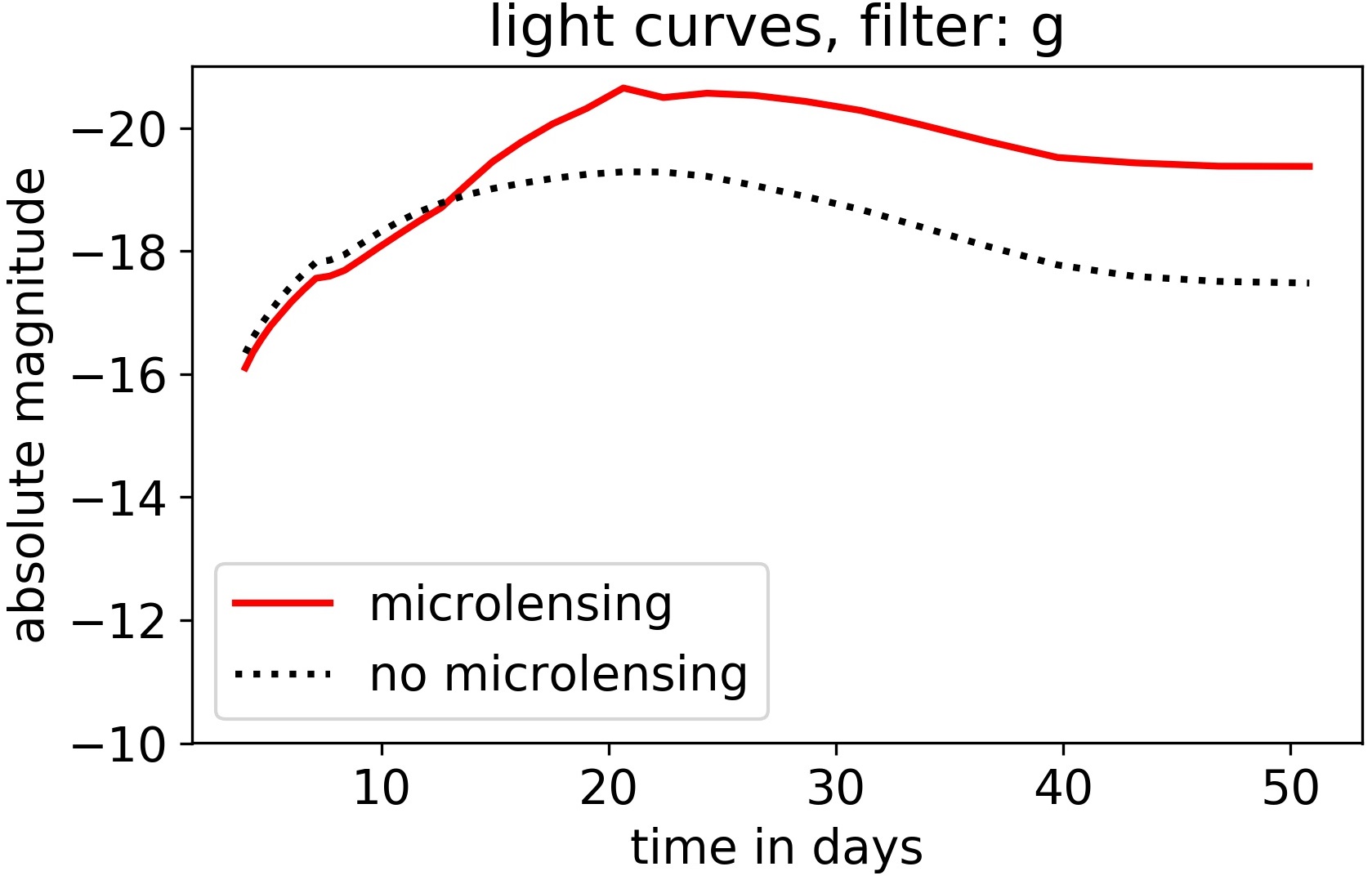}}
\subfigure{\includegraphics[scale=0.4]{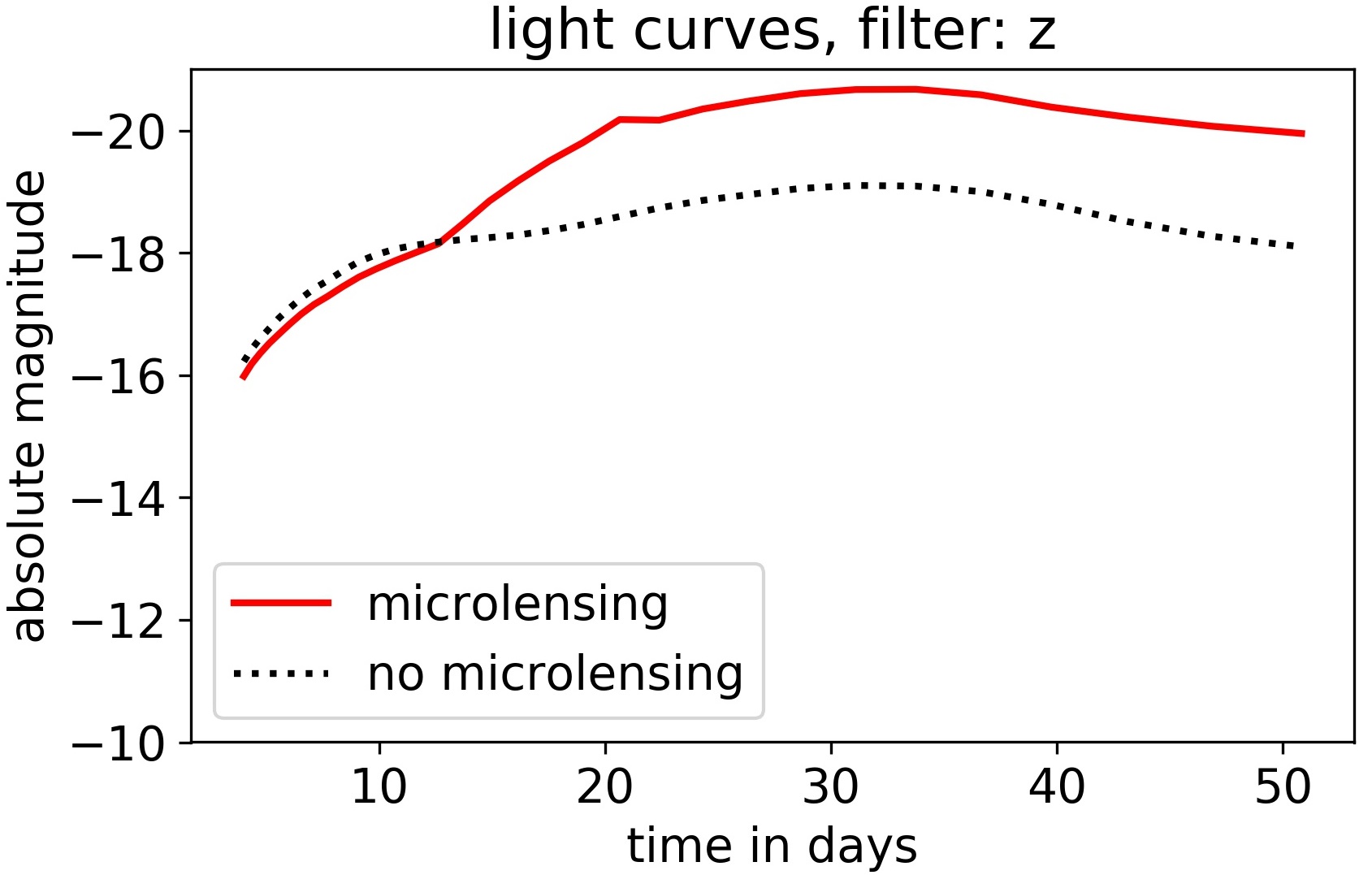}}
\subfigure{\includegraphics[scale=0.4]{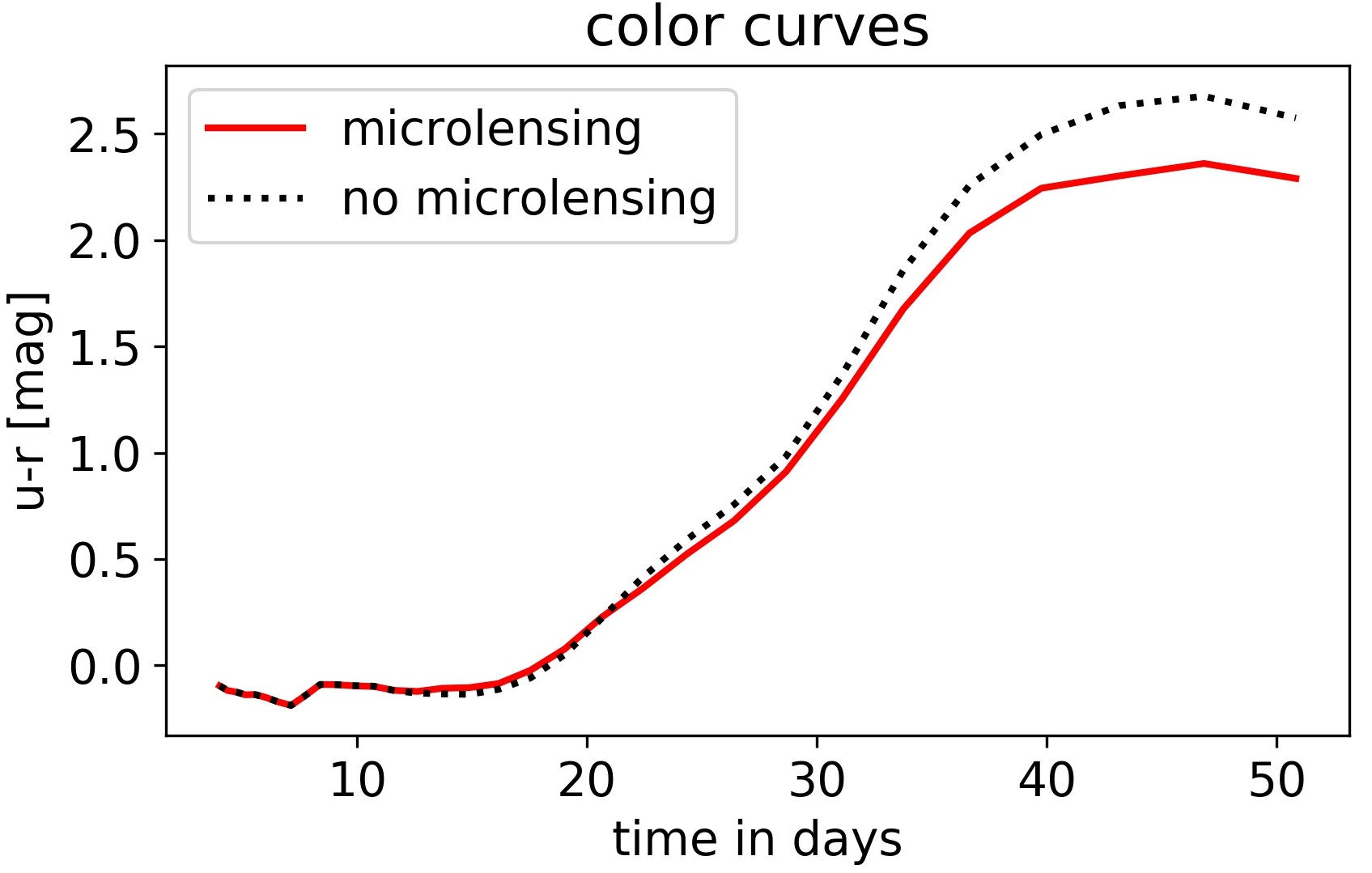}}
\subfigure{\includegraphics[scale=0.4]{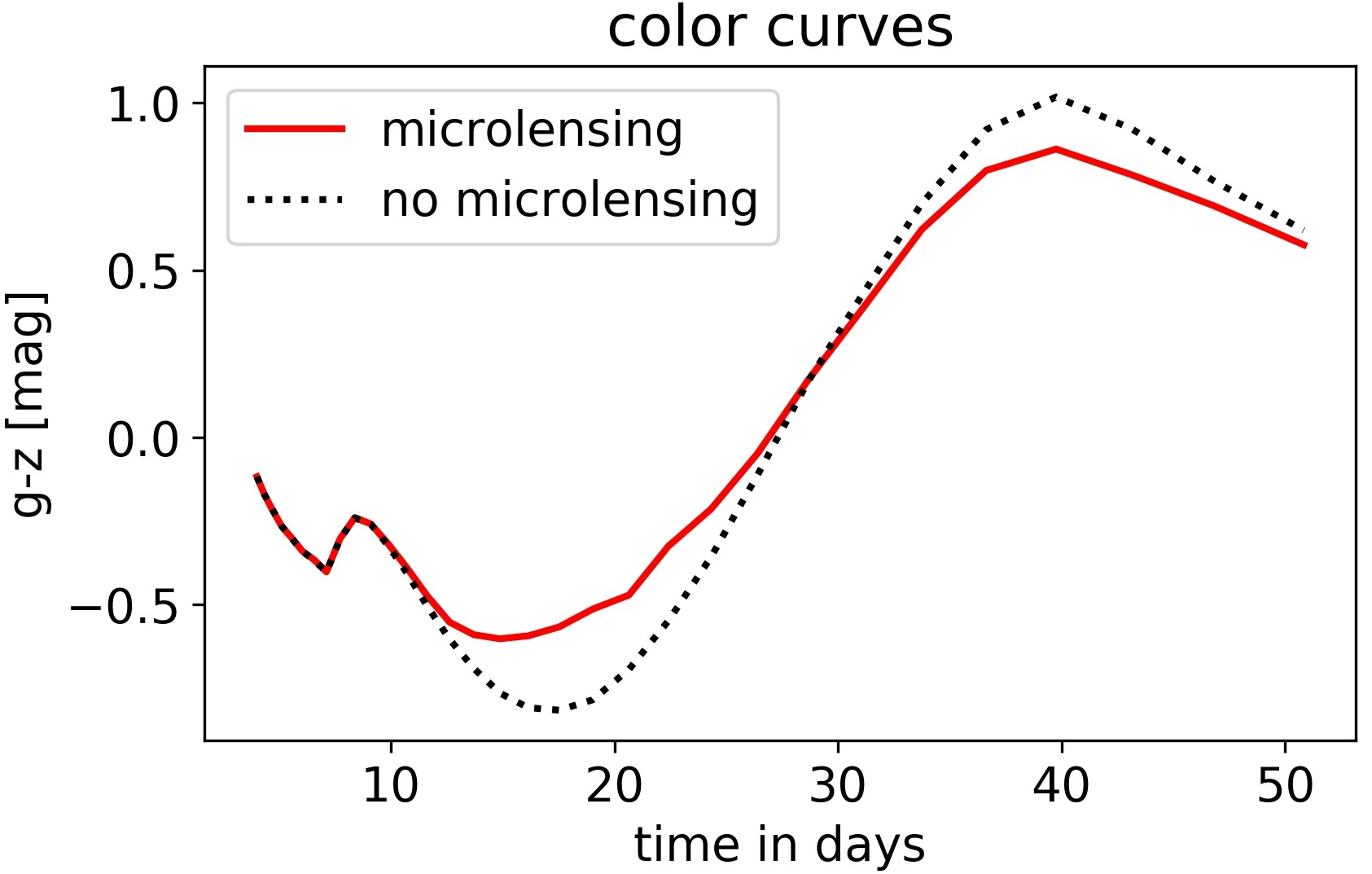}}
\caption{Influence of microlensing on light curves and color curves for
  SN shown in Figure \ref{fig: micro non micro comparison extreme case}. Whereas the light curves are
  highly influenced by microlensing, the color curve \textit{u-r} is very
  similar for the case of microlensing and non-microlensing. This is
  not the case for all color curves, as shown for example by \textit{g-z}.}
\label{fig: micro influence on light curves and color curve extreme case}
\end{figure*}

\begin{figure*}[htbp]
\centering
\includegraphics[scale=0.45]{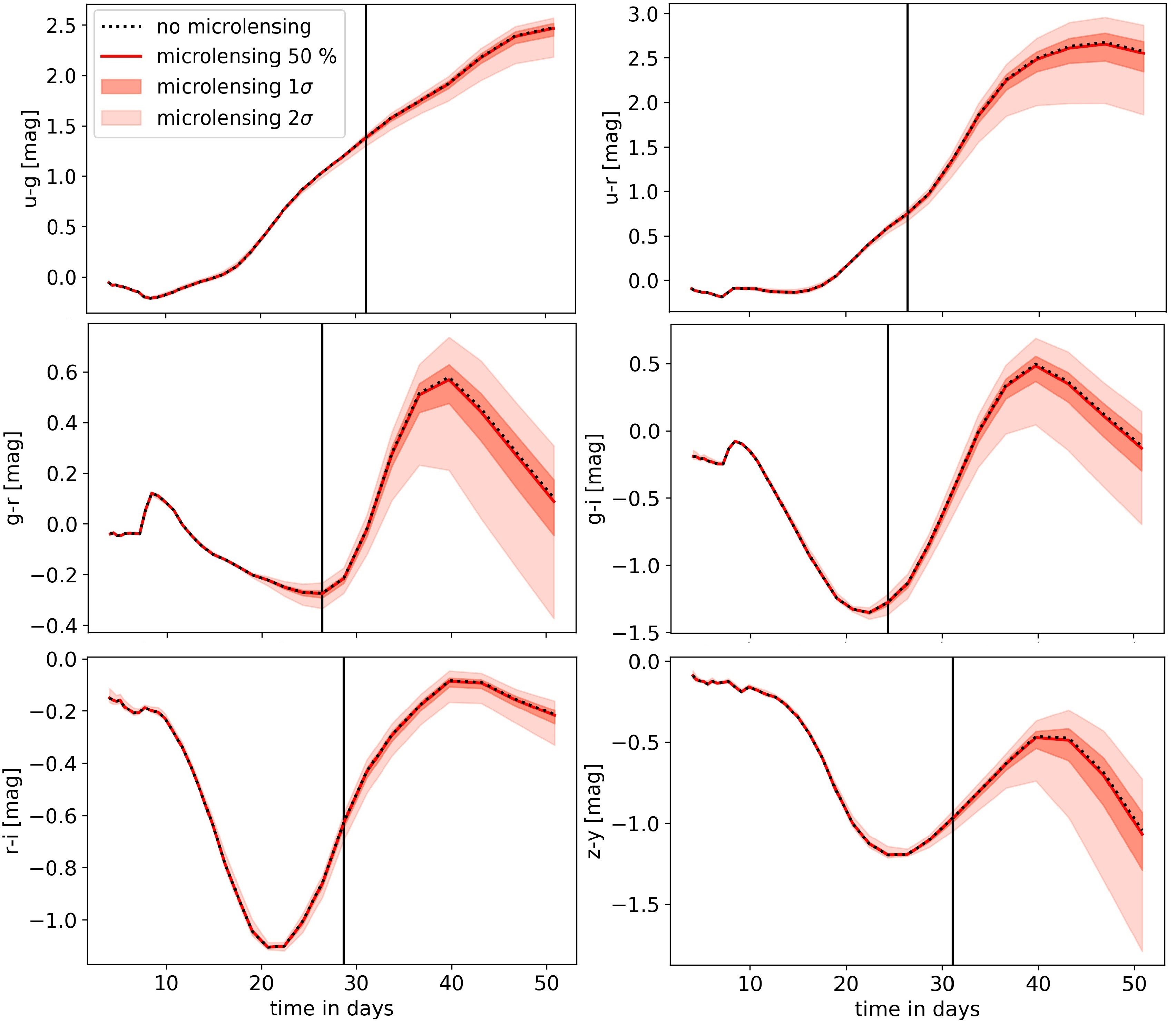}
\caption[Achromatic LSST rest-frame color curves comparing
microlensing and no microlensing for 10000 random SNe positions in a
magnification map corresponding to $\kappa=0.6$, $\gamma=0.6$, and
$s=0.6$.]{Comparison of non-microlensed color curves (dotted
  black) to
  microlensed ones (with median in solid red, and 1$\sigma$
  and 2$\sigma$ range in different shades), for 10000 random SNe
  positions in magnification 
  map. The vertical black line indicates the first time the 2$\sigma$ spread of the microlensed color curves exceeds 0.1 magnitudes. The panels are all rest-frame LSST color curves for a saddle image
  ($\kappa=0.6$, $\gamma=0.6$, and $s=0.6$, see Figure \ref{fig:
    microlensing map}), which show an achromatic phase similar to the
  one reported by \cite{Goldstein:2017bny}, but we find the achromatic
  phase only for combinations of the bands \textit{u, g, r,} and \textit{i} (except \textit{u-i}) and for the
  color curve \textit{z-y} up to approximately $25-30 $ days after explosion.}
\label{fig: color curves statistic achromatic}
\end{figure*}
\begin{figure*}[htbp]
\centering
\includegraphics[scale=0.9]{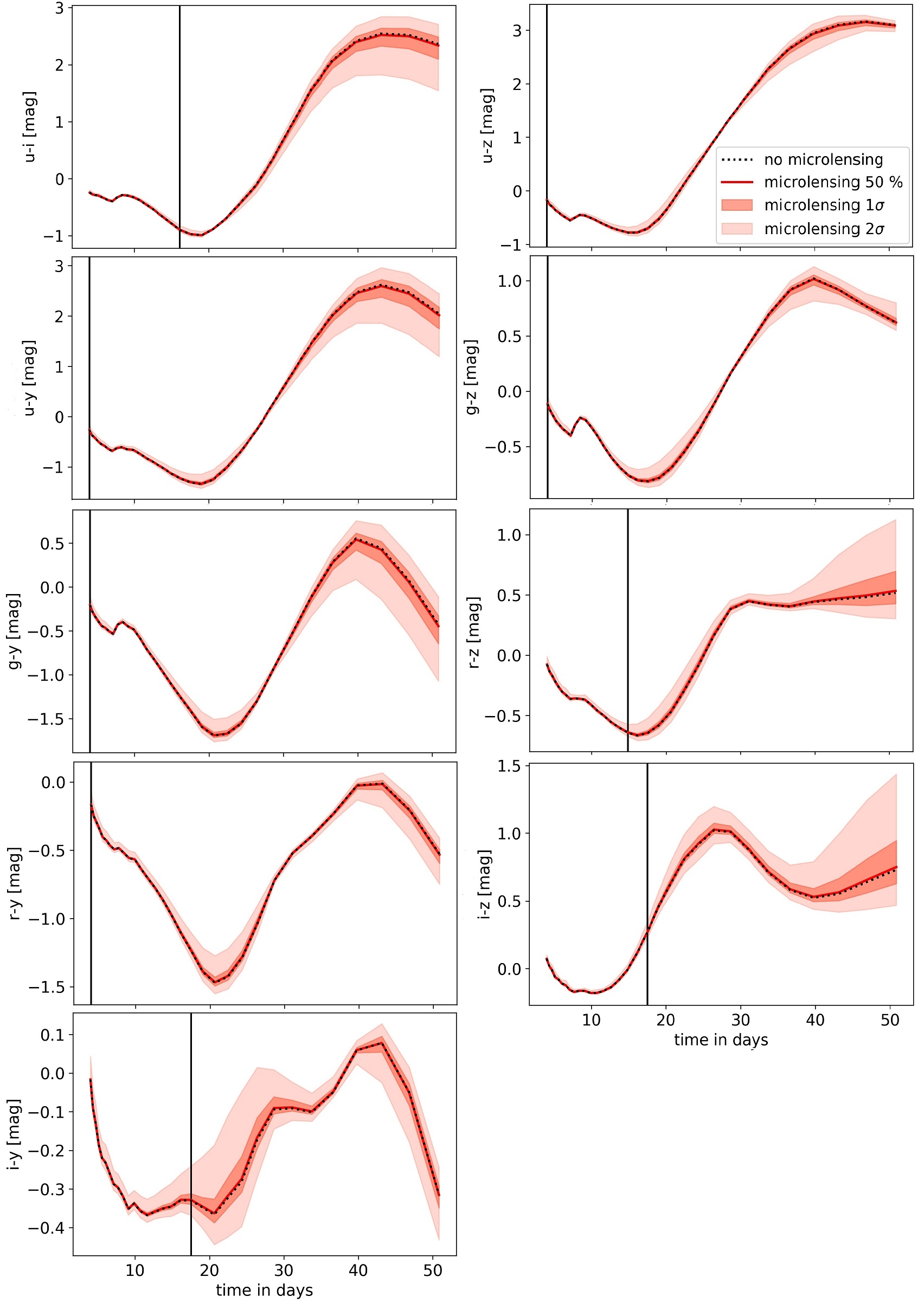}
\caption[Chromatic LSST rest-frame color curves comparing microlensing
and no microlensing for 10000 random SNe positions in a magnification
map corresponding to $\kappa=0.6$, $\gamma=0.6$, and $s=0.6$.]{Panels are produced
  similarly to those in Figure \ref{fig: color curves statistic
    achromatic}, except these colors curves exhibit 
  a shorter or non-existing
  achromatic phase in comparison to those in Figure
  \ref{fig: color curves statistic achromatic}.}
\label{fig: color curves statistic chromatic}
\end{figure*}

First we considered the position $(x,y) = (6.0, 4.3) \, \Rein$ and compared the
non-microlensed flux $F_{\lambda_j,\mathrm{o, cart},\mu=1}(t_i)$ with the
microlensed one $F_{\lambda_j,\mathrm{o,cart}}(t_i)$ for two different
instances in time as illustrated in Figure \ref{fig: micro non micro
  comparison non extreme case}. Panels a) to d) correspond to $t
= \SI{14.9}{\day}$ and e) to h) to $t = \SI{39.8}{\day}$. For both
times, the zoomed-in magnification map (panels a and e) from Figure
\ref{fig: microlensing map} is provided, with the position and
radius of the SN shown by a cyan circle. The radius is defined via the area of
the SN, which contains $99.9\%$ of the total projected specific intensity
$\sum_{j,l,m} I_{\lambda_j,\mathrm{e}}(t_i,x_l,y_m)$. In addition, the
normalized specific intensity profiles (panels b and f) are shown,
where the vertical cyan line corresponds to the radius of the SN and
the dashed black line marks the distance between the center of the SN
and the caustic in the magnification map which separates low and high
magnification regions. The normalized specific intensity of filter
band X is defined as
\begin{equation}
I_\mathrm{X,norm}=\frac{I_\mathrm{X}}{\mathrm{max}(I_\mathrm{X})},
\end{equation}
which corresponds to a radial radiation distribution for a given
filter X. Furthermore the fluxes for the cases with microlensing and
without (panels c and g) are shown together with their relative
strength (panels d and h).

For $t = \SI{14.9}{\day}$, the SN is completely in a homogeneous region
of demagnification as shown in panel a) of Figure \ref{fig: micro
  non micro comparison non extreme case} and therefore the flux is
demagnified by the same amount for all wavelengths, as can be seen in panel
c) and more clearly in panel d)\footnote{We note that the scale difference
  between panels d) and h) is a factor of 600.} independent of the
specific intensity profiles. For the later time, $t =
\SI{39.8}{\day}$, the SN has expanded further and crosses over a
caustic as visible in panel e), such that the outer region of the SN
is partly in a region of high magnification. From the specific
intensity profiles in panel f) we see that the outer ejecta region
emits relatively stronger in the bluer bands (\textit{u} and \textit{g}) than in the red ones (\textit{r,
i, z,} and \textit{y}), because for later days most of the Fe III has combined to Fe II, which is less transparent in the bluer bands than in the red ones \citep{Kasen:2006is,Goldstein:2017bny}. This explains the overall trend that the blue
part of the spectrum is more magnified than the red part, which is
indeed seen in panels g) and h).

For the case constructed in Figure \ref{fig: micro non micro comparison non extreme case} we see a significant impact on the light curves 
due to microlensing
as
shown in red in Figure \ref{fig: micro influence on light curves and color
  curve non extreme case} where the
light curves are highly distorted. For the \textit{u-r} color curve, the effect
of microlensing cancels out up to day 25. Afterward, the crossing of
the micro caustics, separating regions of low and high magnification, in combination with different spatial distributions
of the radiation in \textit{u} and \textit{r} band becomes important. This is an example
for the so-called ``achromatic phase" as reported by
\cite{Goldstein:2017bny}, who find that color curves up to day 20
after explosion are nearly independent of microlensing. They claim this is
due to the similar specific intensity profiles for early days and more
different ones at later days, as we can also see for our case
comparing panel b) and f) in Figure \ref{fig: micro non micro
  comparison non extreme case}.

For further investigation we construct another test case where the
caustic of the magnification map will be crossed during the achromatic
phase, as shown in Figure \ref{fig: micro non micro comparison extreme
  case}. Here the microlensing effect is clearly visible in the flux
ratio, although the specific intensity profiles are more similar as
for later days (compare panels b and f of Figure \ref{fig: micro non
  micro comparison non extreme case}). Also, the influence on the light
curves is visible earlier and more drastic as shown in Figure
\ref{fig: micro influence on light curves and color curve extreme
  case}. The light curves are highly distorted and peaks are 
shifted, which adds large uncertainty to the time-delay
measurement between different images based on light curves that undergo different
microlensing. Even though the \textit{u-r} color curve compensates microlensing
in early phases quite well and is therefore promising for measuring
time delays, this is not true for all color curves as shown for the
case of \textit{g-z}. Here, the microlensed and non-microlensed curves deviate
from each other even though they are in the achromatic phase. 

To explore this further, we
consider a large sample of 10000 random SN positions in the
magnification map shown in Figure \ref{fig: microlensing map}. For each
position, we calculate the light curves using Equation
(\ref{snmicro: light curves ab magnitudes}) and then calculate the color
curves. For each time bin $t_i$, we calculate from the sample the 50th
percentile as well as the 1$\sigma$ and 2$\sigma$ spread. The
results for all rest-frame LSST color curves are shown in Figures
\ref{fig: color curves statistic achromatic} and \ref{fig: color
  curves statistic chromatic}, where the vertical black line marks the time when the 2$\sigma$ spread is the first time beyond $0.1 \mathrm{mag}$. We find the general trend that the achromatic phase in the color curves becomes shorter the further the different bands are apart. As in
\cite{Goldstein:2017bny}, we find an achromatic phase-like behavior until 25 to 30 
days after explosion, but only for rest-frame color curves containing
combinations of \textit{u, g, r,} or \textit{i} bands (except \textit{u-i}) or the color curve \textit{z-y} (Figure
\ref{fig: color curves statistic achromatic}). As soon as we combine
one of the bands \textit{u, g, r,} or \textit{i} with \textit{z} or \textit{y} we see the influence of
microlensing earlier (Figure \ref{fig: color curves statistic
  chromatic}). This behavior can be explained by looking at the
normalized specific intensity profiles for early times as shown in
panel b) of Figure \ref{fig: micro non micro comparison non extreme
  case}: The profiles for the outer region (pixel 150 to 200) are
similar for filters \textit{z} and \textit{y}, but different from \textit{u, g, r,} and \textit{i}. Since
the achromatic phase depends highly on the specific intensity
profiles, the investigation of different explosion models is necessary
to explore this further (S. Huber et al., in preparation). 

In addition to the different durations of the
achromatic phase for the various color curves, we note that some of
our color curves from {\tt ARTIS} are different in shape from those of 
SEDONA in \cite{Goldstein:2017bny}. It is also very important to
emphasize that our results in this section are for rest-frame color
curves, which means that different color curves will be more or less
useful depending on the redshift of the source.

%

\section{Photometric uncertainty of LSST}
\label{sec:Appendix LSST uncertainty}

The photometric uncertainty $\sigma_1$ from Equation \ref{eq:noise realization random mag including error LSST science book} is defined as:
\begin{equation}
\sigma_1^2 = \sigma_\mathrm{sys}^2+\sigma_\mathrm{rand}^2$,
where $\sigma_\mathrm{sys}=0.05
\end{equation}
and 
\begin{equation}
\sigma_\mathrm{rand}^2=(0.04-\gamma^c) x + \gamma^c x^2 (\mathrm{mag}^2).
\end{equation}
The parameter $\gamma^c$ varies from 0.037 to 0.040 for different filters and
$x=10^{4(m-m_5)}$, where $m$ is the magnitude of the SN data
point and $m_5$ is the 5$\sigma$ point-source depth (for more details see \cite{2009:LSSTscience}, Sec. 3.5, p. 67).

\section{Optimistic estimate of the number of LSNe Ia}
\label{sec:Appendix optimistic estimate of the number of LSNe Ia with well measured delay}
The numbers presented Table \ref{tab: total number of LSNe Ia from OM 10} are based on the prediction of OM10 which depends on a sharp magnitude cut, allowing only for systems where the \textit{i}-band peak magnitude of the fainter image for a double or the 3rd brightest image for a quad is brighter than $\SI{22.6}{\mag}$. In our investigation, assuming the W7 model with random microlensing magnifications, we investigated also fainter systems where we find that we can also get well measured time delays for these systems although the fraction is reduced by at least a factor of 1.7. This fraction for faint systems is an overestimation, because bright systems yield better time delay measurements than faint systems and our investigated sample peaks around the cut applied in OM10 and therefore we probe mostly the bright regime of the systems fainter than $\SI{22.6}{\mag}$.

However results presented by \cite{Wojtak:2019hsc} suggest that we can find approximately 440 LSNe Ia over the 10 year survey with the same approach as in OM10 but allowing also for fainter systems. If we assign about 80 of those systems to bright ones as suggested by Table \ref{tab: total number of LSNe Ia from OM 10}, we expect to find 4.5 times more faint ($> \SI{22.6}{mag}$) systems than the bright ($\le \SI{22.6}{mag}$) ones. Further, we calculate with Equation (\ref{eq:res:LSST DATA only}) the fraction of systems with well measured delay, but separately for faint and bright systems, leading to $f_\mathrm{total,faint}$ and $f_\mathrm{total,bright}$. Therefore an optimistic estimate of the number of LSNe Ia with well measured time delay for a given cadence can be calculated as
\begin{equation}
\scalebox{0.9}{$N_\mathrm{GoodDelay,cad} =  N_\mathrm{LSNeIa,cad} f_\mathrm{total,bright} +  4.5 N_\mathrm{LSNeIa,cad} f_\mathrm{total,faint}$,}
\end{equation} 
where $N_\mathrm{LSNeIa,cad}$ are the numbers from Table \ref{tab: total number of LSNe Ia from OM 10}. The results are shown in Figure \ref{fig:LSST only and LSST+follow-up, total number with good delays, optimistic estimate}. Despite the fact that the numbers are a factor of approximately 3.5 higher than in Figure \ref{fig:LSST only and LSST+follow-up, total number with good delays}, we see the same general trend for magenta, orange, and blue cadence strategies, leaving the overall conclusions on cadence strategies unchanged.

\begin{figure*}[h!]
\centering
\includegraphics[width=0.6\textwidth]{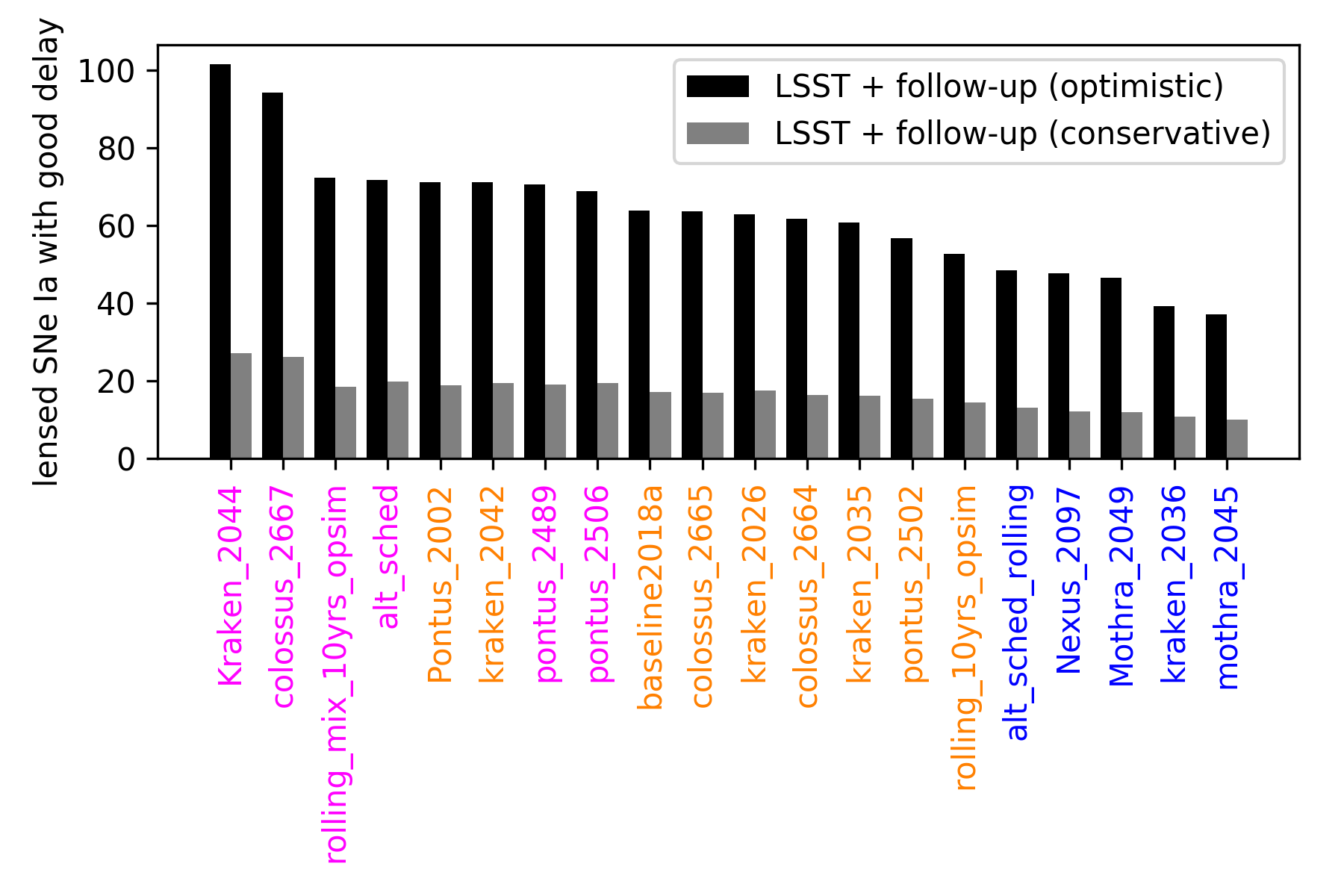}
\caption{Number of LSNe Ia with well measured time delay for 10-year survey 
including faint systems with \textit{i}-band peak magnitude (of fainter image in doubles, or third brightest image in quads) fainter than 22.6 mag (black bars), in comparison to results presented in Figure \ref{fig:LSST only and LSST+follow-up, total number with good delays} (gray bars). We see that the optimistic (black bars) and conservative (gray bars) estimates show the same trend for magenta, orange and blue cadence strategies.}
\label{fig:LSST only and LSST+follow-up, total number with good delays, optimistic estimate}
\end{figure*}





\end{document}

%% file: LSST_cadence_macro.tex
\newcommand{\bd}{\begin{displaymath}}
\newcommand{\ed}{\end{displaymath}}
\newcommand{\be}{\begin{equation}}
\newcommand{\ee}{\end{equation}}
\newcommand{\beaa}{\begin{eqnarray*}}
\newcommand{\eeaa}{\end{eqnarray*}}
\newcommand{\bea}{\begin{eqnarray}}
\newcommand{\eea}{\end{eqnarray}}

\newcommand{\erg}{\mathrm{erg}}
\newcommand{\s}{\mathrm{s}}

\newcommand{\cm}{\mathrm{cm}}

\newcommand{\lens}{D_\mathrm{d}}
\newcommand{\source}{D_\mathrm{s}}
\newcommand{\ls}{D_\mathrm{ds}}
\newcommand{\sourcez}{z_\mathrm{S}}
\newcommand{\lensz}{z_\mathrm{L}}
\newcommand{\lum}{{D_\mathrm{L}}}
\newcommand{\angular}{D_\mathrm{A}}

\newcommand{\dmag}{\Delta d_\mathrm{mag}}

\newcommand{\Rein}{R_\mathrm{Ein}}
\newcommand{\thetaein}{\theta_\mathrm{Ein}}
\newcommand{\matter}{\Omega_{\mathrm{m},0}}
\newcommand{\vac}{\Omega_{\Lambda,0}}

\newcommand{\dd}{\mathrm{d}}
\newcommand{\ds}{\,\mathrm{d}}

\newcommand{\specificE}{I_{\lambda,\mathrm{e}}}

\DeclareSIUnit\parsec{pc}
\DeclareSIUnit\lightyear{ly}
\DeclareSIUnit\year{yr}
\DeclareSIUnit\erg{erg}
\DeclareSIUnit\ster{ster}
\DeclareSIUnit\arcsec{arcsec}
\DeclareSIUnit\rad{rad}
\DeclareSIUnit\mag{mag}

\newcommand{\nickel}{$^{56}$Ni}
\newcommand{\cobalt}{$^{56}$Co}
\newcommand{\iron}{$^{56}$Fe}

\usepackage{color} 
\definecolor{gruen}{cmyk}{0.35,0.01,0.80,0.1}

\definecolor{blue}{rgb}{0.0,0.0,1.0}
\definecolor{magenta}{rgb}{1.0,0.0,1.0}
\definecolor{orange}{rgb}{1.0,0.5,0.0}

\newcommand{\altsched}{\texttt{\textcolor{magenta}{alt\_sched}}} 
\newcommand{\altschedrolling}{\texttt{\textcolor{blue}{alt\_sched\_rolling}}}
\newcommand{\baseline}{\texttt{\textcolor{orange}{baseline2018a}}}
\newcommand{\colossusfour}{\texttt{\textcolor{orange}{colossus\_2664}}}
\newcommand{\colossusfive}{\texttt{\textcolor{orange}{colossus\_2665}}}
\newcommand{\colossusseven}{\texttt{\textcolor{magenta}{colossus\_2667}}}
\newcommand{\krakentwosix}{\texttt{\textcolor{orange}{kraken\_2026}}}
\newcommand{\krakenfive}{\texttt{\textcolor{orange}{kraken\_2035}}}
\newcommand{\krakenthreesix}{\texttt{\textcolor{blue}{kraken\_2036}}}
\newcommand{\krakentwo}{\texttt{\textcolor{orange}{kraken\_2042}}}
\newcommand{\krakenfour}{\texttt{\textcolor{magenta}{Kraken\_2044}}}
\newcommand{\mothrafive}{\texttt{\textcolor{blue}{mothra\_2045}}}
\newcommand{\mothranine}{\texttt{\textcolor{blue}{Mothra\_2049}}}
\newcommand{\nexusseven}{\texttt{\textcolor{blue}{Nexus\_2097}}}
\newcommand{\pontuszerozerotwo}{\texttt{\textcolor{orange}{Pontus\_2002}}}
\newcommand{\pontusfivezerotwo}{\texttt{\textcolor{orange}{pontus\_2502}}}
\newcommand{\pontusnine}{\texttt{\textcolor{magenta}{pontus\_2489}}}
\newcommand{\pontusfivezerosix}{\texttt{\textcolor{magenta}{pontus\_2506}}}
\newcommand{\rollingmixopsim}{\texttt{\textcolor{magenta}{rolling\_mix\_10yrs\_opsim}}}
\newcommand{\rollingopsim}{\texttt{\textcolor{orange}{rolling\_10yrs\_opsim}}}